\documentclass[12pt]{article}
\usepackage{graphicx}
\usepackage{natbib} 
\usepackage{url} 
\usepackage{subfig}
\usepackage{bm}
\usepackage{amsmath,amssymb,amsfonts}

\usepackage{algorithm}
\usepackage{algpseudocode}
\usepackage{setspace}
\usepackage{cancel}

\usepackage{booktabs} 
\usepackage[table,xcdraw]{xcolor}

\newcommand{\blind}{0}

\begin{document}

\bibliographystyle{Perfect}

\def\spacingset#1{\renewcommand{\baselinestretch}%
{#1}\small\normalsize} \spacingset{1}


\if0\blind
{
    \title{\bf Block Vecchia Approximation for Scalable and Efficient Gaussian Process Computations}
    \author{Qilong Pan$^{1}$, Sameh Abdulah$^2$, Marc G. Genton$^{1}$, Ying Sun$^{1}$ 
    \thanks{
    The authors gratefully acknowledge \textit{please remember to list all relevant funding sources in the unblinded version}}\hspace{.2cm}\\
    \\
     $^1$Statistics Program,\\
     $^2$Applied Mathematics and Computational Sciences Program, \\ 
    King Abdullah University of Science and Technology (KAUST),\\ Thuwal, Saudi Arabia.
    }
  \maketitle
} \fi

\if1\blind
{
  \bigskip
  \bigskip
  \bigskip
  \begin{center}
    {\LARGE\bf Title}
\end{center}
  \medskip
} \fi

\bigskip
\begin{abstract}
Gaussian Processes (GPs) are vital for modeling and predicting irregularly-spaced, large geospatial datasets. However, their computations often pose significant challenges in large-scale applications. One popular method to approximate GPs is the Vecchia approximation, which approximates the full likelihood via a series of conditional probabilities. The classical Vecchia approximation uses univariate conditional distributions, which leads to redundant evaluations and memory burdens. To address this challenge, our study introduces block Vecchia, which evaluates each multivariate conditional distribution of a block of observations, with blocks formed using the K-means algorithm. The proposed GPU framework for the block Vecchia uses varying batched linear algebra operations to compute multivariate conditional distributions concurrently, notably diminishing the frequent likelihood evaluations. Diving into the factor affecting the accuracy of the block Vecchia, the neighbor selection criterion is investigated, where we found that the random ordering markedly enhances the approximated quality as the block count becomes large. To verify the scalability and efficiency of the algorithm, we conduct a series of numerical studies and simulations, demonstrating their practical utility and effectiveness compared to the exact GP. Moreover, we tackle large-scale real datasets using the block.
Vecchia method, i.e., high-resolution 3D profile wind speed with a million points. 

(Code: https://github.com/paper-code1/BV-Gaussian)

\end{abstract}

\noindent%
\vfill

\newpage
\spacingset{2} 

\section{Introduction}

Gaussian Process (GP) is an essential tool in spatial statistics, widely used for modeling and predicting geospatial data. Nonetheless, there is a major challenge when handling large datasets from irregularly distributed locations.
\textcolor{black}{
Indeed, assume there are $n$ locations ${\bf s}_1, \ldots, {\bf s}_n \in \mathbb{R}^d$,  where $d$ is the dimension of locations. 
The GPs are specified by a mean function and a covariance function, $GP(\mu({\bf s}), C_{\bm \theta}(\cdot, \cdot))$, where $C_{\bm \theta}$ is usually assumed to have a parametric form and $\bm \theta \in \mathbb R^p$. 
Let $y_i:=y\left({\bf s}_i\right) \in \mathbb{R}$ represent single observation at location ${\bf s}_i$ and denote the observed data vector by $\boldsymbol{y} = \left(y_1, \ldots, y_n\right)^\top$. The data $\boldsymbol{y}$ can be modeled using GPs as $\boldsymbol{y} \sim \mathcal{N}_n\left(\bm \mu, \bm \Sigma_{\boldsymbol\theta}\right)$, where $\bm \Sigma_{\boldsymbol\theta}$ is a covariance matrix with $(i, j)$ entry determined by a given covariance function $C_{\boldsymbol\theta}\left(
{\bf s}_i, {\bf s}_j
\right)$.
}
Without loss of generality, we assume that $\boldsymbol{y}$ has a mean of zero. Statistical inference about $\boldsymbol{\theta}$ often relies on the Gaussian log-likelihood function
\begin{equation}
	\label{eq:likeli}
 \ell({\boldsymbol\theta};\boldsymbol{y})=-\frac{n}{2}\log(2\pi) - \frac{1}{2}\log |{{\boldsymbol \Sigma}_{\boldsymbol\theta}}|-\frac{1}{2}{\boldsymbol{y}}^\top {\boldsymbol \Sigma}_{\boldsymbol\theta}^{-1}{\boldsymbol{y}},
\end{equation}
where the Cholesky decomposition of $\boldsymbol\Sigma_{\boldsymbol\theta}$ imposes memory burden $\mathcal{O}(n^2)$ and computational cost $\mathcal{O}(n^3)$, e.g., the covariance matrix of 200K locations require 160GB memory and 2.6 Petabyte (PB)  flops.

Various studies have focused on the computational and memory challenges of modeling and predicting using large-scale GPs. These efforts primarily explore two strategies: sparse approximation and low-rank approximation of the covariance matrix \citep{furrer2006covariance,kaufman2008covariance,bevilacqua2016covariance}. For instance, sparse approximation techniques such as covariance tapering have been widely studied. This method applies a tapering function that diminishes to zero with increasing distance between two points to the covariance function. This function transforms the original dense covariance matrix into a sparse format, thus reducing the computations and memory burden.
\textcolor{black}{
In low-rank approximation, the full covariance matrix is approximated with two matrices of lower rank, which is achieved by $\bf A \approx UV^\top$, where $\bf U$ and $\bf V$ are matrices with fewer columns than $\bf A$. This reduces the rank of the matrix to the number of columns in $\bf U$ and 
$\bf V$, thus significantly reducing the size of the matrix operations involved \citep{huang2018hierarchical, mondal2023tile}.
Additionally, modern hardware technologies capable of low-precision calculations, such as NVIDIA GPUs, have enhanced the optimization of the sparse covariance matrix.} This improvement is achieved by assigning different precision levels to various sections of the dense covariance matrix, thus reducing computational complexity rather than completely omitting these sections \citep{abdulah2019geostatistical,abdulah2021accelerating,cao2022reshaping}. For low-rank approximations, various methods are employed to enable faster computations and reduced memory compared to the original dense matrix \citep{katzfuss2011spatio,huang2018hierarchical,abdulah2018parallel, mondal2022parallel}.

The Vecchia approximation is among the earliest statistical methods for approximating GPs. It approximates the joint distribution of a GP by decomposing it into a product of independent univariate conditional distributions\citep{vecchia1988estimation}. This approach reduces computational demands and memory burden by using only a limited number of neighboring points in each univariate conditional distribution, improving speed and lowering memory requirements. 
The approximated log-likelihood has a computational complexity of $\mathcal{O}(nm^3)$ and memory complexity of $\mathcal{O}(nm^2)$, compared to the exact $\mathcal{O}(n^3)$ and $\mathcal{O}(n^2)$. Here, $n$ is the total number of spatial locations, and $m$ is the number of neighbors included in the conditioning set ($m \ll n$). 
Nonetheless, the scalability of the Vecchia approximation encounters two primary hurdles. Firstly, the decomposition in univariate conditional distributions may lead to the redundant computation of multiple conditional log-likelihoods, particularly when they involve common neighbors. This redundancy not only increases computational overhead but also decreases the efficiency of processing. Secondly, the main computations in the Vecchia approximation are small matrix operations, which are better suited for execution on CPUs than Graphics Processing Units (GPUs). More importantly, the performance of GPUs cannot be fully utilized for conducting the small matrix operations\citep{pan2024gpuaccelerated}. GPUs are designed for handling computationally intensive tasks that benefit from extensive parallelization, such as matrix-to-matrix multiplications. However, the tasks arising from the Vecchia approximation which is characterized by numerous small and discrete operations are not inherently compatible with the GPU architecture's strengths. For a detailed review of the Vecchia approximation, refer to\citep{pan2024gpuaccelerated, katzfuss2022scaled, zhang2022multi,   guinness2021gaussian, guinness2018permutation, katzfuss2021general, zhang2021fixed, jimenez2023scalable}.

In this work, we introduce a block version of Vecchia approximation using the batched GPU framework, i.e., block Vecchia algorithm, 
\textcolor{black}{
where the full likelihood is approximated by the product of a series of multivariate conditional probabilities represented by blocks and their neighbors, and modern GPU architectures are used to accelerate the proposed algorithm. 
These blocks are created by the K-means algorithm and the classic Vecchia can be viewed as a special case when each block size is 1.
Technically, built upon the MAGMA library \citep{dong2016magma}, the enhancement in our GPU framework lies in applying varying batched matrix operations, which execute the computationally light tasks, e.g., the Cholesky decomposition of the small covariance matrix, in parallel on a single GPU by creating multiple block threads. The batched operations accelerate the algorithm and the consideration of multivariate conditional probability creates computationally intensive tasks, which makes the compute node in GPU more efficient compared to univariate conditional cases in the classic Vecchia.}  
Additionally, further investigation establishes that the block Vecchia approximation algorithm significantly enhances the capability to manage larger problem sizes, outperforming the classic Vecchia algorithm's scalability. In the end, through comprehensive evaluations encompassing numerical studies, simulation experiments, and analyses of real datasets \textcolor{black}{in the context of the Mat\'ern covariance function considered in the GP modeling}, key findings have been identified, underscoring the efficiency and effectiveness of the block Vecchia method:

\begin{enumerate}
    \item \textcolor{black}{\textbf{Choice of block count and conditioning size}: In general, the modeling capabilities, including parameter estimation and prediction, improve with an increase in both the block count and the conditioning size in block Vecchia approximations. This insight is crucial for optimizing the approximation process, ensuring both efficiency and precision.}

    \item \textbf{Importance of Ordering}: The sequence of blocks plays a critical role in the accuracy of the approximation, i.e., the random ordering outperforms others as the number of blocks increases.  This finding highlights the need for smart ordering to enhance the accuracy of the block Vecchia approximation.

    \item \textcolor{black}{\textbf{Performance Enhancements and Scalability}: Remarkably, the block Vecchia method demonstrates an approximately 80X speedup and 40X larger problem size compared to the classic Vecchia algorithm. This scalability enables the algorithm to leverage modern computational resources effectively, making it a powerful tool for addressing large-scale statistical modeling challenges.}
    
     \item \textcolor{black}{\textbf{Efficient and Accurate Prediction}:  Similar to the parameter estimation, the prediction accuracy, such as mean square error and standard deviation, can be improved along with the increase of block count and conditioning size.}
\end{enumerate}

The paper is structured as follows: 
Section~\ref{sec:framework} offers a detailed explanation of our proposed implementation. Section~\ref{sec:numericalresults} presents the evaluation of our implementation from a numerical study and simulation experiments compared to the exact GP. \textcolor{black}{Section~\ref{sec:realdataset} provides results on the real dataset in 3D million-level data points} and we conclude in Section~\ref{sec:conclusion}.

\section{Block Vecchia Framework}
\label{sec:framework}
This section provides an overview of the proposed framework for the block Vecchia algorithm. It begins with an explanation of the preprocessing steps, including location clustering, block reordering, and nearest neighbor selection for each block centroid. Following this, the memory requirements for the block Vecchia algorithm are discussed in detail, alongside a thorough description of the proposed implementation. Next, the prediction process using the block Vecchia algorithm is outlined. Finally, a comparison is presented, highlighting the expected memory usage and computational complexity of the block Vecchia algorithm relative to the classic Vecchia algorithm.

\subsection{Clustering and Block Permutation}
\begin{figure}[htbp]
    \centering
    \includegraphics[width = 0.45\textwidth, height= 0.45\textwidth]{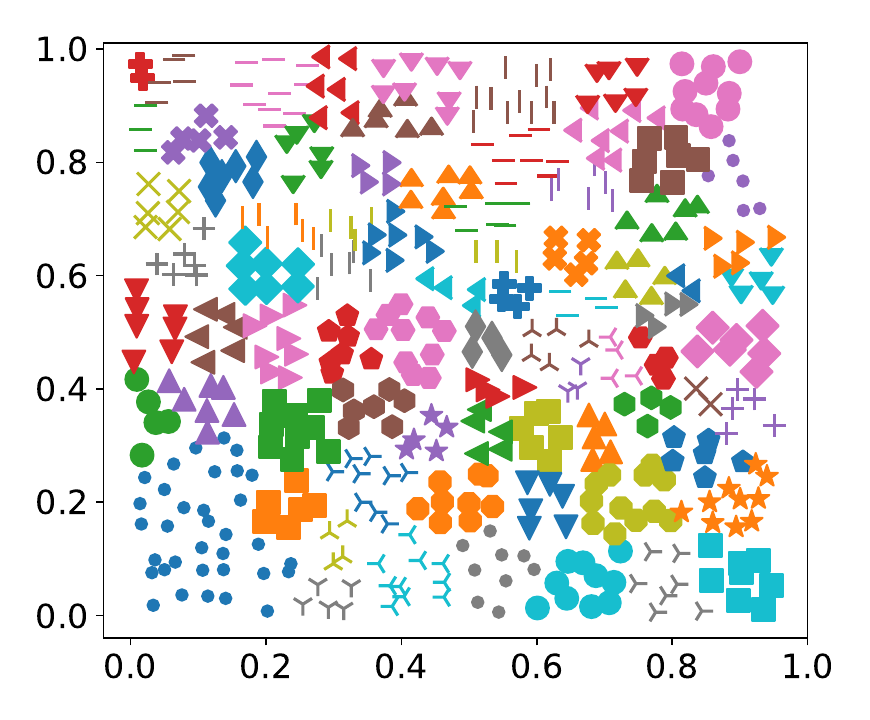}
    \caption{An example illustrating K-means in the block Vecchia, with 500 random locations in $[0, 1] \times [0, 1]$ and 80 blocks. Shape markers represent the blocks. 
    }
    \label{fig:kmeans-examples}
\end{figure}

The first step of the block Vecchia is to discover natural groupings in data based on some similarity measures by clustering methods. 
\textcolor{black}{Clustering involves using algorithms to organize a set of points into groups (or clusters) where points within the same group are more similar to each other than to those in different groups. Examples include K-means, hierarchical clustering, spectral clustering, and density-based spatial clustering \citep{xu2005survey, saxena2017review}. In the context of the block Vecchia, we use the K-means clustering algorithm, considering its simplicity and portability for large-scale computing. For example, 500 locations are randomly generated in the unit area, and 80 blocks are pre-specified. The blocks are clustered based on the coordinates using K-means, and the results are visualized in Figure \ref{fig:kmeans-examples}.
}

\begin{figure}[htbp]
    \centering
    \subfloat[Morton]{\includegraphics[width=0.2\textwidth,height=0.2\textwidth]{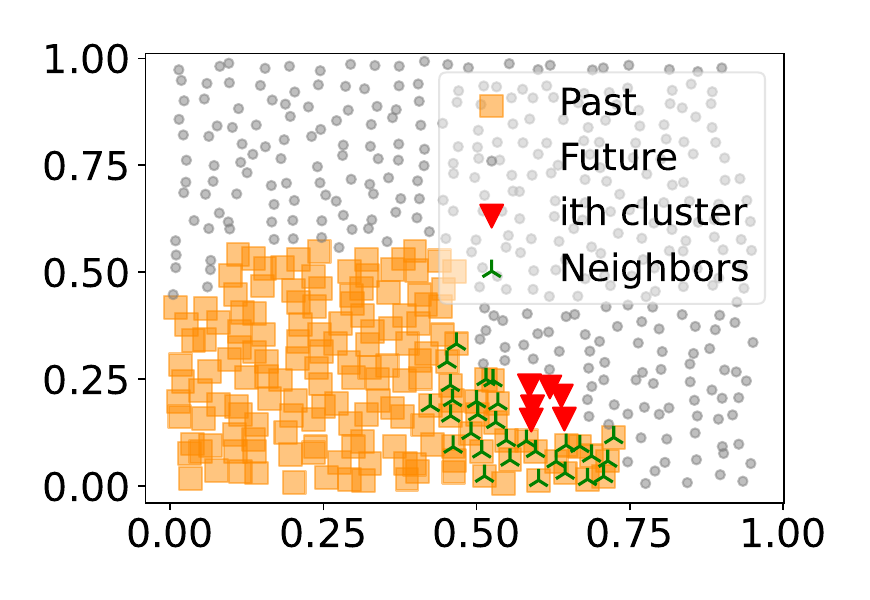}}
    \subfloat[Random]{\includegraphics[width=0.2\textwidth,height=0.2\textwidth]{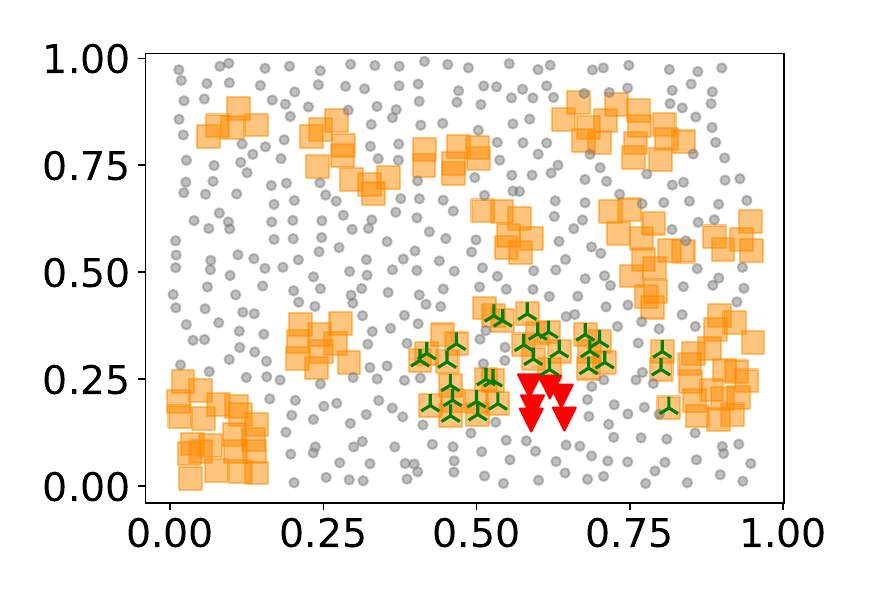}}
    \subfloat[KDtree]{\includegraphics[width=0.2\textwidth,height=0.2\textwidth]{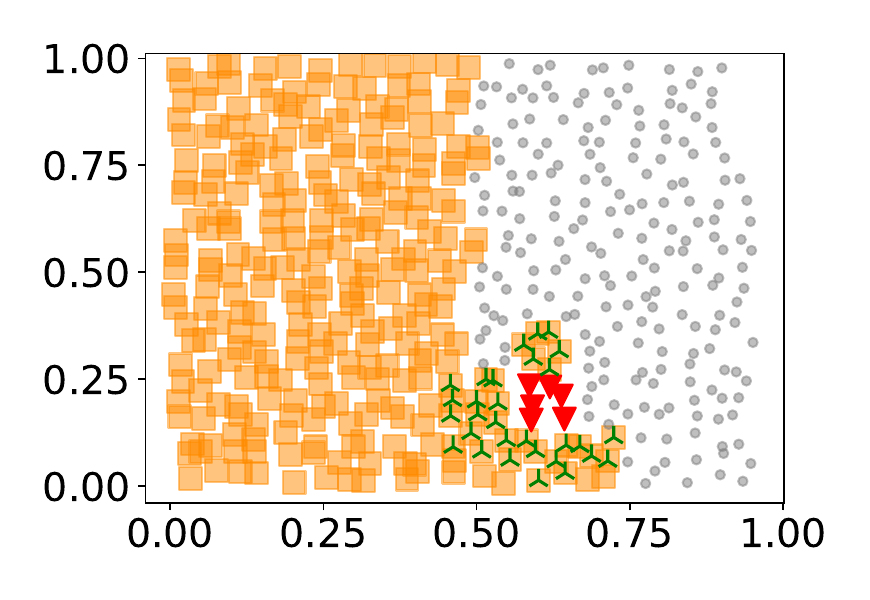}}
    \subfloat[Maxmin]{\includegraphics[width=0.2\textwidth,height=0.2\textwidth]{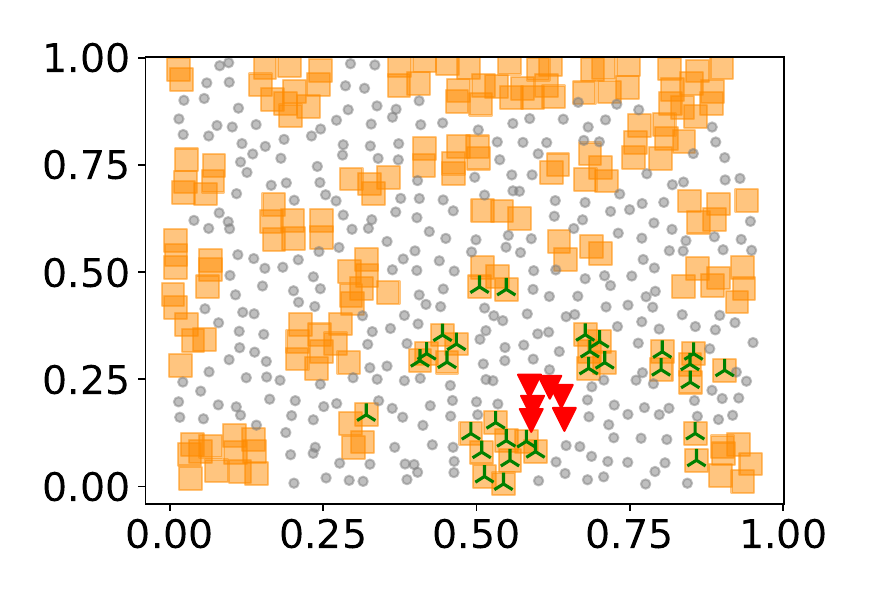}}
    \subfloat[Hilbert]{\includegraphics[width=0.2\textwidth,height=0.2\textwidth]{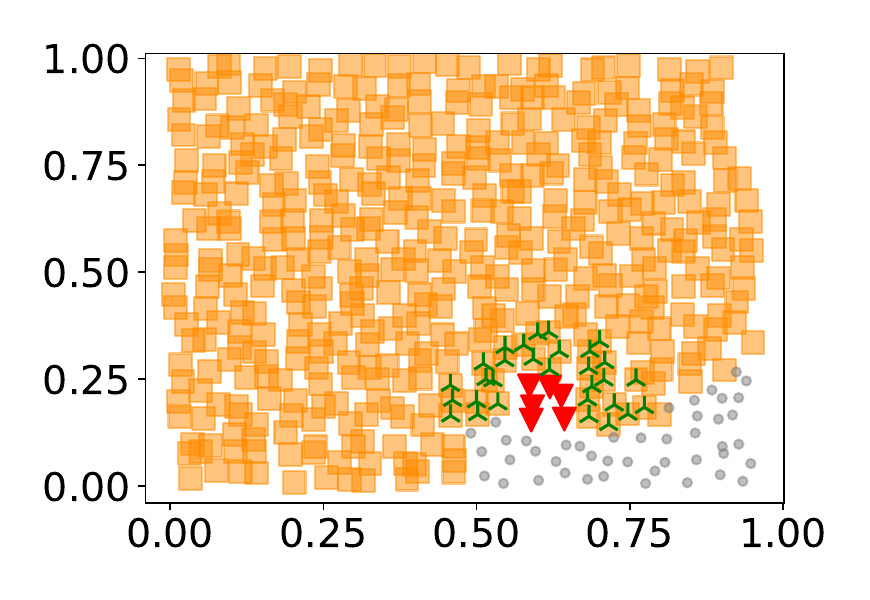}}
    \caption{\textcolor{black}{An example illustrating the impact of orderings on the neighbor selection for block Vecchia,  where we have 500 uniform random locations in $[0, 1]\times [0, 1]$ and 80 blocks, the square, small circle, triangle-down, and tri-up represents past points, future points, blocks, and neighbors, respectively. }
    }
    \label{fig:blockNN}
\end{figure}

Following the K-means clustering, the block Vecchia algorithm requires reordering the blocks. \textcolor{black}{Considering that the conditional probability in the Vecchia approximation is sensitive to its previous points, the ordering is crucial as it influences the selection of candidates for approximating the conditional probability.
Figure \ref{fig:blockNN} presents an illustrative example involving 500 uniformly random locations within $[0, 1]\times[0, 1]$.}
The example demonstrates four permutations and their 30 nearest neighbors using: Morton reordering \citep{walker2018morton}, random reordering \citep{guinness2018permutation}, KDtree reordering \citep{bentley1975multidimensional}
, maxmin reordering (mmd) \cite{guinness2018permutation} and Hilbert reordering \citep{hilbert1935stetige, chen2024impact}. \textcolor{black}{Morton reordering, or the Z-order reordering, is a space-filling curve approach that reduces multi-dimensional data to one dimension while preserving the locality of data points. This technique is beneficial in improving memory access patterns and data compression \citep{walker2018morton}. 
Random reordering rearranges the data points randomly, which can be useful for mitigating biases presented in the original ordering and potentially improving model robustness. Maxmin reordering is achieved by reordering the data in a way that maximizes the minimum distance, typically ensuring that the most informative points are considered earlier in the computational process. 
KDtree reordering is a space-partitioning data structure that helps optimize queries and operations that depend on spatial proximity.
Like Morton, Hilbert reordering tends to preserve a better locality than the Z-order curve, which can enhance performance in specific applications \citep{chen2022hilbert}.
}


\subsection{Block Vecchia Algorithm}

Having a set of geospatial data consisting of $n$ spatial locations and their corresponding observations denoted by $\boldsymbol{y}$, as well as any permutation $\zeta$ of the given centroids of blocks, the likelihood, or the joint density, for the observations $\boldsymbol{y}$ can be represented as a product of a series of multivariate conditional densities:
\begin{align}
    L({\boldsymbol\theta};\boldsymbol{y}) &= p_{\boldsymbol\theta}\left(y_1, \ldots, y_n\right) \nonumber \\
    & = p_{\boldsymbol\theta}\left(\boldsymbol{y}_{B_1}\right) \prod_{i=2}^{bc} p_{\boldsymbol\theta}\left(\boldsymbol{y}_{B_i} \mid \boldsymbol{y}_{B_1}, \ldots,\boldsymbol{y}_{B_{i-1}}\right) \label{eq:bv}, \\
    & = p_{\boldsymbol\theta}\left(\boldsymbol{y}_{B^\zeta_1}\right) \prod_{i=2}^{bc} p_{\boldsymbol\theta}\left(\boldsymbol{y}_{B^\zeta_i} \mid \boldsymbol{y}_{B^\zeta_1}, \ldots,\boldsymbol{y}_{B^\zeta_{i-1}}\right) \label{eq:bv-zeta},    
\end{align}
where arbitrary partitions of the observations and permutations of the partitions will not affect the joint density. \textcolor{black}{In (\ref{eq:bv}), $B_1, B_2, \ldots, B_{bc}$ are the partition of $\{1, 2, \ldots, n\}$, e.g., an integer set $B_i = \{b_{i1}, \ldots, b_{il_i}\}$  \textcolor{black}{where $b_{ij}$ is the $jth$ point in the $ith$ block and $l_i$ is the size of the $ith$ block}, and $bc$ means the block count with $\sum_{i=1}^{bc} l_{i} = n$. Here $\zeta$ in (\ref{eq:bv-zeta}) represents the permutation of the blocks.} The block Vecchia approximation replaces the complete conditioning vectors $(\boldsymbol{y}_{B^\zeta_1}, \ldots,\boldsymbol{y}_{B^\zeta_{i-1}} )$ with a subvector, where the length of subvector is far less than the length of the whole vector. 
\textcolor{black}{Specifically, a subvector, i.e., $m_i$ nearest neighbors $\boldsymbol{y}_{NN, B^{\zeta}_i} = (y_{j_1}, \ldots, y_{j_{m_i}})^\top$, are selected from the labeling of $B^\zeta_1 \cup B^\zeta_2 \cup \cdots \cup B^\zeta_{i-1}$. In the definition of $\boldsymbol{y}_{NN, B^{\zeta}_i}$, $m_i$ represents the size of conditioning set for the block $B^\zeta_i$ and $(y_{j_1}, \ldots, y_{j_{m_i}})$ are the selected points for approximating the $ith$ cluster conditional probability. 
} Then, we define the approximation as 
\begin{align}
    p_{\boldsymbol\theta,  \zeta, NN, B}\left(y_1, \ldots, y_n\right)
    &=
    p_{\boldsymbol\theta}\left(\boldsymbol{y}_{B^\zeta_1}\right) \prod_{i=2}^{bc}
    p_{\boldsymbol\theta}\left(\boldsymbol{y}_{B^\zeta_i}\mid y_{j_1}, \ldots, y_{j_{m_i}}\right) \nonumber \\
    &= 
    p_{\boldsymbol\theta}\left(\boldsymbol{y}_{B^\zeta_1}\right) \prod_{i=2}^{bc} p_{\boldsymbol\theta}\left(\boldsymbol{y}_{B^\zeta_i} \mid \boldsymbol{y}_{NN, B^{\zeta}_i} \right).\label{eq:vecchia-vector}
\end{align}
\textcolor{black}{
Here, $NN$ represents the nearest neighbor selection, and $B = \{B_1, \ldots, B_{bc}\}$ is the set of blocks. In the block Vecchia algorithm, the approximation quality relies on the partition $B$ of observations and the permutation $\zeta$ of the partition. 
}

\begin{figure}[htbp]
    \centering
    \includegraphics[width=.75\textwidth]{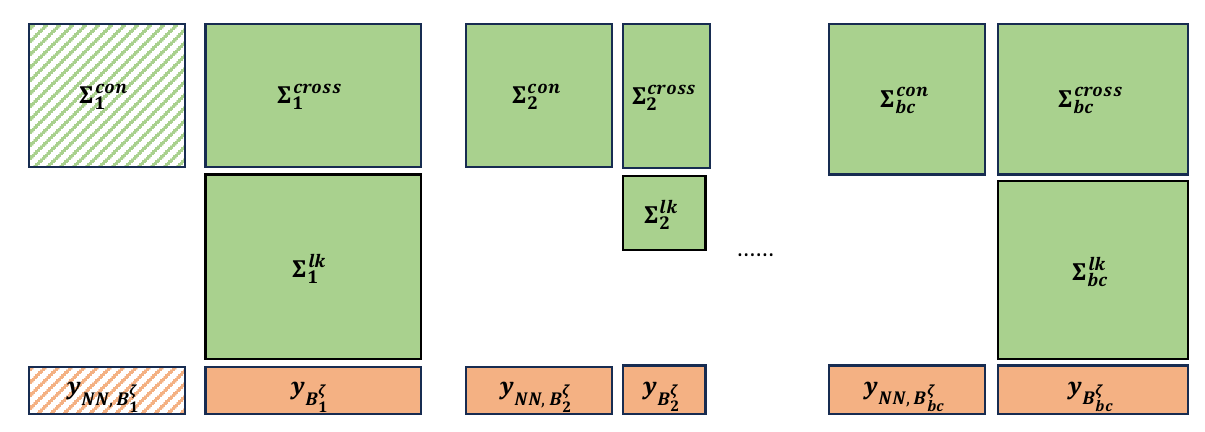}
    \caption{Block Vecchia algorithm (Notations as shown in Algorithm 1 in \ref{spp:alg}).
    }
    \label{fig:archblockvecchia}
\end{figure}

To implement the block Vecchia algorithm, for each spatial block, we should compute three covariance matrices, i.e., the conditioning covariance matrix constructed by its nearest neighbors $\mathbf \Sigma^{con}_{i}$, the cross-covariance matrix between the block and its nearest neighbors $\mathbf \Sigma^{cross}_{i}$, and \textcolor{black}{the covariance matrix constructed by points in the $ith$ block $\mathbf \Sigma^{lk}_{i}$.} Figure \ref{fig:archblockvecchia} depicts the required vector/matrix for each spatial block. Here \textcolor{black}{$\mathbf{y}_{B^{\zeta}_i}$ and $\mathbf y_{ NN, B^{\zeta}_i}$} are the $i$th block's and its neighbors' observations, respectively, which exactly match the representation in equation (\ref{eq:vecchia-vector}) (the dashed $\mathbf \Sigma^{con}_{1}$ and $\mathbf y_{NN, B^{\zeta}_1}$ are supplementary position for batched operations). For every pair $(\mathbf y_{ NN, B^{\zeta}_i}, \mathbf{y}_{B^{\zeta}_i}, \mathbf \Sigma^{con}_{i}, \mathbf \Sigma^{cross}_{i}, \mathbf \Sigma^{lk}_{i})$, their log-likelihoods, $p_{\boldsymbol\theta}\left(\mathbf{y}_{B^{\zeta}_i} \mid \mathbf y_{ NN, B^{\zeta}_i} \right)$, are independent of each other. That is to say, the task of computing log-likelihood in (\ref{eq:vecchia-vector}) can be divided into $bc$ independent small tasks using batched operations, 
\textcolor{black}{
    including batchedPOTRF (batched Cholesky decomposition), batchedTRSM (batched triangular linear solver), batchedGEMM (batched matrix-to-matrix multiplication), batchedGEMV (batched matrix-to-vector multiplication), batchedDotProduct (batched inner product).
}
The details of the implementation are summarized in the Algorithm \ref{alg:blockvecchia-cluster} in \ref{spp:alg}, including the clustering, block permutation, nearest neighbor searching, covariance-related matrix construction, and varying batched operations for Vecchia computation \citep{dong2016magma}.

\subsection{Block Vecchia Prediction}

Let us consider the task of predicting values at new spatial locations $\boldsymbol{S}_* = \{ {\bf s}_{n+1}, {\bf s}_{n+2}, \ldots, {\bf s}_{n+n^*} \}$ based on observed data $\boldsymbol{y} = (y_1, y_2, \ldots, y_n)^\top$ at locations $\boldsymbol{S} = \{ {\bf s}_1, {\bf s}_2, \ldots, {\bf s}_n \}$. In the standard GP framework, the joint distribution of the observed and new data is multivariate normal:
\begin{equation*}
\begin{pmatrix}
\boldsymbol{y} \\
\boldsymbol{y}_*
\end{pmatrix}
\sim
\mathcal{N}_{n+n^*} \left(
\begin{pmatrix}
\boldsymbol{\mu}_y \\
\boldsymbol{\mu}_{y_*}
\end{pmatrix},
\begin{pmatrix}
\mathbf{\Sigma}_{yy} & \mathbf{\Sigma}_{y y_*} \\
\mathbf{\Sigma}_{y_* y} & \mathbf{\Sigma}_{y_* y_*}
\end{pmatrix}
\right),    
\end{equation*}
where $\boldsymbol{y}_* = (y_{n+1}, y_{n+2}, \ldots, y_{n+n^*})^\top$ are the values to be predicted, and $\mathbf{\Sigma}_{yy}$, $\mathbf{\Sigma}_{y y_*}$, and $\mathbf{\Sigma}_{y_* y_*}$ are covariance matrices computed using a specified covariance function parameterized by $\boldsymbol{\theta}$. The exact conditional distribution of $\boldsymbol{y}_*$ given $\boldsymbol{y}$ is then 
$ \boldsymbol{y}_* \mid \boldsymbol{y} \sim \mathcal{N}_{n^*} \left(
\boldsymbol{\mu}_{y_* \mid y}, \mathbf{\Sigma}_{y_* \mid y}\right)$
with
$
\boldsymbol{\mu}_{y_* \mid y} = \boldsymbol{\mu}_{y_*} + \mathbf{\Sigma}_{y_* y} \mathbf{\Sigma}_{yy}^{-1} (\boldsymbol{y} - \boldsymbol{\mu}_y)$, $\mathbf{\Sigma}_{y_* \mid y} = \mathbf{\Sigma}_{y_* y_*} - \mathbf{\Sigma}_{y_* y} \mathbf{\Sigma}_{yy}^{-1} \mathbf{\Sigma}_{y y_*}.
$
Computing $\mathbf{\Sigma}_{yy}^{-1}$ is computationally expensive for large $n$. To alleviate this, the block Vecchia approximation approximates the joint conditional probability by limiting the conditioning to a subset of observations. The block Vecchia approximation involves the following steps: 1) Divide the new locations $\boldsymbol{S}_*$ into $bc$ blocks $B^*_1, B^*_2, \ldots, B^*_{bc}$, each containing $l$ nearby locations; 2) For each block $B^*_i$, select a conditioning set $\boldsymbol{y}_{NN, B^*_i}$ consisting of $m$ nearest neighbor observations from $\boldsymbol{y}$; 3) Approximate the joint conditional probability as
    \begin{equation}
        p_{\boldsymbol{\theta}}(\boldsymbol{y}_* \mid \boldsymbol{y}) \approx \prod_{i=1}^{bc} p_{\boldsymbol{\theta}} \left( \boldsymbol{y}_{B^*_i} \mid \boldsymbol{y}_{NN, B^*_i} \right),
    \end{equation}
where $\boldsymbol{y}_{B^*_i}$ are the predictions for block $B^*_i$. For each block, the conditional distribution is:
    $
    \boldsymbol{y}_{B^*_i} \mid \boldsymbol{y}_{NN, B^*_i} \sim \mathcal{N}_{\#B^*_i} \left(
    \boldsymbol{\mu}_{B^*_i}, \mathbf{\Sigma}_{B^*_i}
    \right),
    $
    with
    $
    \boldsymbol{\mu}_{B^*_i} = \mathbf{\Sigma}^{\text{cross}}_i \left( \mathbf{\Sigma}^{\text{con}}_i \right)^{-1} \boldsymbol{y}_{NN, B^*_i},
    $
    $
    \mathbf{\Sigma}_{B^*_i} = \mathbf{\Sigma}^{\text{lk}}_i - \mathbf{\Sigma}^{\text{cross}}_i \left( \mathbf{\Sigma}^{\text{con}}_i \right)^{-1} \left( \mathbf{\Sigma}^{\text{cross}}_i \right)^\top,
    $
    where $\mathbf{\Sigma}^{\text{con}}_i$ is the covariance matrix of the conditioning set $\boldsymbol{y}_{NN, B^*_i}$,
    $\mathbf{\Sigma}^{\text{cross}}_i$ is the cross-covariance matrix between $\boldsymbol{y}_{B^*_i}$ and $\boldsymbol{y}_{NN, B^*_i}$,
    $\mathbf{\Sigma}^{\text{lk}}_i$ is the covariance matrix among the locations in block $B^*_i$. This approximation reduces the computational cost from $\mathcal{O}(n^3)$ to $\mathcal{O}(bc \times m^3)$, where $m \ll n$. The prediction using the block Vecchia approximation can be formalized in Algorithm~\ref{alg:block-vecchia-prediction} in \ref{spp:alg}.

The output, $\boldsymbol{y}_*$ and $\{ \mathbf{\Sigma}_{B^*_i} \}_{i=1}^{bc}$, helps the univariate (and multivariate) conditional simulation and the prediction interval. Considering the simplicity and computational efficiency, the univariate configuration is adopted. The variance $\boldsymbol{\sigma} = \left( \sigma_1, \sigma_2, \cdots, \sigma_{n*} \right)^\top$ is extracted from the diagnal element from the $\{ \mathbf{\Sigma}_{B^*_i} \}_{i=1}^{bc}$ for the predicted values $\boldsymbol{y}_*$. Then the conditional simulations are conducted using the $\boldsymbol{y}_*$ and $\boldsymbol{\sigma}$. We calculate the sample mean and variance $(\Tilde{\boldsymbol{\mu}}, \Tilde{\boldsymbol{\sigma}}^2)$ and then the 95\% confidence interval is $\left(\Tilde{\mu} - z_{\alpha/2}\Tilde{\sigma}_i, \Tilde{\mu} + z_{\alpha/2}\Tilde{\sigma}_i\right)$ with $\alpha = 0.05$. \textcolor{black}{For general applications, we propose the following scheme. Firstly, it is recommended to use a small block count and conditioning size to estimate the parameters. Then, these estimated parameters serve as initial values for accurate settings with larger block counts and conditioning sizes, continuing until a clear convergence trend emerges. This approach eliminates the need to determine the optimal block count and conditioning size explicitly, as real-world applications often vary in problem size and required precision. Employing this automated estimation method provides a more adaptable and practical solution.}

\subsection{GPU and Batched Linear Algebra}

\textcolor{black}{
    GPUs are specialized hardware designed to handle parallel operations at a massive scale, which makes them particularly well-suited for tasks that involve large-scale numerical computations, e.g., GEMM and elementwise operations. Unlike CPUs, GPUs have hundreds of thousands of cores designed for handling multiple operations simultaneously. This parallelism can be leveraged to significantly speedup the computation of complex mathematical models, such as those used in the Vecchia approximation \citep{pan2024gpuaccelerated}.} 

\textcolor{black}{   
    Batched operations involve processing multiple data points or operations simultaneously, thereby reducing the overhead associated with individual computations \citep{haidar2015batched}. In the context of the block Vecchia approximation, which consists of abundant independent and computationally light conditional probabilities, batched operations can be employed to improve computational efficiency by processing those covariance matrices and vectors at the same time, e.g., lines 17 - 21 and 24 - 27 in the Algorithm \ref{alg:blockvecchia-cluster} in \ref{spp:alg}. In other words, when implemented on a GPU, these matrices and vectors can be processed in parallel batches rather than sequentially as on a CPU, which dramatically reduces the computation overheads. By utilizing GPU acceleration through batched operations, the computational burden of the block Vecchia approximation is significantly reduced, allowing for faster processing times, especially when dealing with large datasets. This not only improves the feasibility of applying such methods in practice but also opens the door to analyzing more complex models and larger datasets that would be computationally prohibitive on traditional CPU-based systems.
}

\subsection{Computational and Memory Complexity} 

We analyze the memory usage and computational complexity of the block Vecchia implementation, comparing it against the traditional Vecchia approach. The memory footprint of the classic Vecchia algorithm is $\mathcal{O}(nm^2)$ for many small symmetric covariance matrices \textcolor{black}{$\boldsymbol{\Sigma}_{i}, i=1, \ldots, n$ and $\mathcal{O}(nm)$ for the conditioning vectors $\boldsymbol{y}_{NN, B^{\zeta}_i}, i=1, \ldots, n$ where $m$ and $n$ stands for the conditioning size and observations, respectively.} 
For the block Vecchia algorithm, each block requires three covariance matrices $\boldsymbol{\Sigma}_i^{lk}$, $\boldsymbol{\Sigma}_i^{cross}$, $\boldsymbol{\Sigma}_i^{con}$, \textcolor{black}{$i=1, \ldots, bc$ where the average block size is approximately $n/bc$},  and two observation vectors $\boldsymbol{y}_{NN, B^{\zeta}_i}$ and $\boldsymbol{y}_{B_i}$, $i=1, \ldots, bc$, with memory complexities of $\sim{bc(n/bc)^2/2}$, $\sim{bc*m(n/bc)/2}$ $\sim{bc*m^2/2}$, $\sim{bc*m}$, and $\sim{bc*(n/bc)}$, respectively. Therefore, the memory for the classic Vecchia is approximately $\sim{nm^2/2 + nm}$ and for the block Vecchia is $\sim{n^2/(2bc)+mn/2}$ ${+ bc*m^2/2 + m*bc + n}$. In terms of the arithmetic complexity, for the block Vecchia, the complexity primarily stems from the Cholesky factorization of the covariance matrix, $\sim{bc(n/bc)^3/3}$ and $\sim{bc(m)^3/3}$, and matrix-to-matrix multiplication $\sim{2*bc*m(n/bc)^2}$. In contrast, with the classic Vecchia approximation, the complexity for the Cholesky factorization operations is $\sim{n(m^3/3)}$.

\begin{figure}[htbp]
    \centering
    \subfloat[Memory complexity]{\includegraphics[width=0.5\textwidth]{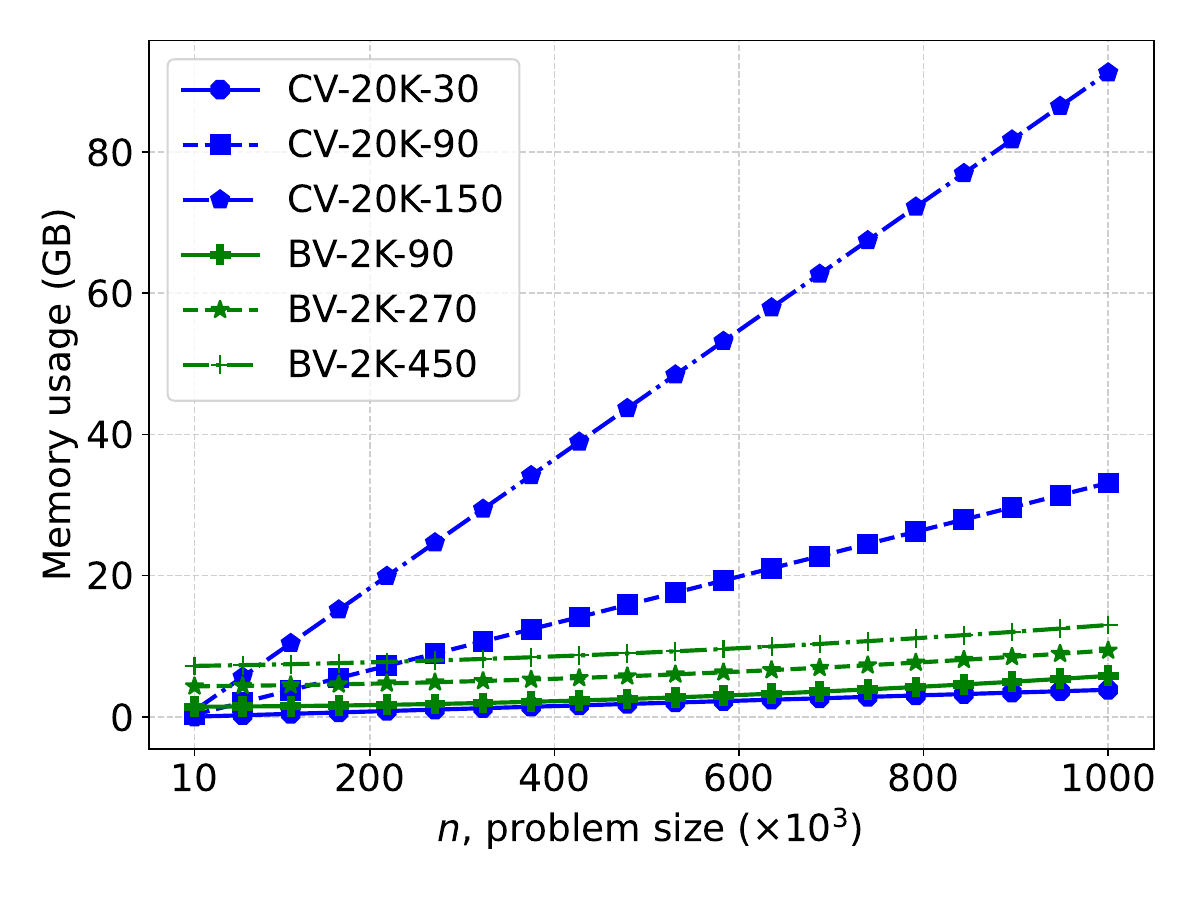}}
    \hfill
    \subfloat[Computational complexity]{\includegraphics[width=0.5\textwidth]{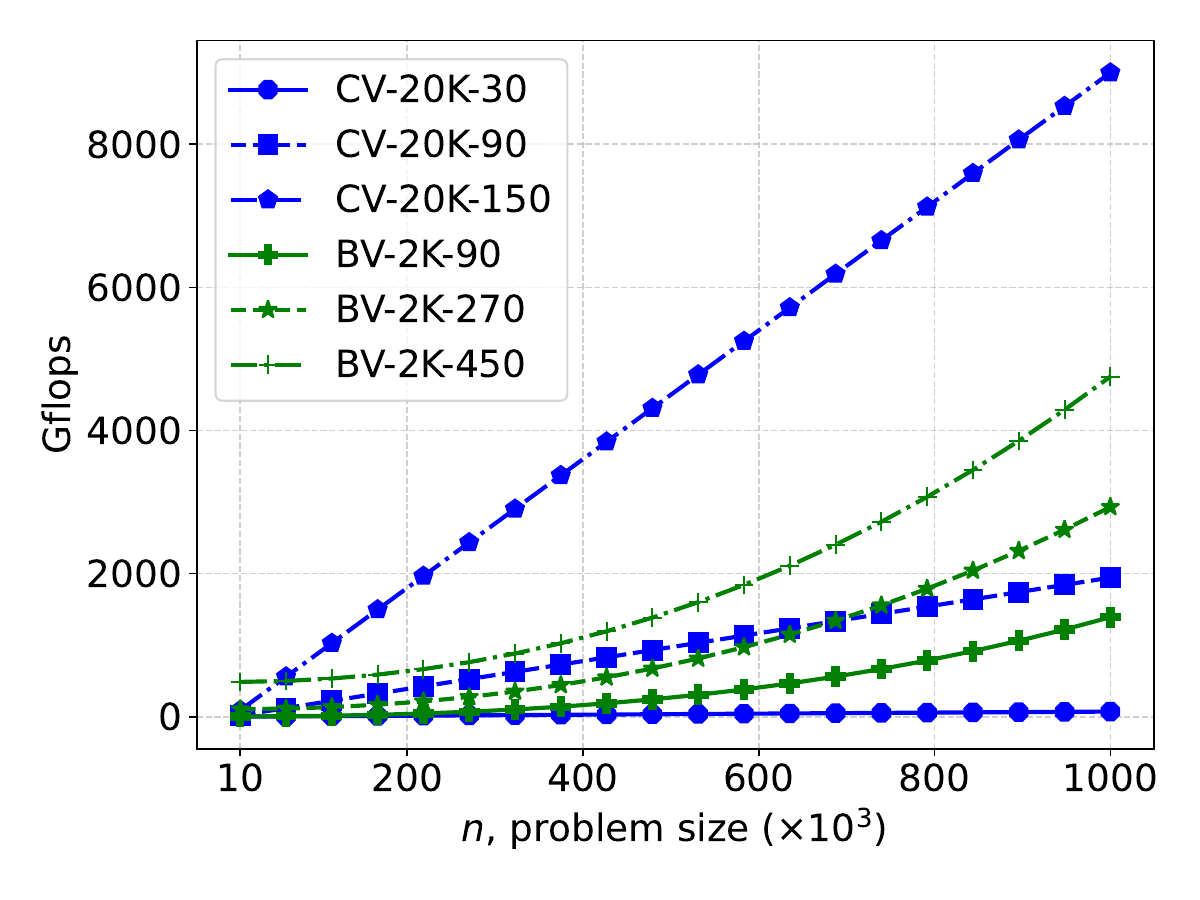}}
    \caption{\textcolor{black}{Comparison of Arithmetic complexity: block Vecchia (BV) versus classic Vecchia (CV) algorithms. The format of the legend is Method-ConditioningSize, e.g., CV-20K-30 represents the classic Vecchia with conditioning size 30; BV-2K-90 represents the block Vecchia with block count 2,000 and conditioning size 90. 
    }}
    \label{fig:complexity}
\end{figure}

In Figure \ref{fig:complexity}, to have a fair comparison between classic Vecchia and block Vecchia, we set the approximate block size 100, i.e., $n/bc \approx 100$ and nearest neighbor 3X larger than the classic Vecchia.  Figure \ref{fig:complexity} (a) shows the memory footprint in gigabytes (GB) for increasing problem sizes when employing the classic Vecchia and block Vecchia algorithms. The block Vecchia notably lowers the memory burden. Figure ~\ref{fig:complexity} (b) demonstrates the floating-point operations (flops) in Gflops when using the block Vecchia and classic algorithms. The figure shows a comparable computation in the number of required flops for the Vecchia algorithm. The observations underscore that the block Vecchia creates computationally intensive tasks compared to the classic Vecchia and then it inspires us to utilize the modern GPU architecture to accelerate the computations.

\section{Numerical Studies}
\label{sec:numericalresults}

This section assesses the accuracy and computational performance of the proposed block Vecchia algorithm compared to the classical Vecchia algorithm. The block Vecchia code was compiled using GCC version 10.2.0 (or 12.2.0) and CUDA  v11.4 (or v11.8), and linked with Intel MKL v2022.2.1, MAGMA v2.6.0 (or v2.7.2), GSL v2.6, and NLopt v2.7.1 optimization libraries. Our machine has 40 CPU cores (Intel(R) Xeon(R) Gold 6230R CPU @ 2.10GHz) and NVIDIA V100 GPU 32GB. Each experiment was repeated five times to ensure repeatability and consistency in the time-to-solution metric. Accuracy was evaluated using the deterministic KL divergence metric alongside qualitative analysis, with simulation studies conducted under various parameter settings. Furthermore, the block Vecchia algorithm was compared with exact GPs in terms of parameter estimation and predictive performance across two real-world applications. 
\subsection{KL Divergence}

In this subsection, we calculate the deterministic metric, KL divergence, to assess the accuracy of the block Vecchia algorithm with the help of {\em ExaGeoStat} \citep{abdulah2018exageostat}, on Gaussian random fields with $n=20000$ spatial locations \textcolor{black}{ within $[0, 1] \times [0, 1]$}. We also include additional small-scale examples in \ref{spp:numerical-smaller} and \ref{spp:soomthness-smaller} for consumer-grade GPUs, using a dataset size of $n = 8000$.

The KL divergence is defined as:
\begin{equation}
    D_{\text{KL}}({\mathcal {N}}_{0}\parallel {\mathcal {N}}_{1})={\frac{1}{2}}\left\{\operatorname {tr} \left({ {\boldsymbol \Sigma }}_{1}^{-1}{ {\boldsymbol \Sigma }}_{0}\right)-n+\log {|{ {\boldsymbol \Sigma }}_{1}| \over |{{\boldsymbol \Sigma }}_{0}|}\right\},
    \label{eq:kl-gaussian}
\end{equation}
where ${\mathcal {N}}_{0}, {\mathcal {N}}_{1}$ represents two \textcolor{black}{$n-$dimension Gaussian distributions with zero-mean and covariance matrices $\boldsymbol \Sigma_0,  \boldsymbol \Sigma_1$, representing the exact and Vecchia-approximated covariance matrix.
Let ${\mathcal {N}}_{0}$ be the exact distribution and ${\mathcal {N}}_{1}$ be the approximate distribution by the block Vecchia. Then (\ref{eq:kl-gaussian}) simplifies as:
\begin{equation}
    D_{\text{KL}}({\mathcal {N}}_{0}\parallel {\mathcal {N}}_{1})= \ell_0({\boldsymbol\theta};\boldsymbol{0}) - \ell_a({\boldsymbol\theta};\boldsymbol{0}),
    \label{eq:kl-gaussian-vecchia}
\end{equation}
where $\ell_0({\boldsymbol\theta};\boldsymbol{0})$ represents the exact log-likelihood at the $\mathbf{y} = \boldsymbol{0}$ which is calculated by {\em ExaGeoStat} and $\ell_a({\boldsymbol\theta};\boldsymbol{0})$ represents the Vecchia-approximated log-likelihood at $\mathbf{y} = \boldsymbol{0}$.}
In addition, we rely on the isotropic Matérn covariance function as in (\ref{eq:maternkernel}), and \cite{gneiting2002nonseparable} provided other kernels.
\begin{align}
    C_{\bm \theta}({\bf s}_i, {\bf s}_j)&=\sigma^2 \frac{2^{1-\nu}}{\Gamma(\nu)}\left(\frac{\|{\bf s}_i-{\bf s}_j\|}{\beta}\right)^\nu {\cal K}_\nu\left(\frac{\|{\bf s}_i-{\bf s}_j\|}{\beta}\right),
    \label{eq:maternkernel}
\end{align}
where ${\boldsymbol\theta} = \left(\sigma^2, \beta, \nu \right)^\top$, $\sigma^2$ is the variance, ${\cal K}_\nu(\cdot)$ is the Bessel function of the second kind of order, $\nu$$\Gamma(\cdot)$ is the gamma function, and $\beta>0$ and $\nu>0$ are range and smoothness parameters, respectively. Next, we select the parameter for smoothness  $\nu$ across values of 0.5, 1.5, and 2.5, representing varying degrees: low, medium, and high, respectively. Furthermore, we adjust the effective range for each smoothness level to 0.1, 0.3, and 0.8, corresponding to low, medium, and high dependency values in the unit square, respectively. Then, the corresponding parameter $\beta$ is calculated and reported in Table \ref{tab:accuracy-setting}. These adjustments are crucial as they influence the correlation within the data \citep{pan2024gpuaccelerated}.

\begin{table}[htbp]
    \caption{The cross combinations of low/medium/high smoothness and low/medium/high effective range. Each entry in the table represents $\beta$, and the $0.1, 0.3, 0.8$ are the statistical effective range, i.e., the distance at which spatial correlations fall to 5\% \citep{huang2021competition}.}
    \centering
    \label{tab:accuracy-setting}
        \begin{tabular}{llll}
        \toprule
                              & $\nu=0.5$                        & $\nu=1.5$                       & $\nu=2.5$                        \\
                              \midrule
        effective range=$0.1$ & $0.026270$  & $0.017512$ & $0.014290$  \\
        effective range=$0.3$ & $0.078809$ & $0.052537$ & $0.042869$  \\
        effective range=$0.8$ & $0.210158$ & $0.140098$ & $0.114318$ \\\bottomrule
        \end{tabular}
\end{table}

\begin{figure}[t!]
    \centering
    \subfloat[$bc=2500$]{
    \includegraphics[width=0.5\textwidth]{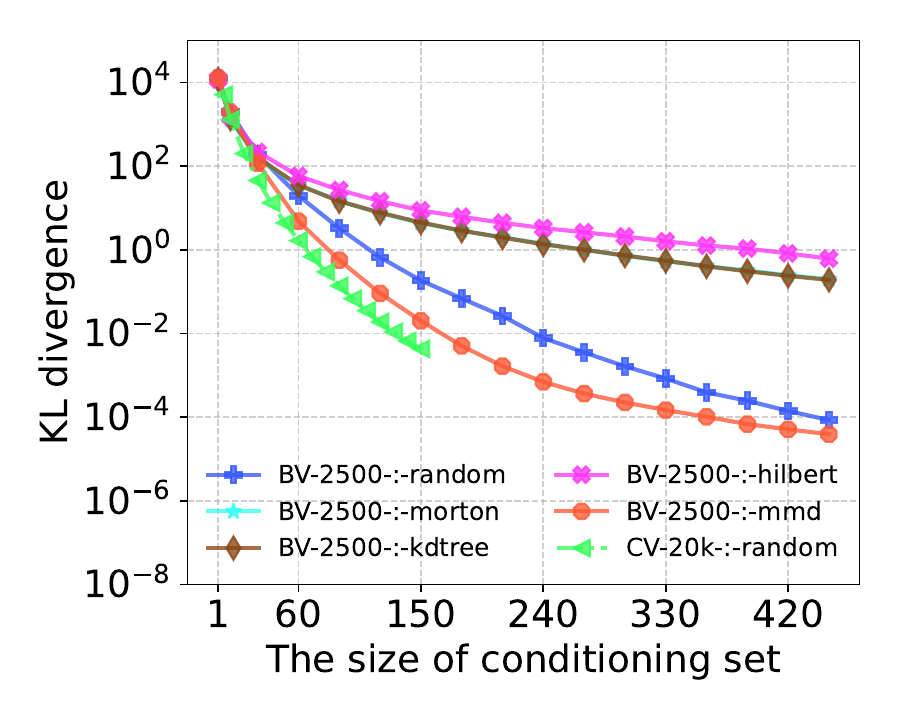}
    \label{fig:block-perm-bc2500}
    }
    \subfloat[Random reodering]{
    \includegraphics[width = 0.5\textwidth]{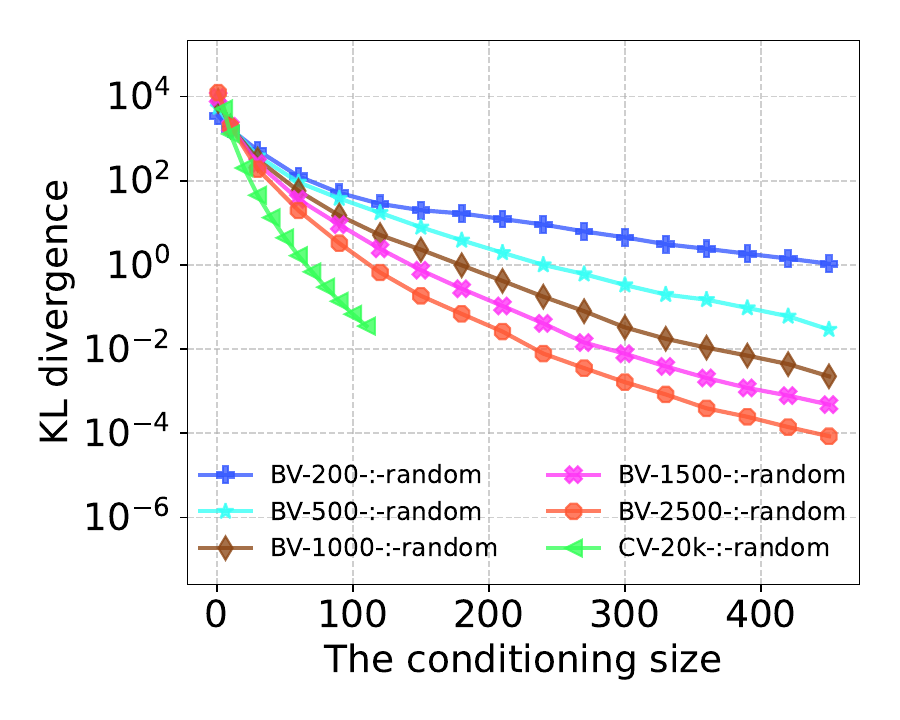}
    \label{fig:block-random-bc}
    }
    \caption{KL divergence and conditioning size along with increasing block count and different permutations under $\beta=0.052537$, $\nu=1.5$ and $\log_{10}$ scale. 
    }
    \label{fig:20-kl-bc-15}
\end{figure}

\begin{figure}[h!]
    \centering
    \subfloat[$\beta=0.026270, \nu = 0.5$]{\includegraphics[width=0.5\textwidth]{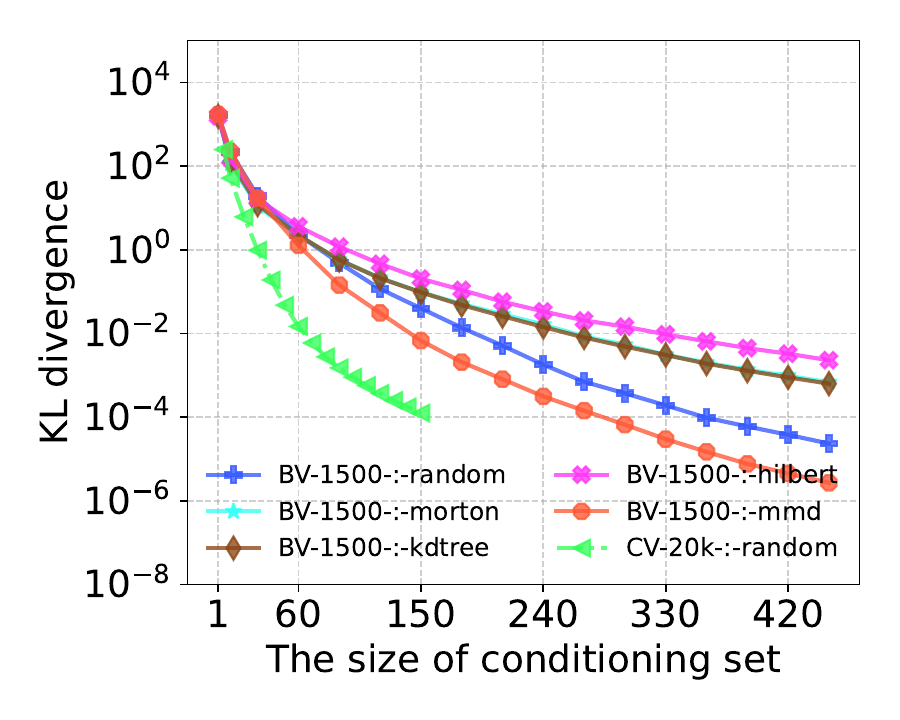}}
    \subfloat[$\beta=0.014290, \nu = 2.5$]{\includegraphics[width=0.5\textwidth]{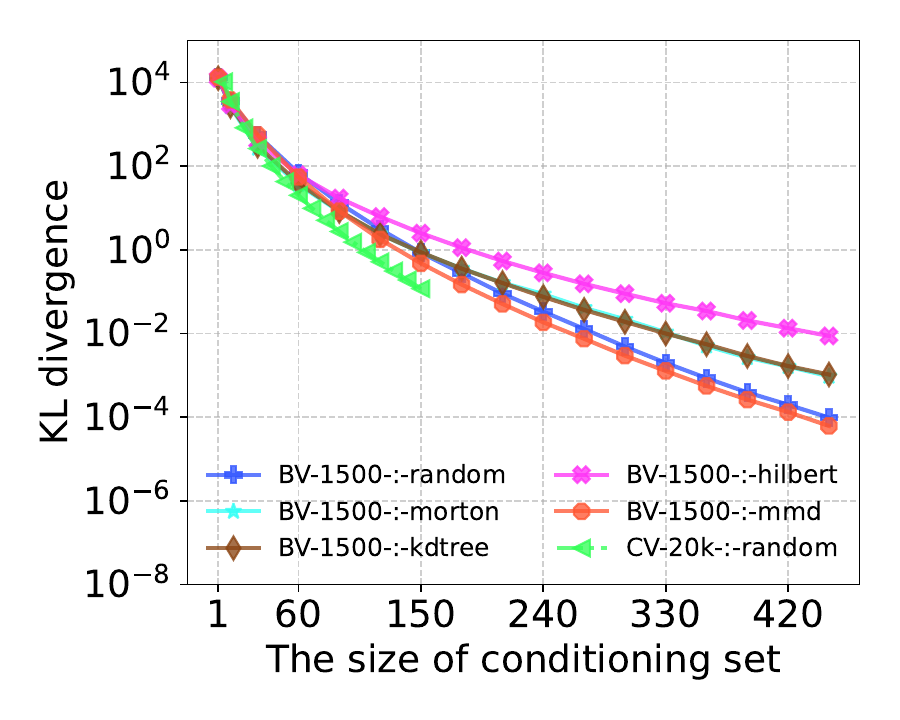}}
    \caption{KL divergence and conditioning size under 20K locations with $\log_{10}$ scale under the low effective range. 
    }
    \label{fig:20k-kl-1500}
\end{figure}

\begin{figure}[htbp]
    \centering
    \includegraphics[width=0.55\textwidth]{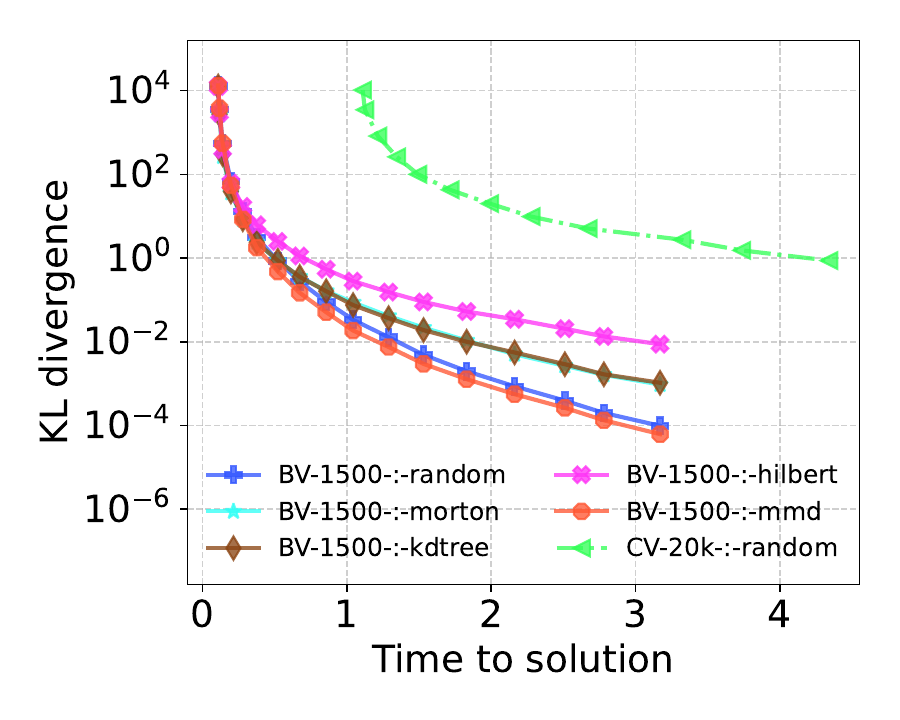}
    \caption{KL divergence and time-to-solution (second) under 20K locations with $\log_{10}$ scale under $\nu=2.5, \beta=0.014290$.
    }
    \label{fig:20k-kl-1500-time}
\end{figure}

In the following experiments, the shorthand format is employed, i.e., \textit{block count (bc) - conditioning size (cs) - ordering method}. For instance, \textit{BV-1000-60-Morton} means the Block Vecchia (BV) with $bc=1000$, $cs=60$, and Morton ordering; \textit{CV-20K-60-random} means the Classic Vecchia (CV) with $bc=20{,}000$ and $cs=60$, and random ordering. In the classic Vecchia algorithm, random ordering is adopted as the baseline due to its superior \textcolor{black}{accuracy within large-scale spatial scenarios \citep{guinness2018permutation, pan2024gpuaccelerated}.}
\textcolor{black}{The calculation of the KL divergence is performed as in (\ref{eq:kl-gaussian-vecchia})}, and the outcomes are visualized in Figures \ref{fig:20-kl-bc-15}, \ref{fig:20k-kl-1500}, \ref{fig:20k-kl-1500-time}. In all subfigures, the y-axis represents the logarithm of the KL divergence value.
The numerical studies are divided into three parts, presenting some of the main results (please refer to \ref{spp:numerical}, \ref{spp:soomthness}, \ref{spp:faster} in the supplementary materials for the comprehensive results).

\textcolor{black}{We investigate the impact of permutation and block count of the block Vecchia algorithm.  As illustrated in Figure \ref{fig:block-perm-bc2500}, the permutation plays an essential role in enhancing the accuracy of the block Vecchia algorithm, where the maxmin permutation (mmd in figures) achieves the
highest accuracy, random permutation yields near-optimal accuracy, while other permutations
fail to produce promising results. In Figure \ref{fig:block-random-bc}, we
set random reordering as the default, recognizing that it achieves near-optimal accuracy
while demanding fewer computational resources than the maxmin reordering algorithm \citep{guinness2018permutation}. The KL divergence is plotted against different block counts, consistently confirming that a larger block count leads to more accurate results. This enhancement can be attributed to two factors: firstly, a smaller block size (or a larger block count) ensures that the neighbors are representative of the points within the block; secondly, random ordering potentially yields better neighbor candidates for blocks. Figure \ref{fig:20-kl-bc-15} only presents the case of $\beta=0.052537$ and $\nu=1.5$. Refer to Figure \ref{fig:20-kl-bc-15-appendix} for more parameter settings.}

    In our analysis of two approximation methodologies, the classic Vecchia exhibits a pronounced decrease in accuracy as smoothness increases. At the same time, the block Vecchia demonstrates considerable robustness, maintaining accuracy across a range of smoothness parameters.
    For the evaluation of the block Vecchia approximation capabilities concerning smoothness parameters, we focus on the case where $bc=1500$, identified as both relatively efficient and accurate. 
    Figure \ref{fig:20k-kl-1500} provides quantitative evidence of this phenomenon: for the \textit{CV-20K-150-random} configuration, KL divergence escalates from $10^{-4}$ to $10^{-1}$, indicating a substantial loss in approximation precision. Conversely, the \textit{BV-1500-450-random} configuration shows only a slight decrease in KL divergence, from $10^{-4.6}$ to $10^{-4}$, highlighting the effectiveness of the BV method to manage the complexities associated with high smoothness levels effectively. \textcolor{black}{Furthermore, as $\nu$ increases, the gap between the block Vecchia and classic Vecchia methods diminishes. Refer to Figure \ref{spp:soomthness} for more parameter settings.}

The block Vecchia algorithm outperforms the classic Vecchia algorithm in both computational efficiency and accuracy. To evaluate the efficiency and accuracy of the block Vecchia, we focus on the case at $bc=1500$, which is the same as above.
    In Figure \ref{fig:20k-kl-1500-time}, the time-to-solution is defined as the cumulative duration required to compute the log-likelihood for a single iteration in MLE. This includes the time for matrix generation and batched BLAS operations while explicitly excluding the time spent on nearest neighbor searching and K-means clustering. These latter operations are omitted from consideration due to their one-time execution at the outset of MLE and their minimal impact on the overall computation time across numerous iterations.
    The \textit{BV-1500-:-random} achieves equivalent levels of accuracy approximately 5X faster than its classic counterpart, \textit{CV-20K-:-random}. Additionally, it offers a more precise KL approximation. Notably, the block Vecchia provides a more accurate approximation within the same time frame compared to the classic Vecchia, \textit{CV-20K-:-random}. Refer to Figure \ref{spp:faster} for more parameter settings. \textcolor{black}{In addition, we provide the accuracy of block Vecchia approximation along with the increasing number of locations in \ref{spp:increasingnumber}.}

\subsection{Simulations for Parameter Estimation and Prediction}

In this section, we assess the accuracy of statistical parameter estimation and the prediction uncertainty of the block Vecchia.  Using the \textit{ExaGeoStat} framework \citep{abdulah2018exageostat}, a high-performance unified framework for computational geostatistics on many-core systems, we generate 50 datasets on irregular 2D spatial locations within the unit square $[0,1] \times [0,1]$ to investigate the parameter uncertainty numerically, which has been studied for exact GP asymptotically \citep{wang2023parameterization}. These datasets are based on Gaussian random fields with a problem size of $n=20,000$ observations and the Matérn covariance function, which adheres to the parameter configuration outlined in Table \ref{tab:accuracy-setting}. We estimate the parameters set $(\sigma^2, \beta, \nu)$ using the BOBYQA optimization algorithm \citep{powell2009bobyqa}, which is a gradient-free method and has great power to search the global optimal value. For our approximation methodologies, we select the classic Vecchia approach with a conditioning size of 30 as our baseline, following the recommendation made by \citep{guinness2018permutation, pan2024gpuaccelerated}. This conditioning size is also chosen based on its demonstrated capability to provide comparable approximation accuracy in terms of KL divergence, as evidenced in Figure \ref{fig:20k-kl-1500} for the classic Vecchia algorithm. In contrast, for the block Vecchia approximation, we set the block count, $bc=1500$, identified as offering a balance between accuracy and computational efficiency relative to the classic approach, with varying conditioning sizes of $(10, 30, 60, 120, 180)$ explored.  

\textcolor{black}{
Following the parameter estimation, we  evaluate the predictive performance of the block Vecchia method. The experimental setup involves the simulated dataset of $20,000$ points used in the parameter estimation, where 90\% of the data is used for training and 10\% for testing. The number of blocks in the block Vecchia method is set to 200. Using the true parameter values from the model in Table \ref{tab:accuracy-setting}, we conduct 1,000 rounds of conditional simulation and calculate the 95\% confidence interval. Then, we report the Prediction Interval Coverage Probability (PICP) based on these simulations \citep{zhao2008statistical, nag2023bivariate}}:
\begin{equation}
    PICP = \frac{1}{n^*}\sum_{i=1}^{n^*} \boldsymbol{1}_{\hat y_{i} \in [L_i, U_i]},
\end{equation}
where $\hat y_i$ is the predicted value at $\bm s_i$, $n^*$ is the total number of new locations, $L_i$ is the lower and $U_i$ is the upper prediction bounds at $\bm s_i$. 

\begin{figure} [htbp]
    \centering
    \subfloat[$\hat \sigma^2$]{\includegraphics[width=0.33\textwidth]{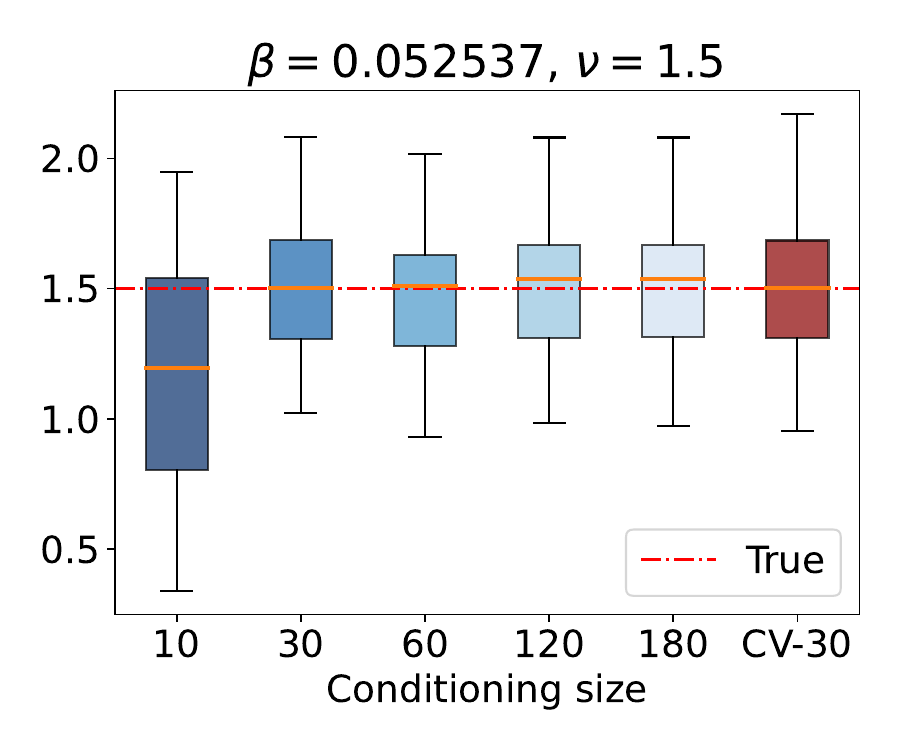}} 
    \subfloat[$\hat \beta$]{\includegraphics[width=0.33\textwidth]{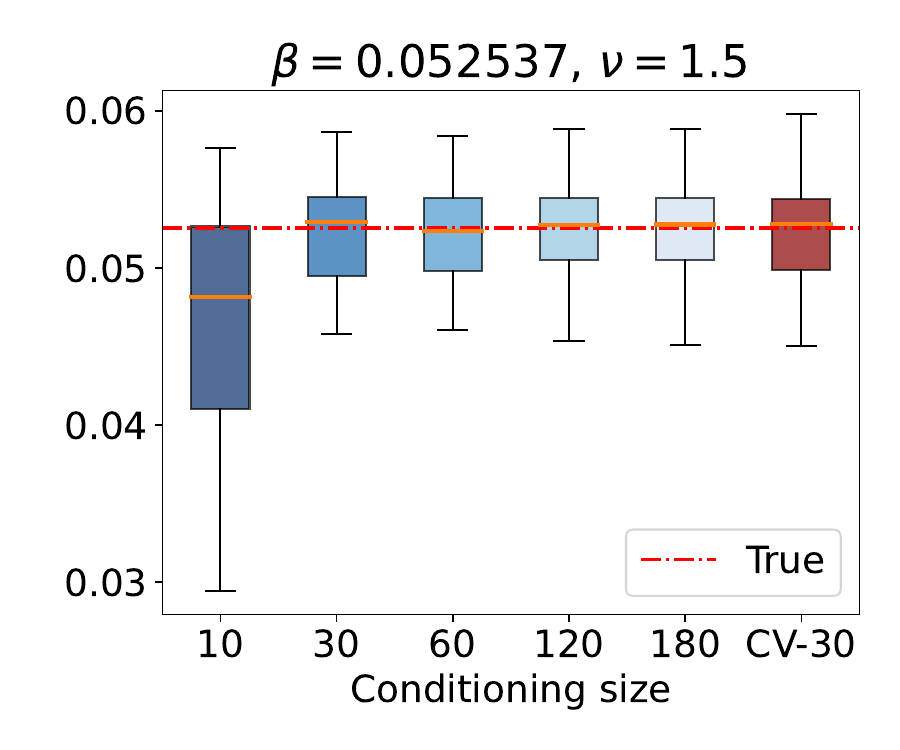}} 
    \subfloat[$\hat \nu$]{\includegraphics[width=0.33\textwidth]{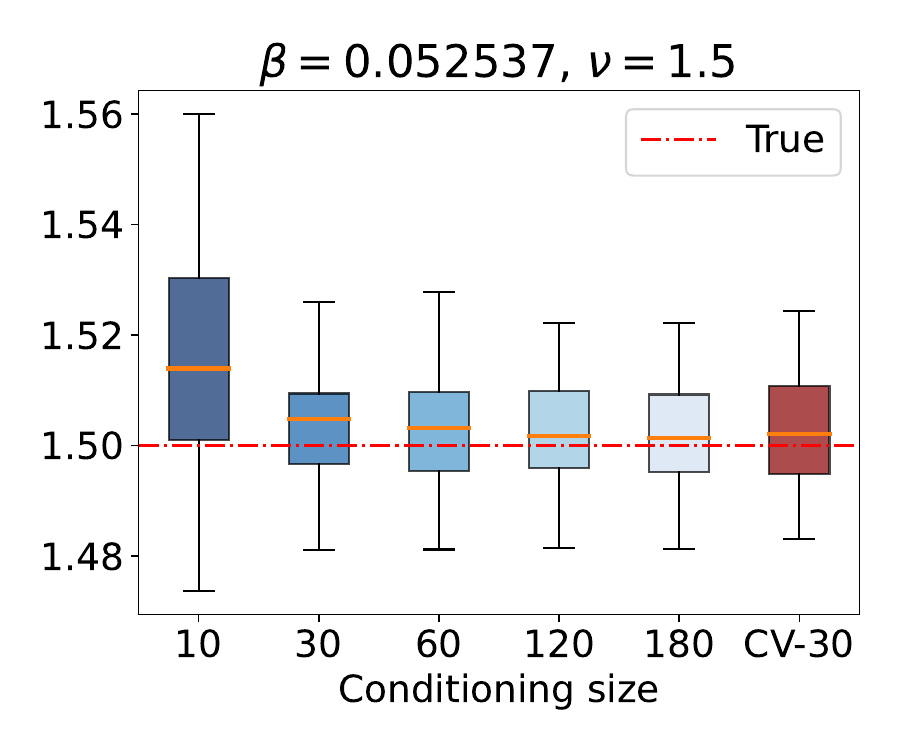}} 
    \caption{Simulations of $n=20,000$ (50 samples) on parameter estimation at $\nu=1.5$ and middle effective ranges. The number of the block is set as 1500 for every block Vecchia settings.}
    \label{fig:simu-20k-15}
\end{figure}

Results of parameter estimation about the case of $\nu=1.5$ and the middle effective range are depicted in Figure \ref{fig:simu-20k-15}, and results of PICP are shown in Figure \ref{fig:predrate}. The additional outcomes sharing the same trend are detailed in the supplementary materials \ref{spp:simu}.

\begin{figure}
    \centering
    \includegraphics[width=0.5\linewidth]{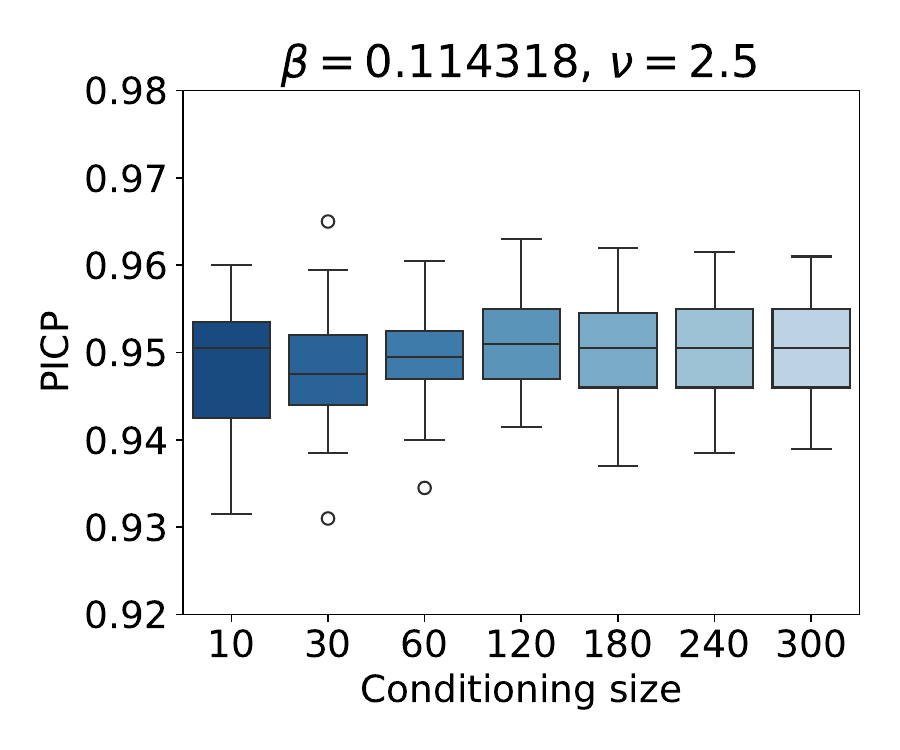}
    \caption{PICP for the simulated datasets.}
    \label{fig:predrate}
\end{figure}

As illustrated in Figure \ref{fig:simu-20k-15}, there is a clear trend toward convergence of the median values of all estimators to the true parameter values with the increase in the number of nearest neighbors. Concurrently, an observable reduction in the variance of all estimators occurs with the increased neighbor count.
In addition, when configured with $30$ nearest neighbors, the block Vecchia algorithm attains parameter estimation accuracy comparable to that of the classic Vecchia method with $30$ nearest neighbors. Notably, this equivalence in estimation accuracy is achieved with a significant increase in computational efficiency, approximately a $14$X speedup. 
   \textcolor{black}{Figure \ref{fig:predrate} assesses the predictive performance of the block Vecchia approximation. The ratio of the block count and the number of predicted locations is 1:10, representing a 10X speedup. As we increase the conditioning size, the PICP becomes higher and its variance becomes smaller, other parameter setups have a similar trend;  see Section \ref{spp:simu}.}

\subsection{Block Vecchia Performance on Large-Scale Real Datasets}

\begin{figure}[htbp]
    \centering
    \subfloat[Residual of Observations (soil)]{\label{fig:soil}\includegraphics[width=0.5\textwidth]{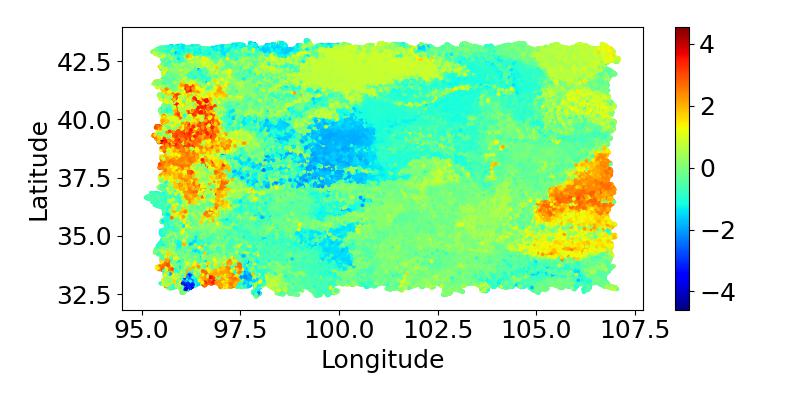}}
    \subfloat[Residual of Observations (wind)]
    {\label{fig:wind}\includegraphics[width=0.5\textwidth]{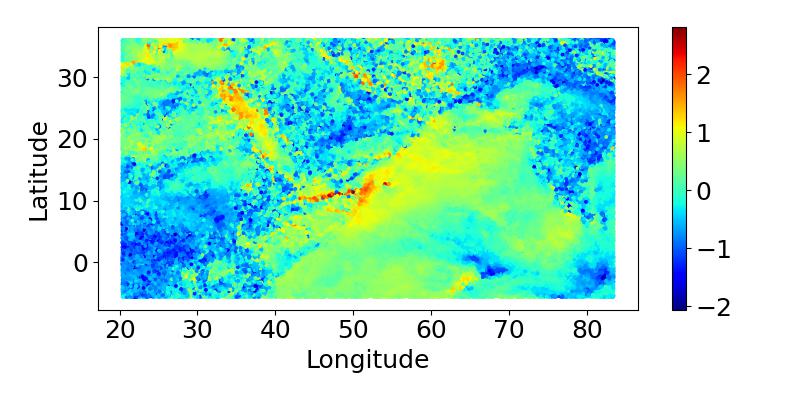}}
    \caption{Real datasets residuals: soil moisture and wind speed.}
    \label{fig:dataset}
\end{figure}

We further compare our block Vecchia algorithm with the exact GP on two subsampled real datasets: a soil moisture dataset from the Mississippi River Basin region and a wind speed dataset from the Middle East region \citep{pan2024gpuaccelerated}. The soil moisture data cover the Mississippi River basin in the United States on January 1, 2004, as reported in \citep{chaney2016hydroblocks}. The dataset, previously used in \citep{huang2018hierarchical, abdulah2018exageostat}, involves Gaussian field modeling and contains 2 million irregularly distributed locations. 
To mitigate computational expenses, we randomly selected 250K locations for the training dataset and 25K for the testing dataset. This subsampling enables us to compare the estimated parameters obtained through the Vecchia approximation with those from the exact modeling at an affordable expense, considering that employing all 2 million locations would pose significant computational burdens \citep{pan2024gpuaccelerated}.
The residuals, shown in Figure \ref{fig:dataset} (a), 
are fitted using a zero-mean Gaussian process model, which incorporates a Mat\'ern covariance function as in (\ref{eq:maternkernel}).  For the block Vecchia methods, using random ordering, seven different conditioning sizes (10, 30, 60, 90, 120, 180, 210) across four block counts (1K, 5K, 15K, 25K) are considered. In the Vecchia approximation and  {\it ExaGeostat}, BOBYQA \citep{powell2009bobyqa} is adopted as the optimization algorithm with the same configuration.
Besides, we use {\it ExaGeostat} \citep{abdulah2018exageostat} to estimate the parameters for the exact Gaussian process. Finally, the estimated parameters are used to perform exact spatial prediction, known as kriging, and the Mean Square Prediction Error (MSPE) is calculated \citep{abdulah2018exageostat}.

The second set of real data addressed in this study was generated using the WRF-ARW (Weather Research and Forecasting - Advanced Research WRF) model, which produced a regional climate dataset for the Arabian Peninsula in the Middle East, as described \citep{powers2008description}. The model has a 5 km horizontal grid spacing, spanning 51 vertical levels, with the highest level at 10 hPa. 
This dataset spans 37 years, providing daily data. Each file records 24 hours of hourly wind speed measurements across 17 atmospheric layers. This study focuses on the dataset on September 1, 2017, starting at 00:00 AM. Our interest is in wind speed measurements at a height of 10 meters above the ground, corresponding to layer 0. Distance calculations in the wind speed dataset match those in the soil moisture dataset \citep{pan2024gpuaccelerated}. The residuals are modeled in the same way as the previous dataset.

\begin{figure}[htbp]
    \centering
    \subfloat[Soil $\hat\sigma^2$]{\includegraphics[width=0.33\textwidth]{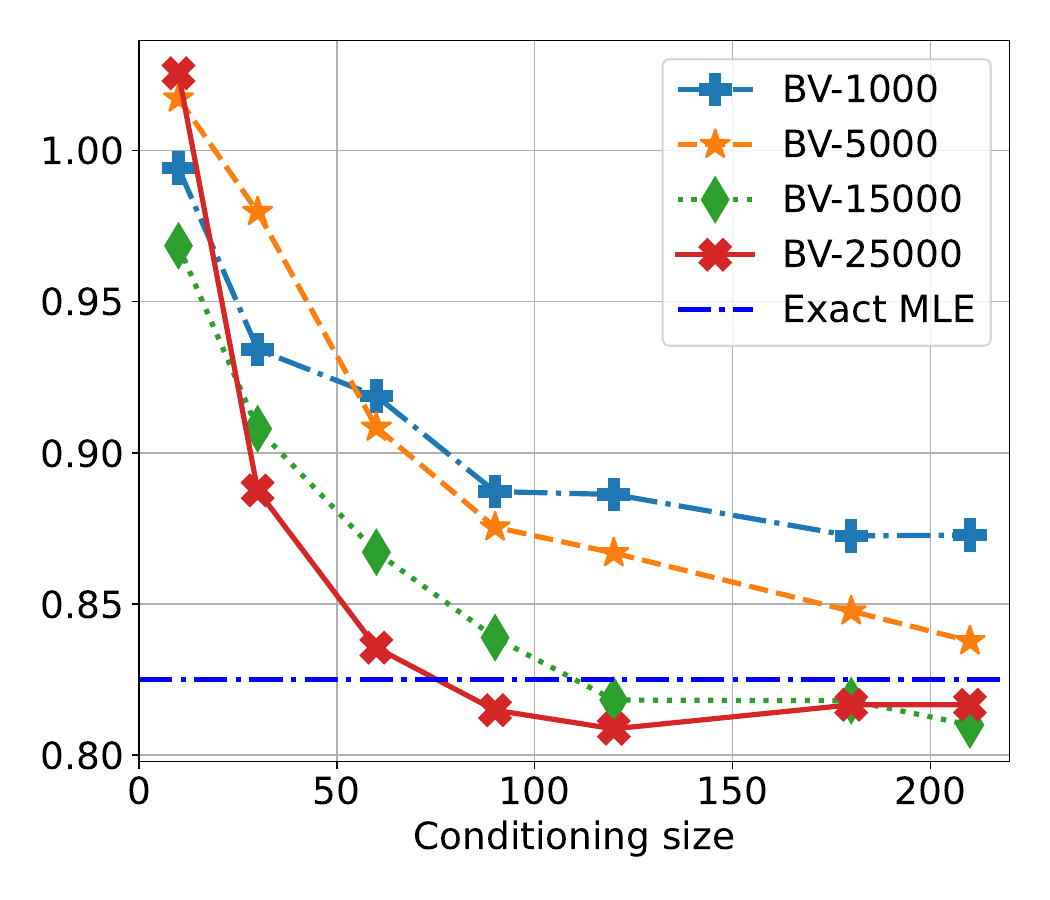}}
    \subfloat[Soil $\hat\beta$]{\includegraphics[width=0.33\textwidth]{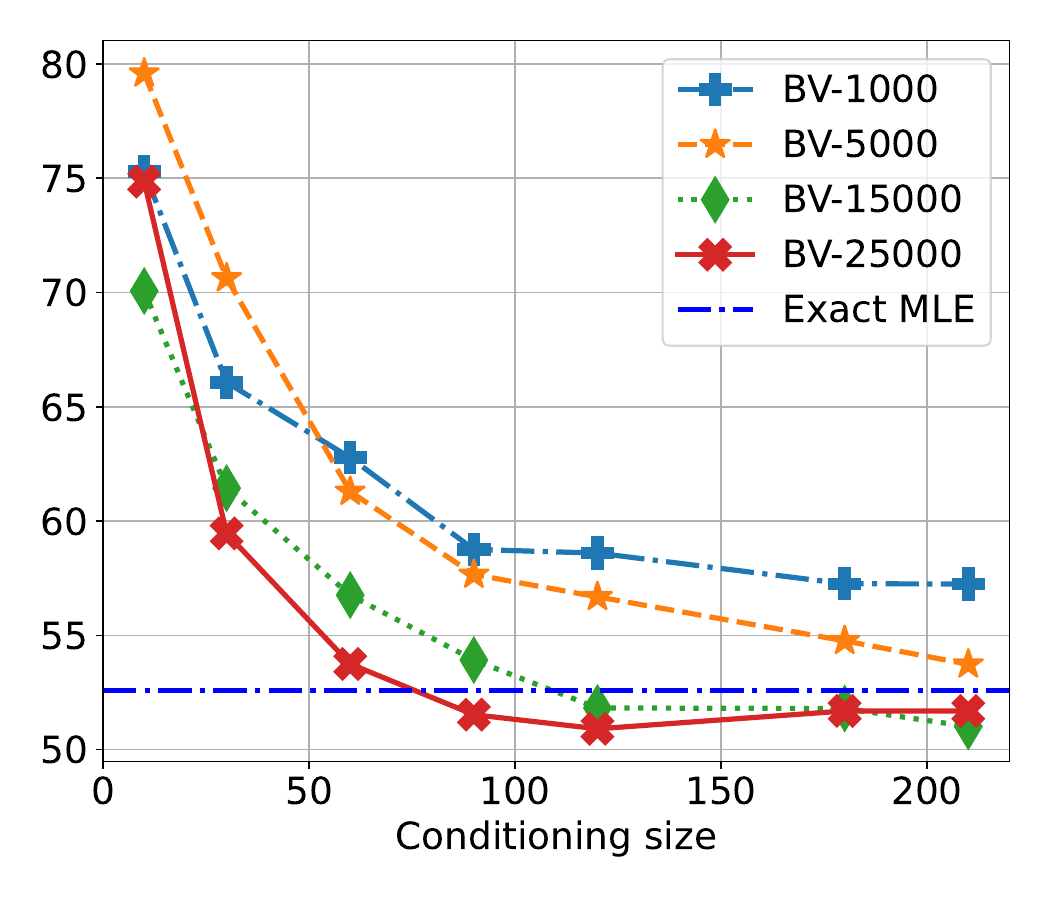}}
    \subfloat[Soil $\hat\nu$]{\includegraphics[width=0.33\textwidth]{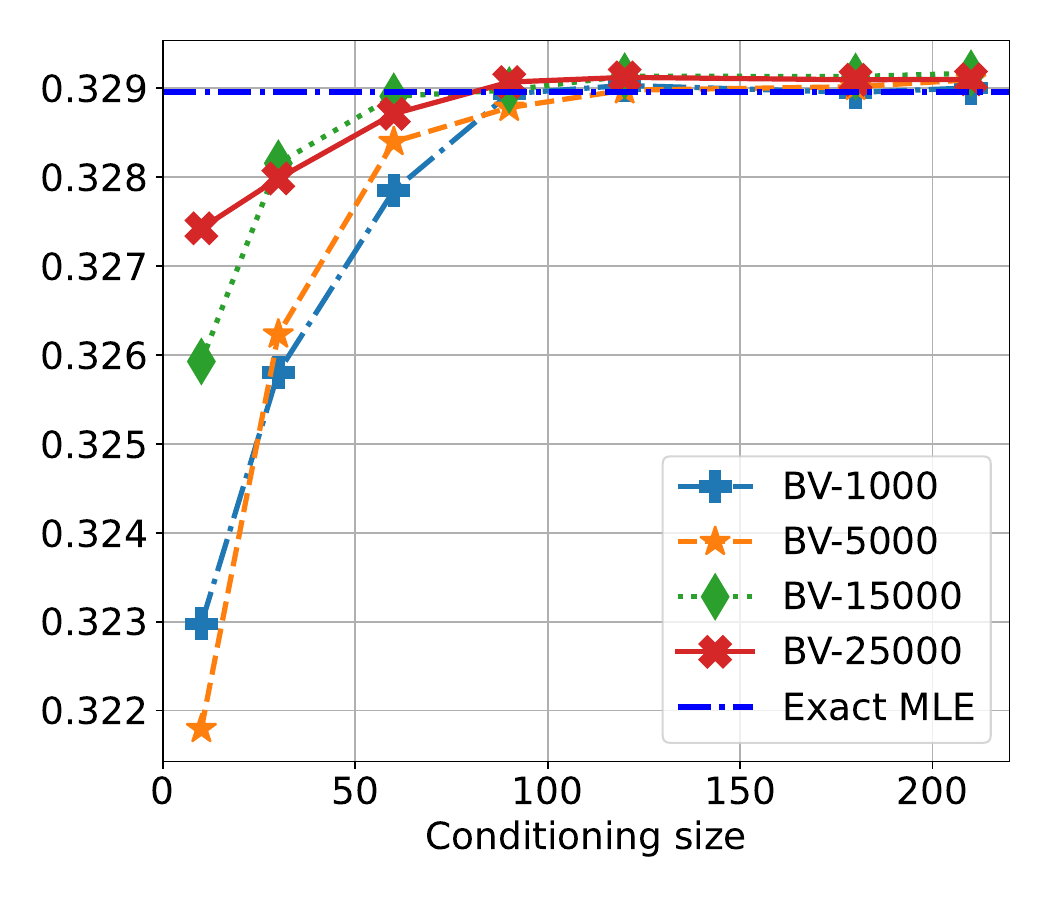}}
    \\
    \subfloat[Wind $\hat\sigma^2$]{\includegraphics[width=0.33\textwidth]{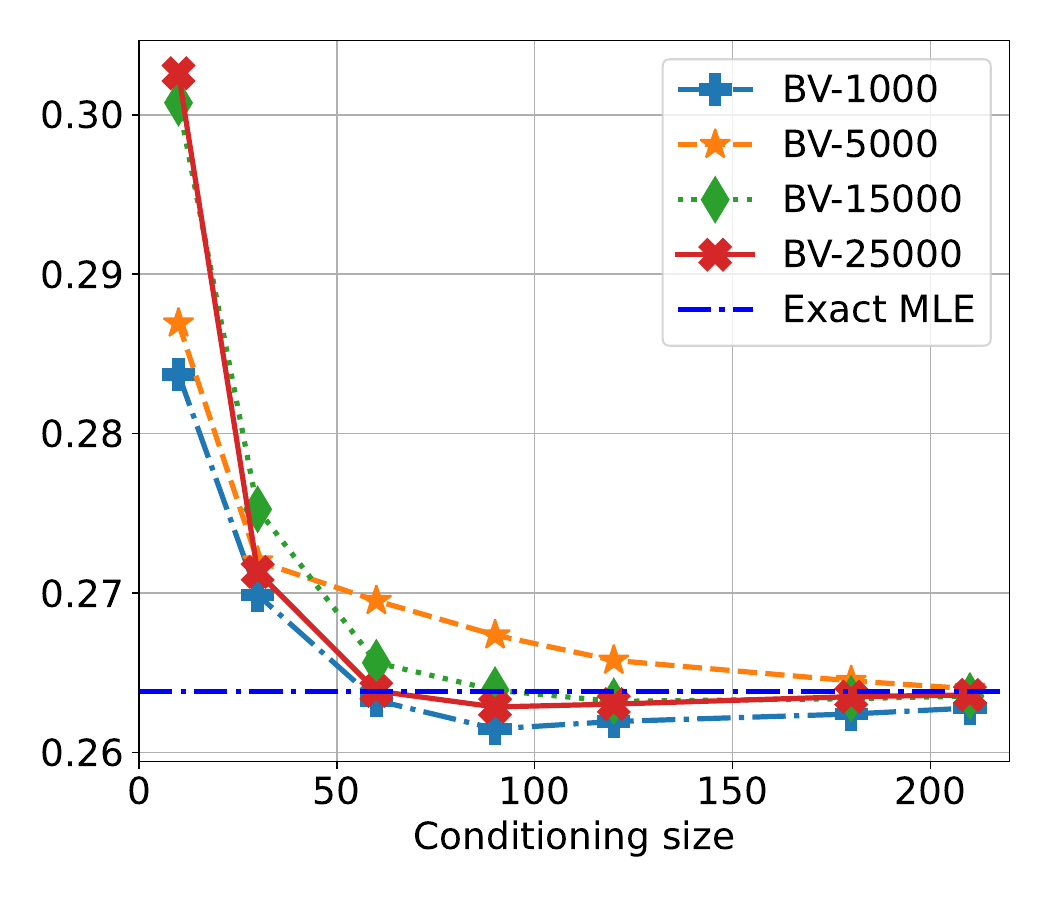}}
    \subfloat[Wind $\hat\beta$]{\includegraphics[width=0.33\textwidth]{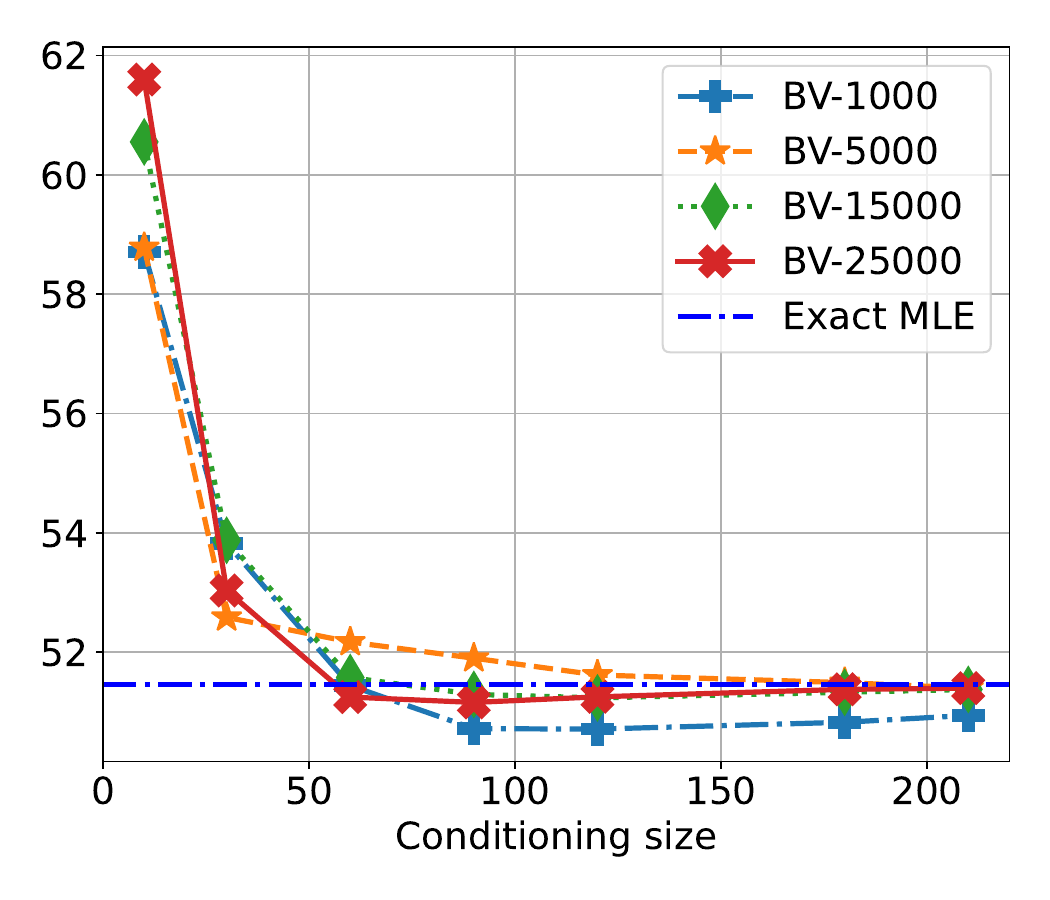}}
    \subfloat[Wind $\hat\nu$]{\includegraphics[width=0.33\textwidth]{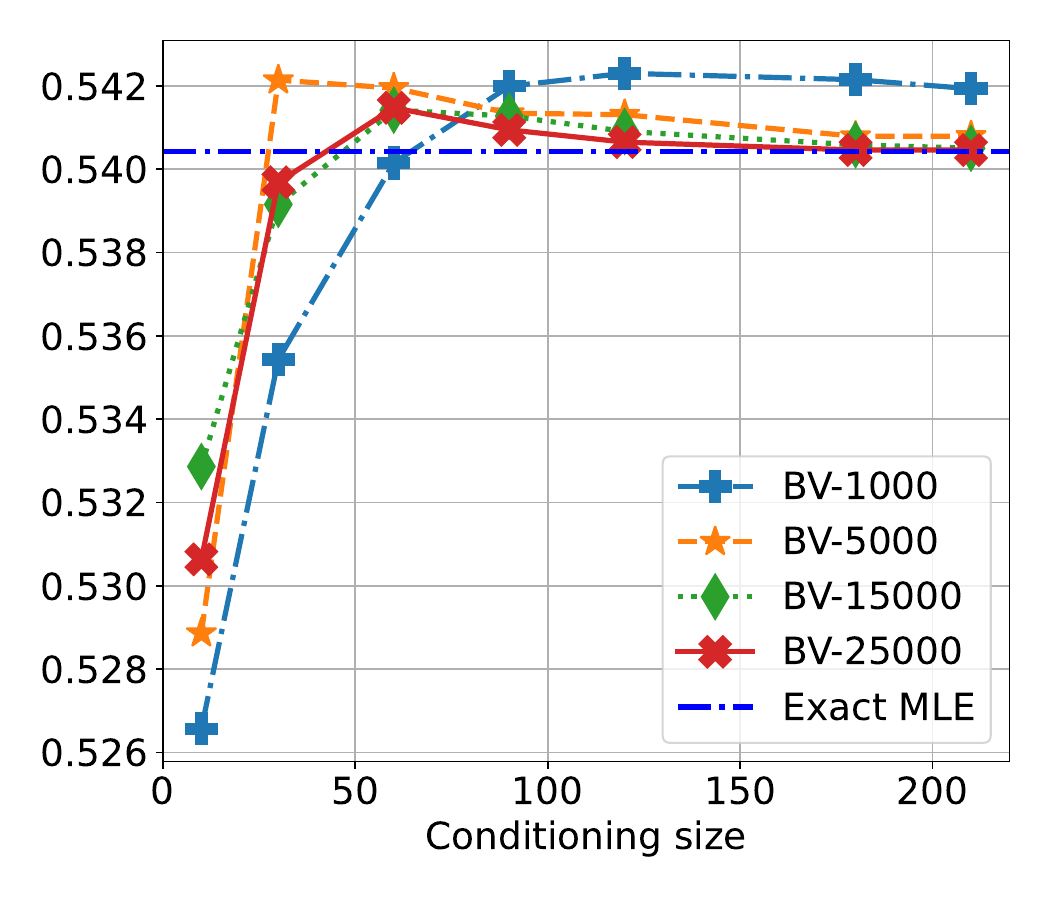}}
    \caption{The estimated parameters using block Vecchia with different block counts compared to  {\it ExaGeoStat} (exact MLE). The first row is the parameter vector for soil moisture, and the second for wind speed.}
    \label{fig:realdataset}
\end{figure}

\begin{figure}[htbp]
    \centering
    \subfloat[Soil]{\includegraphics[width=0.4\textwidth]{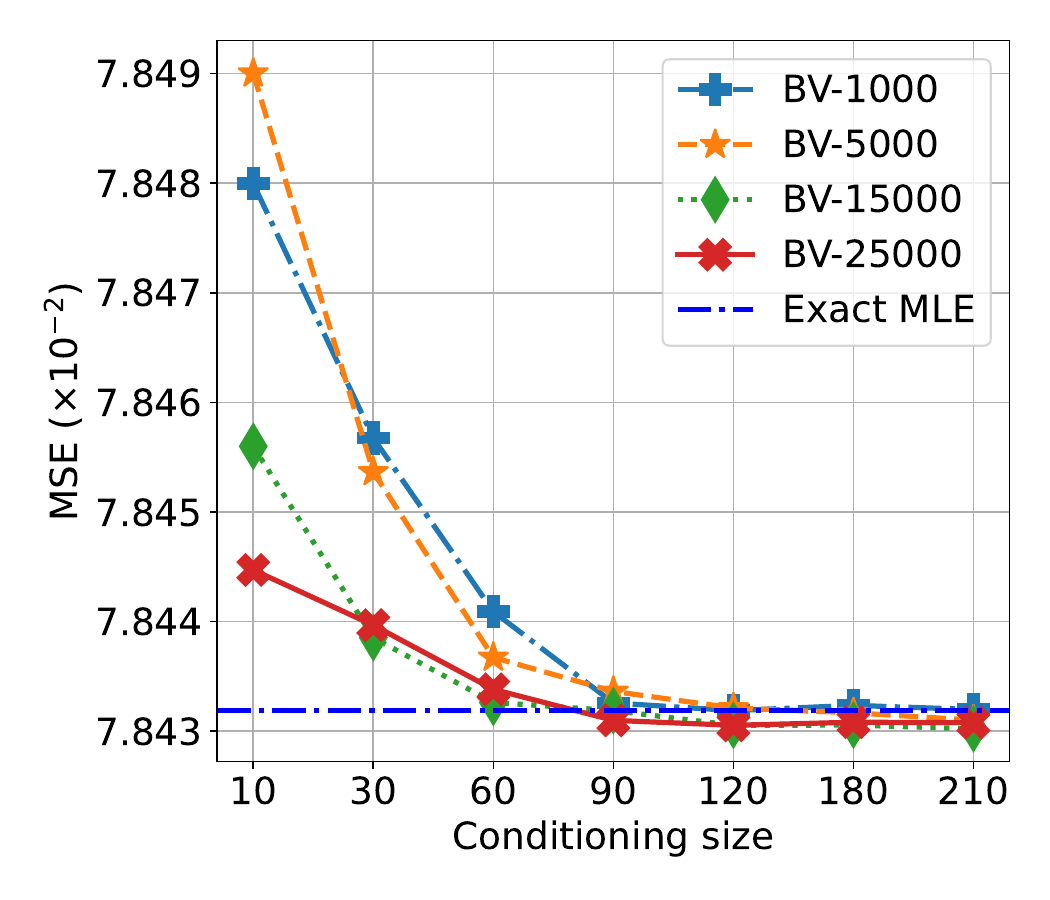}}
    \hspace{10mm}
    \subfloat[Wind]{\includegraphics[width=0.4\textwidth]{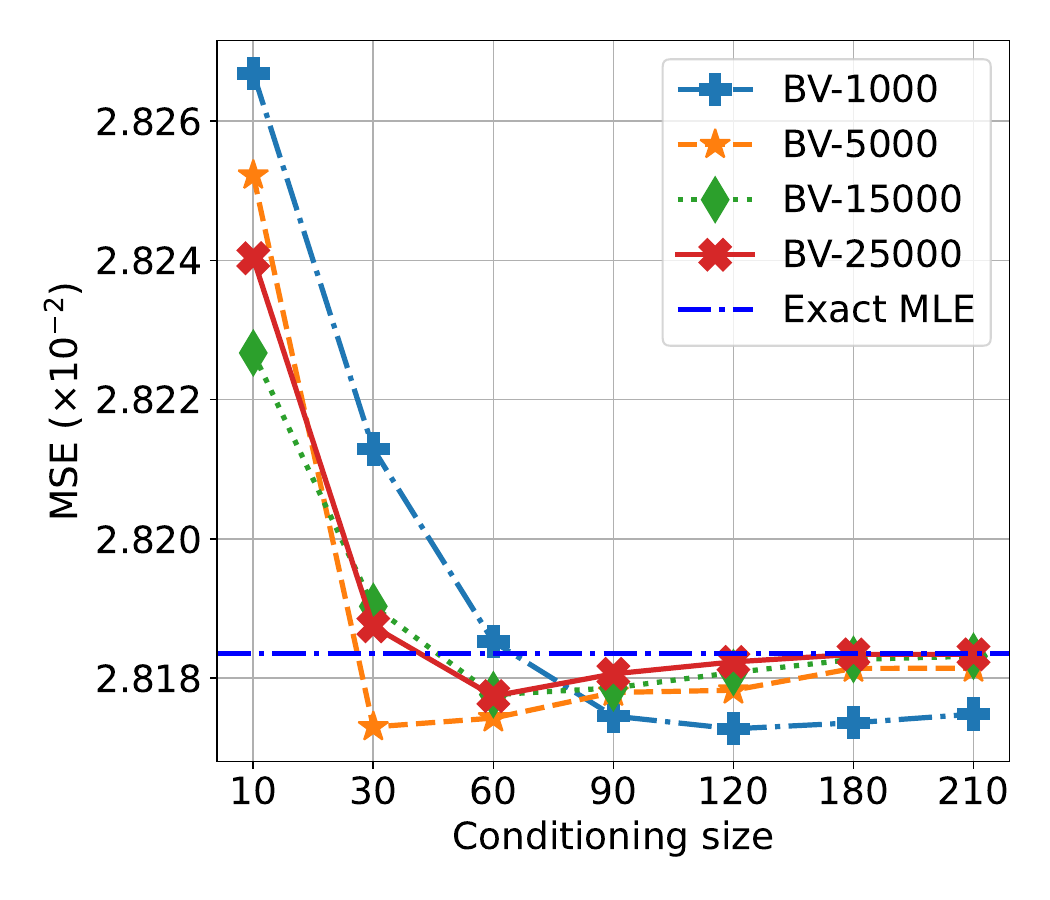}}
    \caption{The MSPE of block Vecchia with different block counts compared to  {\it ExaGeoStat} (exact MLE). The first is the MSPE for soil moisture, and the second for wind speed.}
    \label{fig:realdataset-mse}
\end{figure}

Figure \ref{fig:realdataset} shows the estimated parameters for both datasets, while Figure \ref{fig:realdataset-mse} highlights the MSPE associated with the prediction. We found that the parameter vector $\boldsymbol\theta$, as estimated through the block Vecchia approximation, closely aligns with that obtained via {\it ExaGeoStat} (exact MLE), particularly as the number of conditioning neighbors increases. Figure \ref{fig:realdataset} illustrates that, for both datasets, a conditioning size of 60 with 25,000 block counts is optimal for achieving an estimation close to the exact MLE. Figure \ref{fig:realdataset-mse} further demonstrates that the block Vecchia approximation achieves a prediction error remarkably close to the actual values when predicting missing data.

\section{Application to 3D Wind Speed Profiles}
\label{sec:realdataset}

\textcolor{black}{
Understanding 3D wind speed profiles is vital across multiple disciplines due to their significant impact on various environmental and human activities. 
For example, 
in meteorological forecasting, accurate 3D wind profiles are essential for initializing and running numerical weather prediction models. Winds at different altitudes influence the development and movement of weather systems, such as cyclones and anticyclones. Upper-level winds, like the jet stream, steer weather systems and affect their intensity \citep{kalnay2003}; winds transport heat and moisture vertically and horizontally, affecting temperature distributions and humidity levels. This transport is crucial for predicting phenomena like heatwaves, cold fronts, and precipitation patterns \citep{holton2012}.
In aviation safety and efficiency, turbulence, often caused by wind shear and atmospheric instability, poses safety risks and discomfort. Detailed wind profiles enable pilots and flight dispatchers to anticipate and avoid turbulent areas, enhancing passenger safety \citep{storer2019aviation}.
In renewable energy, the efficiency and feasibility of wind turbines depend on wind speeds at different heights. 3D wind profiles help in selecting optimal turbine hub heights and in designing turbines that maximize energy capture \citep{manwell2010}; detailed vertical wind data allow for precise estimation of potential energy yields, where investors and engineers use this information for planning and development \citep{burton2011}. 
In climate modeling, high-resolution 3D wind data enhance climate models' ability to project future climate scenarios, including temperature and precipitation changes. This information is vital for developing mitigation and adaptation strategies \citep{ipcc2021}.
}

\textcolor{black}{
Traditional GP models are computationally infeasible for such large datasets, particularly when handling the vertical dimension in 3D profiles. However, by breaking down the computations into smaller blocks, the block Vecchia approximation enables the processing of large-scale 3D wind profiles, capturing fine vertical and horizontal wind patterns with much greater efficiency and scalability. This scalability is crucial for analyzing vast datasets across multiple altitudes and spatial regions, which were previously inaccessible due to computational constraints.
In this section, we focus on modeling high-resolution 3D wind speed profiles. The scalability of the block Vecchia method is evaluated on a single GPU (NVIDIA V100 with 32 GB memory) with problem sizes at the million level. The 17 atmospheric layers of the 3D profile are included as an additional input dimension in the 3D profile wind speed data, e.g., Figure \ref{sppfig:3dprofile} illustrates the residuals. The wind speed measurements across the 17 layers span 37 million irregular locations, posing a significant computational challenge for modern computers. We employ a subsampling technique to address this, randomly select 1 million locations and scale the coordinates to $(x,y,z) \in [0,1]^3$. The rest of the experimental configuration remains consistent with the previous section, and we then apply the block Vecchia approximation to estimate the parameters. 
}

\begin{figure}[htbp]
    \centering
    \subfloat[3D Wind $\hat\sigma^2$]{\includegraphics[width=0.45\textwidth]{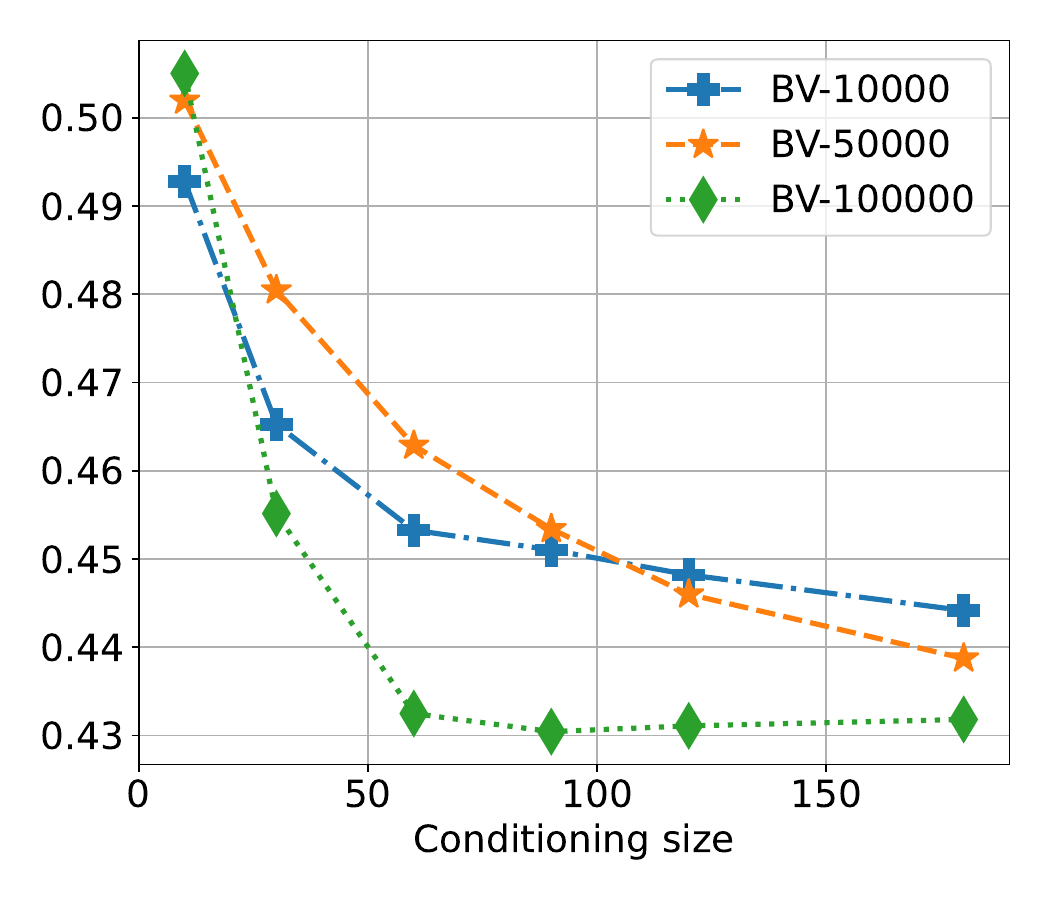}} \hfill
    \subfloat[3D Wind $\hat\beta$]{\includegraphics[width=0.45\textwidth]{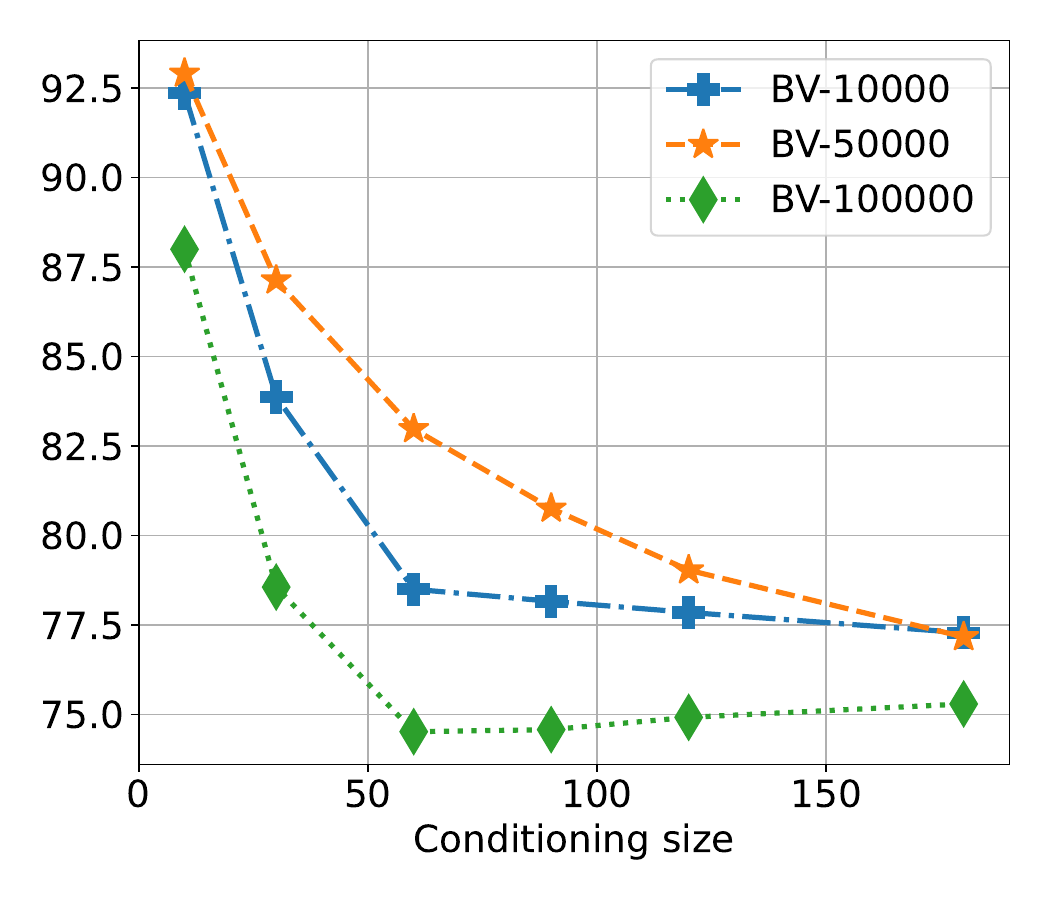}}
    \\
    \subfloat[3D Wind $\hat\nu$]{\includegraphics[width=0.45\textwidth]{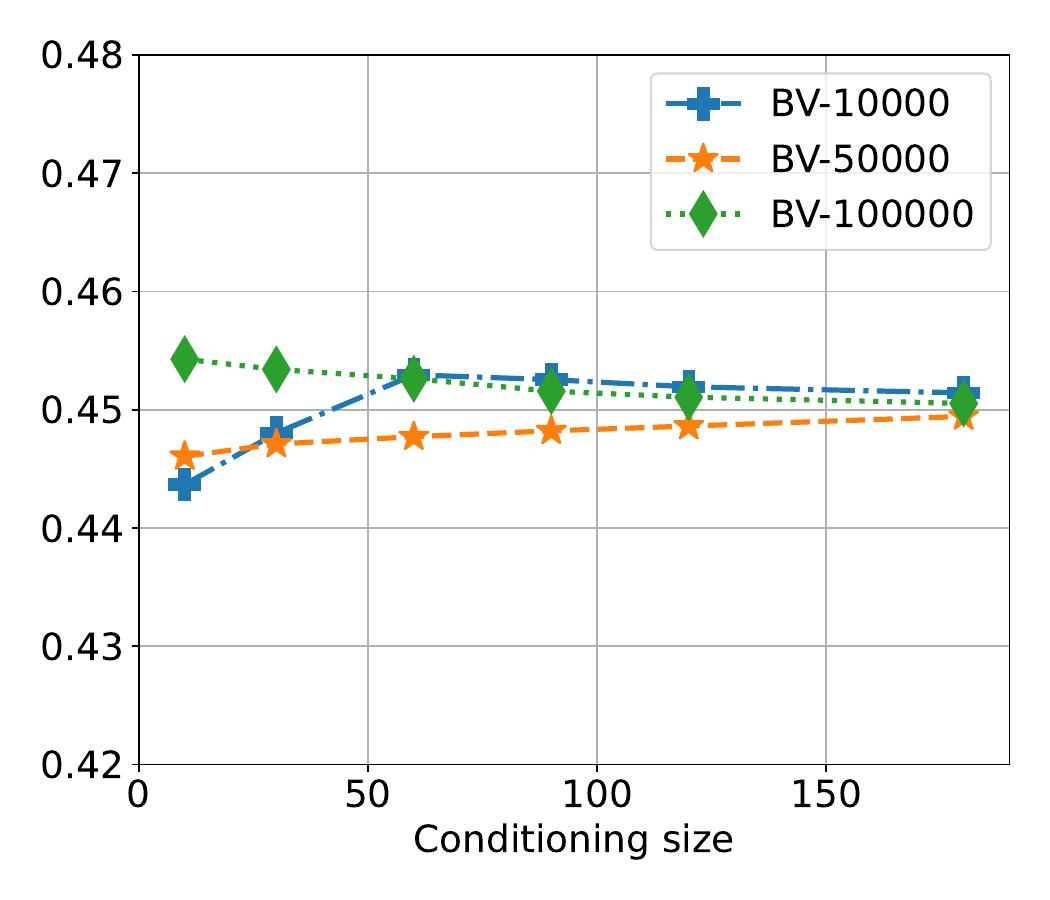}} \hfill
    \subfloat[MSPE]{\includegraphics[width=0.50\linewidth]{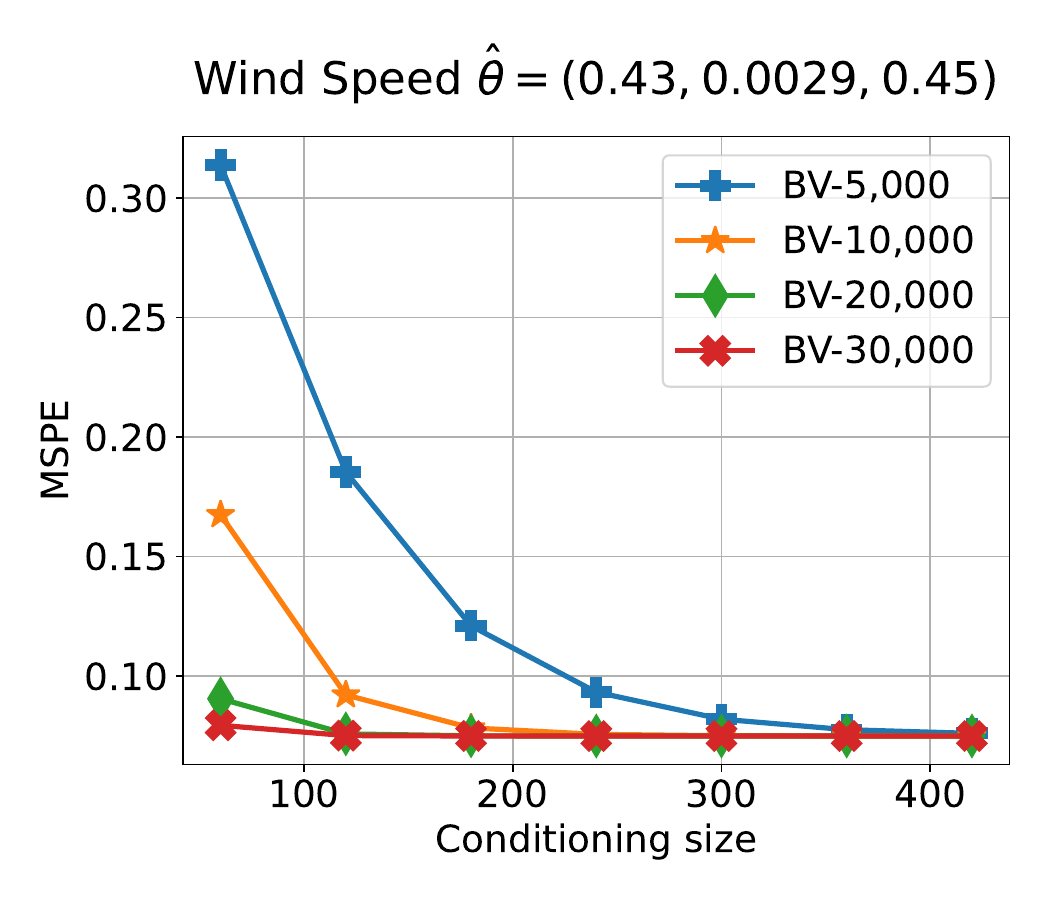}}
    \caption{The estimated parameters using block Vecchia with different block counts (the range parameters in (b) are scaled); and the MSPE of block Vecchia with different block counts for the residuals of 3D wind speed in (d).}
    \label{fig:realdataset-scale}
\end{figure}

\textcolor{black}{
Figure \ref{fig:realdataset-scale} presents the results of parameter estimation for modeling the residuals of 3D wind speed in a million-level context and the MSPE. The number of blocks increases (e.g., 10K, 50K, and 100K), reflecting our interest in improving accuracy in the approximation of the block Vecchia method. The findings indicate that: 1) parameter estimation gradually converges to a specific value as the conditioning size increases; 2) the convergence rate improves with a larger block count. The experiment encourages extending the Vecchia approximation to larger-scale problems. These results demonstrate that larger conditioning sizes and block counts enhance the accuracy of parameter estimation. More importantly, the block Vecchia method can handle much larger problem sizes than the classic Vecchia, facilitating large-scale modeling.
We also assessed the predictive ability of the block Vecchia method. We plug in the converged estimated parameters $\hat \theta$ into the block Vecchia approximation, then conduct 1000 rounds of the conditional simulations, and finally report the MSPE and the standard deviation. Specifically, Figure \ref{fig:realdataset-scale} (d) illustrates the MSPE of the block Vecchia approximation for the residuals of 3D wind speed. We have used different block counts (BV-5000, BV-10000, BV-20000, BV-30000) and varied the conditioning size for prediction in the approximated GP framework, where the block Vecchia method is applied for scalable approximations. It is observed that 1) increasing the conditioning size improves the prediction accuracy, and using a larger block count ($10{,}000$-$30{,}000$) yields more stable results even at smaller conditioning sizes; 2) the gains in prediction accuracy increases at a certain level of conditioning size for all types of block counts. \textcolor{black}{In addition, we provide a full dataset application on soil moisture with 2M points in \ref{spp:realdataset}. }
}

\section{Conclusion}
\label{sec:conclusion}

This study introduces the block Vecchia algorithm and its GPU framework based on batched operations provided by the MAGMA linear algebra library. The algorithm evaluates in batch, simultaneously processing multiple operations or data points, and simultaneously computes the multivariate conditional likelihood of all location blocks to improve efficiency, reduce storage requirements, and enhance scalability in large-scale scenarios. Our numerical study and real datasets analysis provide a deep insight into the block Vecchia algorithm: 1) The analysis reveals that a larger conditioning size and block count can improve modeling accuracy compared to the classic Vecchia algorithm, ensuring efficiency and precision; 2) The sequence of blocks plays a critical role in the accuracy of the approximation, i.e., the random ordering markedly enhances the approximation as the number of blocks increases; 3) The block Vecchia method demonstrates an approximately $80$X speedup compared to the classic Vecchia algorithm without compromising the accuracy of the approximations. This significant enhancement in computational speed, coupled with improved accuracy, represents a substantial advancement in applying Gaussian process models; and 4) scalability: The block Vecchia method allows handling problem sizes $40$X larger than those accommodated by the classic Vecchia algorithm. This scalability enables the algorithm to leverage existing GPUs effectively, making it a powerful tool for addressing large-scale statistical modeling challenges. \textcolor{black}{In a recent study \citep{hazra2024exploring}, the Vecchia approximation (represented by the GpGp package \citep{guinness2021gpgp}) was shown to outperform many popular methods for fitting Gaussian Processes across five large spatial datasets under various parameter settings. This confirms the superiority of the classic Vecchia approach. Accordingly, we applied our proposed block Vecchia algorithm to all datasets from this study, evaluating its performance against the results reported in the study and the local approximation Gaussian Process (laGP, \citep{gramacy2016lagp}). Our results, presented in Table \ref{tab:comparison} in the supplementary material \ref{spp:experiments}, demonstrate that the block Vecchia algorithm is as accurate as GpGp/GpGp0,
while our GPU implementation achieves an average of  $1.5$X speedup in modeling the five given datasets.  
}
\textcolor{black}{According to the asymptotic properties of Vecchia approximation \citep{kang2024asymptotic, zhang2021fixed}, the asymptotic properties of the MLEs from the block Vecchia method will be investigated in our future research. Besides, considering the promising results of the classic Vecchia approximation on high-dimension problems \citep{katzfuss2022scaled, jimenez2023scalable}, it is worth studying computer experiments with block Vecchia.}

\textcolor{black}{
Additionally, we apply the block Vecchia method to large-scale real datasets, 3D wind speed profiles, consisting of millions of data points not addressed in previous studies using other GP methods. In parameter estimation and prediction, the block Vecchia method demonstrates an efficient and accurate approach, with parameter convergence to consistent values and high prediction accuracy across different block counts. The results further show that the block Vecchia approximation, which decomposes the likelihood function into smaller, manageable conditional distributions for efficient parallel processing on modern GPUs, facilitates the use of high-resolution 3D geospatial data in applications such as agriculture, urban planning, and environmental monitoring.
}

\section{Acknowledgment}

This research was supported by King Abdullah University of Science and Technology (KAUST). We gratefully acknowledge the funding and resources provided by KAUST, which made this work possible and has facilitated the research and analysis presented in this study.

\bibliography{Bibliography}

\newpage
\setcounter{page}{1} 
\renewcommand{\thepage}{S\arabic{page}} 
\setcounter{figure}{0} 
\renewcommand{\thefigure}{S\arabic{figure}} 
\setcounter{table}{0} 
\renewcommand{\thetable}{S\arabic{table}} 

\appendix
\section*{Supplementary Materials}

\setcounter{subsection}{0}
\renewcommand{\thesubsection}{S\arabic{subsection}}

\subsection{Block Vecchia Algorithm}
\label{spp:alg}
\begin{algorithm}[H]
\caption{Block Vecchia algorithm with clustering using Batched BLAS}\label{alg:blockvecchia-cluster}
\begin{algorithmic}[1]
\State \textbf{Input:} $n$, $bc$, $g$, $(m_1, \ldots, m_{bc})$, $\zeta$, $\mathcal{K}$
\State \textbf{Initialization:} $(NN, B^{\zeta}_1) = \emptyset$
\State \textbf{Output:} $\ell$ (log-likelihood)

\State \Comment{\textit{(Preprocessing)}}
\State $g:(\boldsymbol{s}_1, \boldsymbol{s}_2, \ldots, \boldsymbol{s}_n) \rightarrow$ $B_1, B_2, \ldots, B_{bc}$ \Comment{Clustering}
\State $\zeta: B_1, B_2, \ldots, B_{bc} \rightarrow B^\zeta_1, B^\zeta_2, \ldots, B^\zeta_{bc}$ \Comment{Block permutation}
\While{$2 \leq j \leq bc$} \Comment{Nearest neighbors}
\State $(NN, B^{\zeta}_{j}) \gets mNearstNeighborsForBlock(\boldsymbol{y}_{B^\zeta_j} \mid \boldsymbol{y}_{B^\zeta_1}, \ldots,\boldsymbol{y}_{B^\zeta_{j-1}}; m_j)$
\EndWhile
\State
\textbf{For} $1\leq i \leq bc$ \textbf{do} \Comment{Kernel vector and matrix}
\State \;\;\;\; $\bm \Sigma^{lk}_{i} \gets \mathcal{K}(\boldsymbol{s}_{B^\zeta_i}, \boldsymbol{s}_{B^\zeta_i})$
\State \;\;\;\; $\bm \Sigma^{con}_i \gets \mathcal{K}(\boldsymbol{s}_{NN, B^{\zeta}_i}, \boldsymbol{s}_{NN, B^{\zeta}_i})$
\State \;\;\;\; $\bm \Sigma^{cross}_i \gets \mathcal{K}(\boldsymbol{s}_{NN, B^{\zeta}_i}, \boldsymbol{s}_{B^\zeta_i})$ 
\State \;\;\;\; $\boldsymbol{y}_{NN, B^{\zeta}_i}$, $\boldsymbol{y}_{B^\zeta_i}$ 
\State
\textbf{End For}

\hrulefill\hrulefill\hrulefill\hrulefill\hrulefill\hrulefill \hrulefill\hrulefill\hrulefill\hrulefill\hrulefill\hrulefill\hrulefill\Comment{\textbf{Correction item computation}}
\State  $\bm\Sigma^{old}_{1:bc} \gets \left(\bm \Sigma^{lk}_{1}, \ldots, \bm \Sigma^{lk}_{bc}\right)^\top$
\textcolor{blue}{
    \State $ \bm L_{1:bc} \gets batchedPOTRF(\bm \Sigma^{con}_{1:bc})$ \Comment{Batched operations}
    \State $ \bm \Sigma^{'cross}_{1:bc} \gets batchedTRSM(\bm L_{1:bc}, \bm \Sigma^{cross}_{1:bc})$
    \State $ \boldsymbol{y}^{'\tau}_{NN, B^{\zeta}_{1:bc}} \gets batchedTRSV(\bm L_{1:bc}, \boldsymbol{y}_{NN, B^{\zeta}_{1:bc}})$
    \State $ \bm \Sigma^{cor}_{1:bc} \gets batchedGEMM \left(transpose(\bm \Sigma^{'cross}_{1:bc}), \bm \Sigma^{'cross}_{1:bc}\right)$
    \State $ \bm \mu^{cor}_{1:bc} \gets batchedGEMV \left( transpose(\bm \Sigma^{'cross}_{1:bc}), \boldsymbol{y}^{'\tau}_{NN, B^{\zeta}_{1:bc}} \right)$
}
\State  $\bm\Sigma^{new}_{1:bc} \gets \bm\Sigma^{old}_{1:bc} - \bm \Sigma^{cor}_{1:bc}$ \Comment{Elememntwise}
\State  $\bm\mu^{new}_{1:bc} \gets \bm \mu^{cor}_{1:bc}$ \Comment{Elememntwise}

\hrulefill\hrulefill\hrulefill\hrulefill\hrulefill\hrulefill \hrulefill\hrulefill\hrulefill\hrulefill\hrulefill\hrulefill\hrulefill\Comment{\textbf{Independent computation
}}

\textcolor{blue}{
    \State $ \bm L'_{1:bc} \gets batchedPOTRF(\bm\Sigma^{new}_{1:bc})$ \Comment{Batched operations}
    \State $ \bm v_{1:bc} \gets batchedTRSV(\bm L'_{1:bc}, \boldsymbol{y}_{B^{\zeta}_{1:bc}} - \bm\mu^{new}_{1:bc})$
    \State $ \bm u_{1:bc} \gets batchedDotProduct \left( transpose(\bm v_{1:bc}), \bm v_{1:bc} \right)$
    \State $ \bm d_{1:bc}  \gets 2\times \log(determinant(\bm L'_{1:bc}))$
}
\While{ $1 \leq i \leq bc$}
    \State $\ell \gets  \ell - \frac{1}{2}\left( u_i  + d_i + l_i\log 2\pi \right)$
\EndWhile

\end{algorithmic}
\end{algorithm}

\begin{algorithm}[H]
\caption{Prediction Using Block Vecchia Approximation}
\label{alg:block-vecchia-prediction}
\begin{algorithmic}[1]
\State \textbf{Input:} $\boldsymbol{y}$, $\boldsymbol{S}$, $\boldsymbol{S}_*$, $bc$, $m$, $\boldsymbol{\theta}$
\State \textbf{Output:}  $\boldsymbol{y}_*$, $\Sigma_{**}$
\State
\State  $g: \boldsymbol{S}_* \rightarrow \{ B^*_1, B^*_2, \ldots, B^*_{bc} \}$. \Comment{Clustering}
\For{$i = 1$ to $bc$} \Comment{Nearest neighbors}
    \State $(NN, B^*_{i}) \gets mNearstNeighborsForBlock(\boldsymbol{y}; m)$
        \State $cov(\boldsymbol{S}_{NN, B^*_i}, \boldsymbol{S}_{NN, B^*_i}) \rightarrow \mathbf{\Sigma}^{\text{con}}_i$  \Comment{Covariance matrices generation}
        \State $cov(\boldsymbol{S}_{NN, B^*_i}, \boldsymbol{S}_{B^*_i}) \rightarrow \mathbf{\Sigma}^{\text{cross}}_i$
        \State $cov(\boldsymbol{S}_{B^*_i}, \boldsymbol{S}_{B^*_i}) \rightarrow \mathbf{\Sigma}^{\text{lk}}_i$
    \State Conditional mean:
    $
    \boldsymbol{\mu}_{B^*_i} = \mathbf{\Sigma}^{\text{cross}}_i \left( \mathbf{\Sigma}^{\text{con}}_i \right)^{-1} \boldsymbol{y}_{NN, B^*_i}
    $
    \State Conditional covariance:
    $
    \mathbf{\Sigma}_{B^*_i} = \mathbf{\Sigma}^{\text{lk}}_i - \mathbf{\Sigma}^{\text{cross}}_i \left( \mathbf{\Sigma}^{\text{con}}_i \right)^{-1} \left( \mathbf{\Sigma}^{\text{cross}}_i \right)^\top
    $
    \State Predicted values: $\boldsymbol{y}_{B^*_i} = \boldsymbol{\mu}_{B^*_i}$
\EndFor
\State $\boldsymbol{y}_* = \left( \boldsymbol{y}_{B^*_1}^\top, \boldsymbol{y}_{B^*_2}^\top, \ldots, \boldsymbol{y}_{B^*_{bc}}^\top \right)^\top$

\State \textbf{Return} $\boldsymbol{y}_*$ and $\{ \mathbf{\Sigma}_{B^*_i} \}_{i=1}^{bc}$ for conditional simulations.
\end{algorithmic}
\end{algorithm}

\subsection{Numerical study with increasing block count}
\label{spp:numerical}

\begin{figure}[htbp]
    \centering
    \subfloat[$\beta=0.052537,bc=200$]{\includegraphics[width=0.33\textwidth]{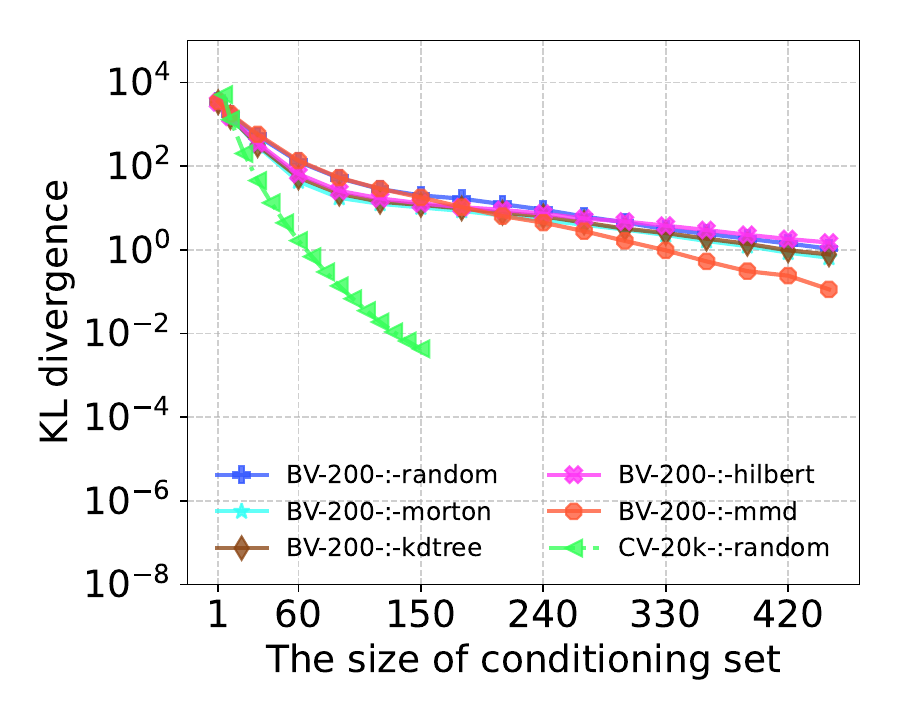}}
    \subfloat[$\beta=0.052537,bc=1500$]{\includegraphics[width=0.33\textwidth]{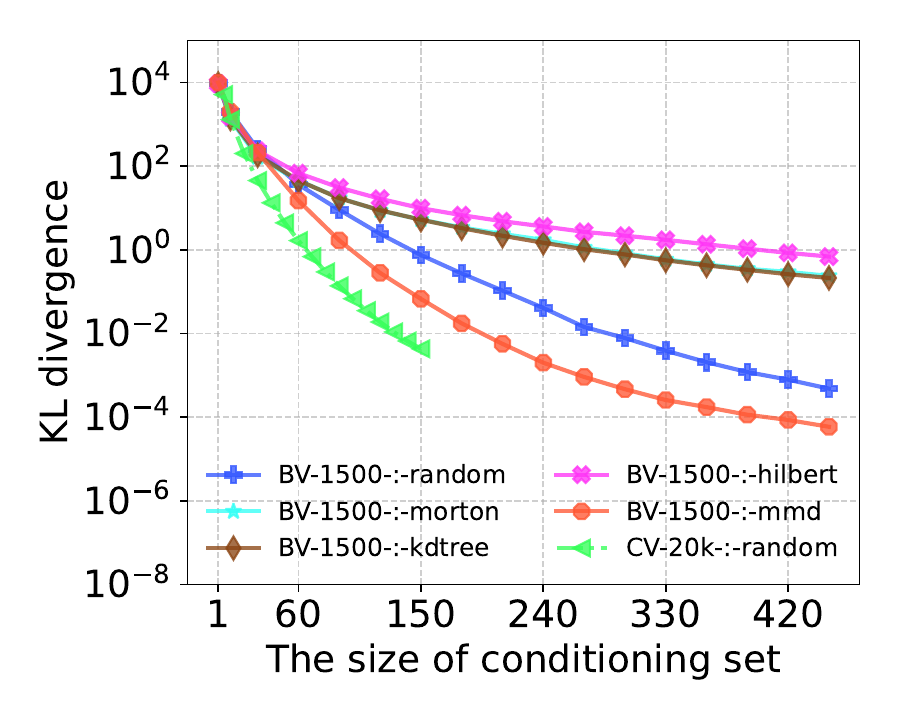}}
    \subfloat[$\beta=0.052537,bc=2500$]{\includegraphics[width=0.33\textwidth]{fig/20k-kl-2500/0.052537_1.500000.pdf}}
    \caption{KL divergence and conditioning size along with different reorderings under medium range/smoothness $\beta=0.052537/\nu=1.5$ and log10 scale.
    }
    \label{fig:20-kl-bc-15-appendix}
\end{figure}

\begin{figure}[htbp]
    \centering
    \subfloat[$\beta=0.026270, \nu = 0.5$]{\includegraphics[width=0.33\textwidth]{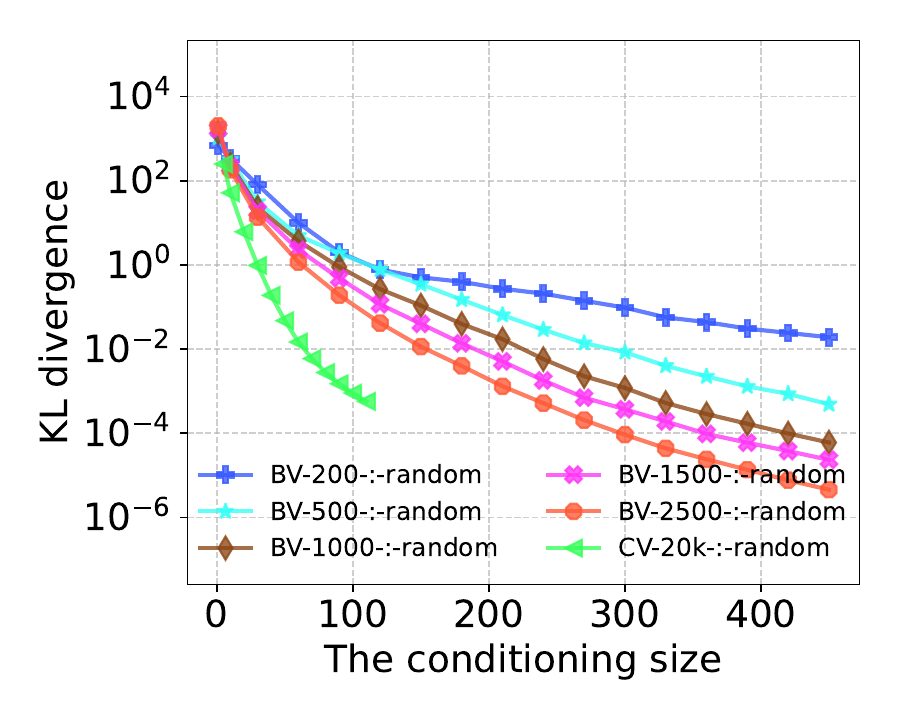}}
    \subfloat[$\beta=0.078809, \nu = 0.5$]{\includegraphics[width=0.33\textwidth]{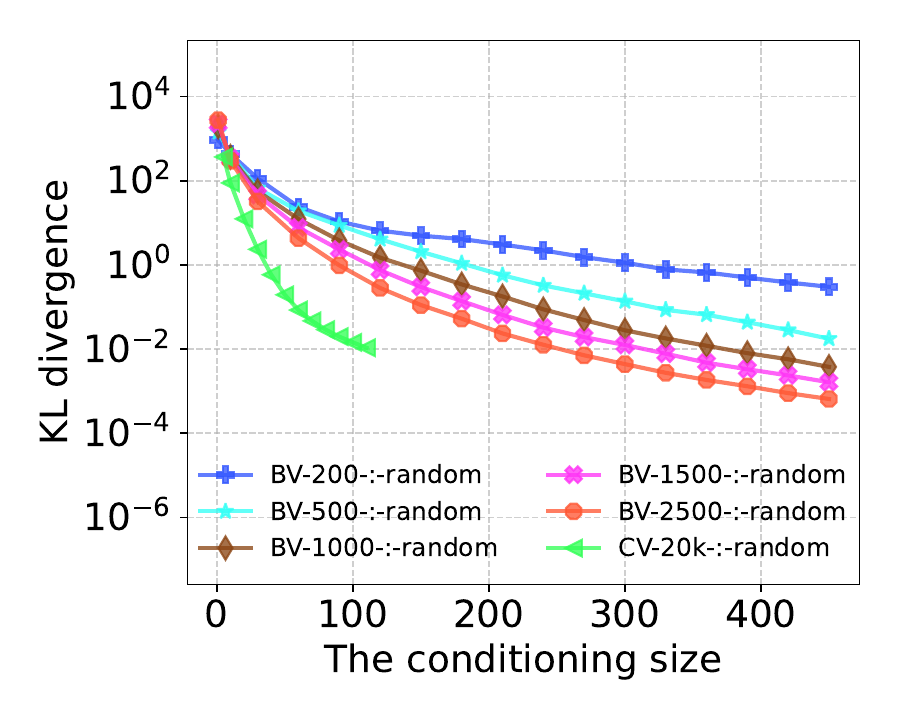}}
    \subfloat[$\beta=0.210158, \nu = 0.5$]{\includegraphics[width=0.33\textwidth]{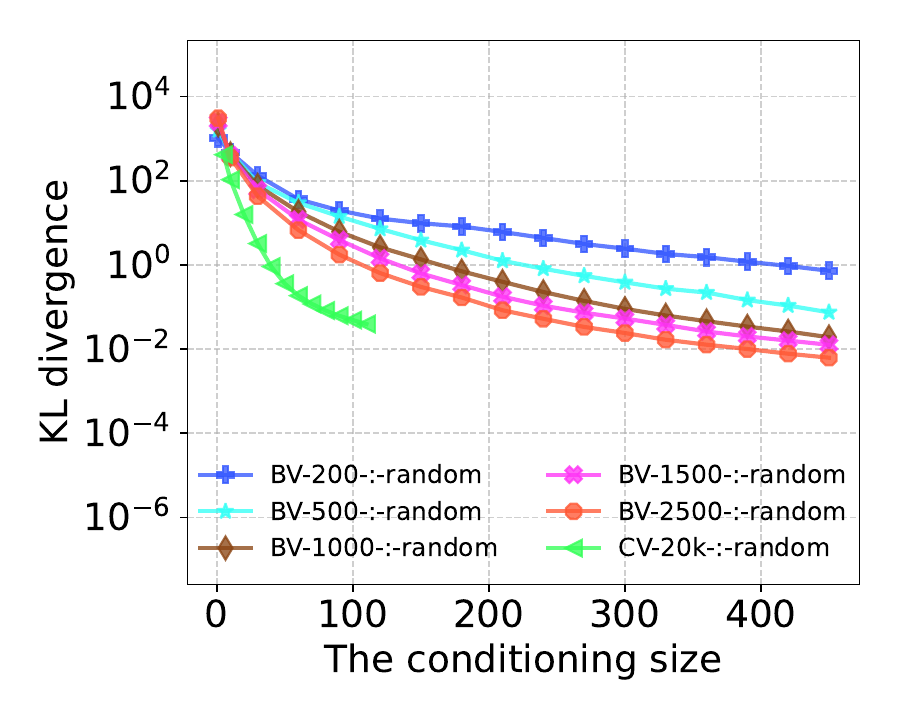}}
    \\
    \subfloat[$\beta=0.017512, \nu = 1.5$]{\includegraphics[width=0.33\textwidth]{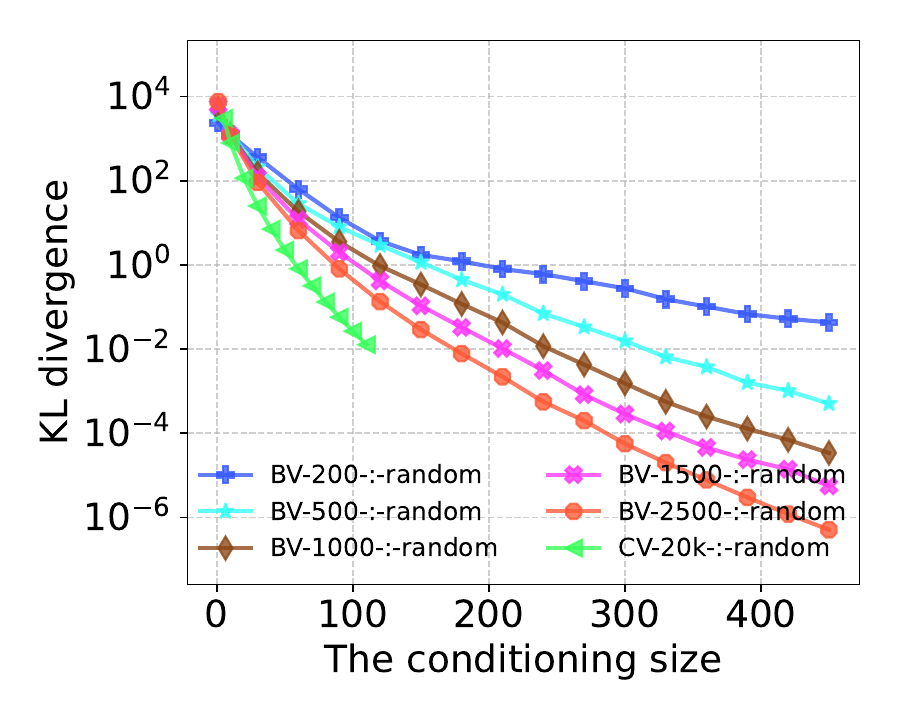}}
    \subfloat[$\beta=0.052537, \nu = 1.5$]{\includegraphics[width=0.33\textwidth]{fig/20k-kl-random/0.052537_1.500000.pdf}}
    \subfloat[$\beta=0.140098, \nu = 1.5$]{\includegraphics[width=0.33\textwidth]{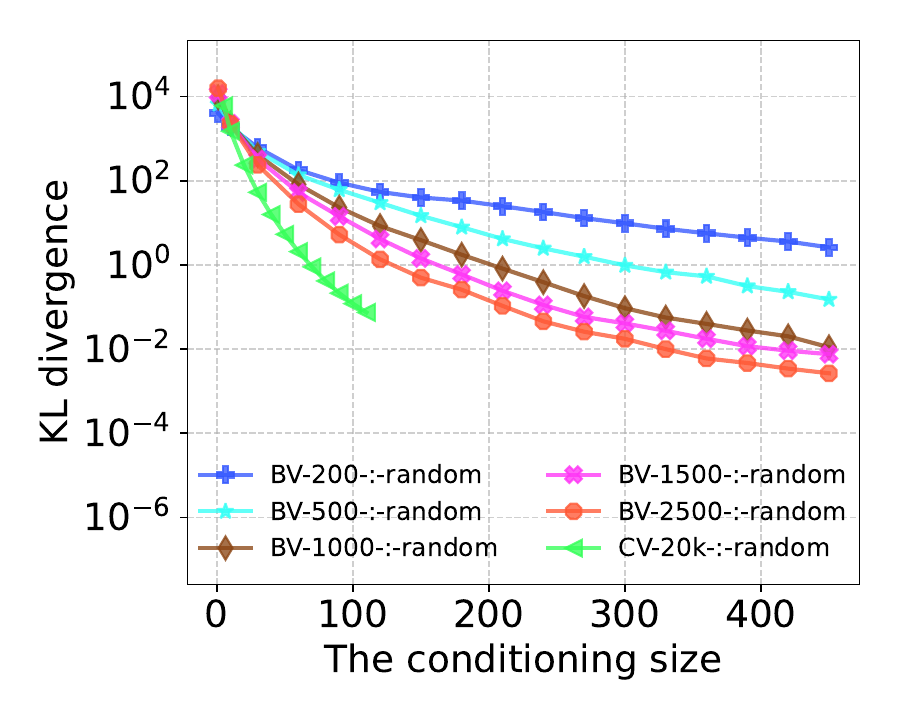}}
    \\
    \subfloat[$\beta=0.014290, \nu = 2.5$]{\includegraphics[width=0.33\textwidth]{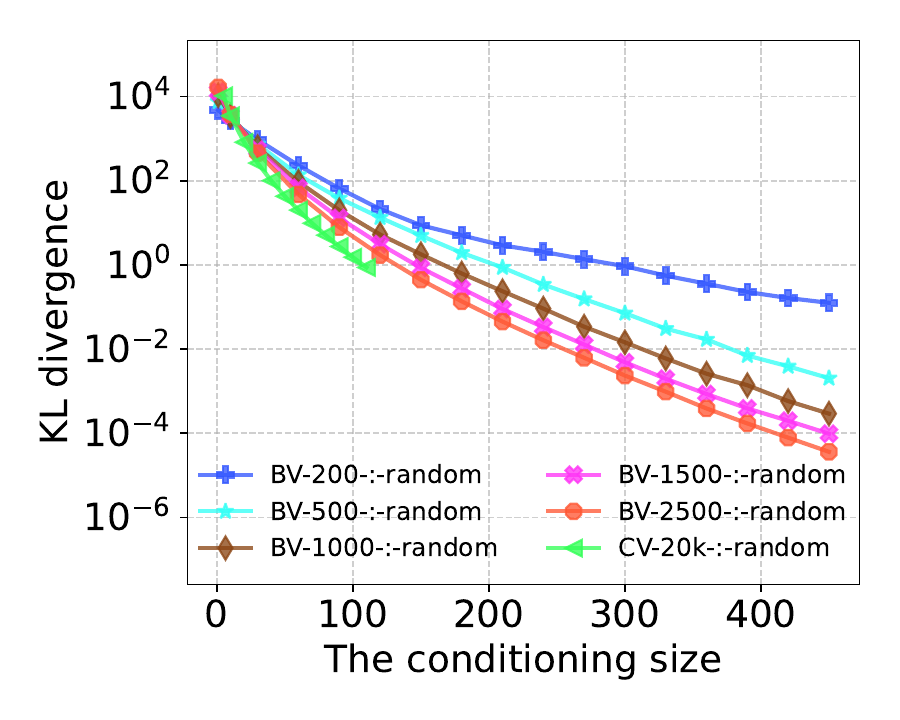}}
    \subfloat[$\beta=0.042869, \nu = 2.5$]{\includegraphics[width=0.33\textwidth]{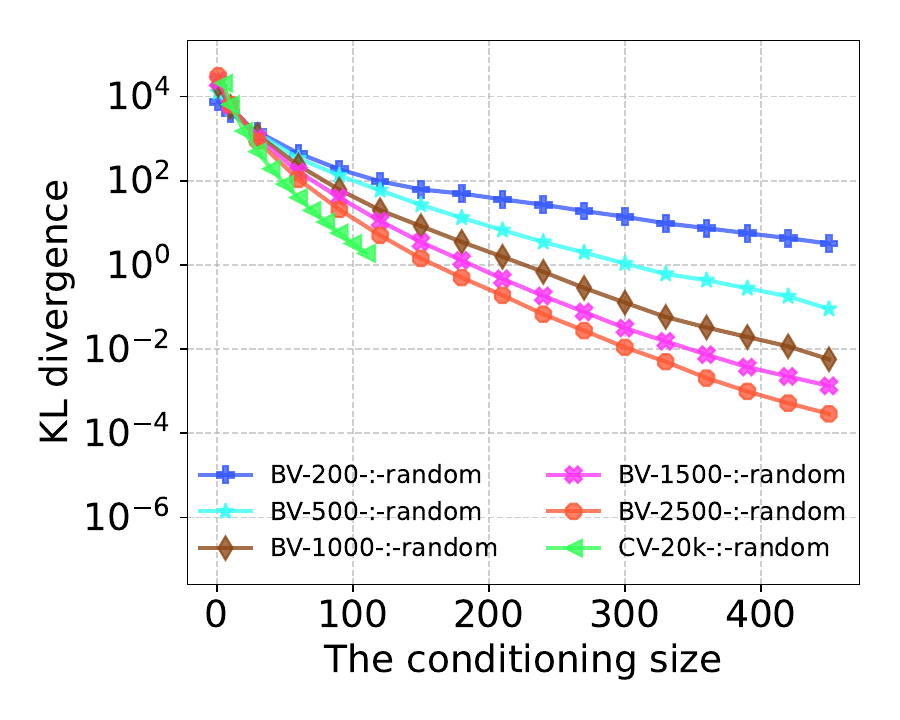}}
    \subfloat[$\beta=0.114318, \nu = 2.5$]{\includegraphics[width=0.33\textwidth]{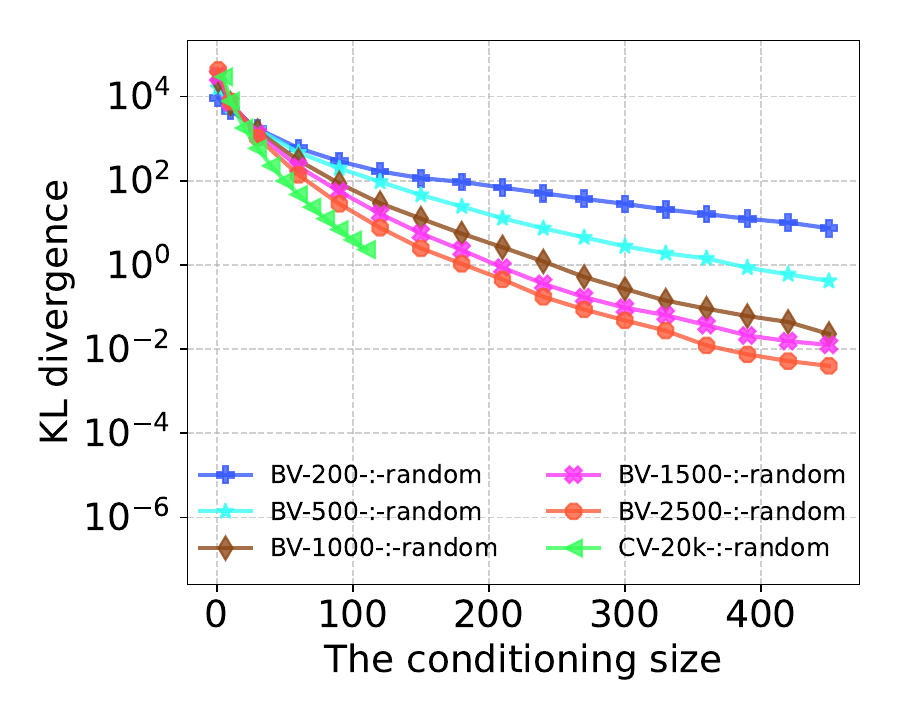}}
    \caption{KL divergence and conditioning size along with increasing block count under log10 scale and random reordering. 
    }
    \label{fig:20-kl-random-appendix}
\end{figure}

\textcolor{black}{
We investigate the numerical accuracy of the block Vecchia algorithm, focusing on the impact of block count and different reorderings. Figure \ref{fig:20-kl-bc-15-appendix} illustrates the KL divergence as the block count increases across three different configurations, with fixed parameters $\beta=0.052537$ and $\nu=1.5$. Our findings indicate that maxmin reordering (mmd in figures) achieves the highest accuracy, random reordering yields near-optimal accuracy, while other reorderings fail to produce promising results. Additionally, a higher block count improves the accuracy of the block Vecchia method for both random and maxmin reorderings. In Figure \ref{fig:20-kl-random-appendix}, we use random reordering as the default, recognizing that it achieves near-optimal accuracy while demanding fewer computational resources compared to maxmin reordering. Across various parameter settings, the KL divergence is plotted against different block counts, consistently confirming that a larger block count leads to more accurate results.
}

\newpage
\subsection{Accuracy of Block Vecchia at Varying Smoothness Levels}
\label{spp:soomthness}

\begin{figure}[htbp]
    \centering
    \subfloat[$\beta=0.026270, \nu = 0.5$]{\includegraphics[width=0.33\textwidth]{fig/20k-kl-1500/0.026270_0.500000.pdf}}
    \subfloat[$\beta=0.078809, \nu = 0.5$]{\includegraphics[width=0.33\textwidth]{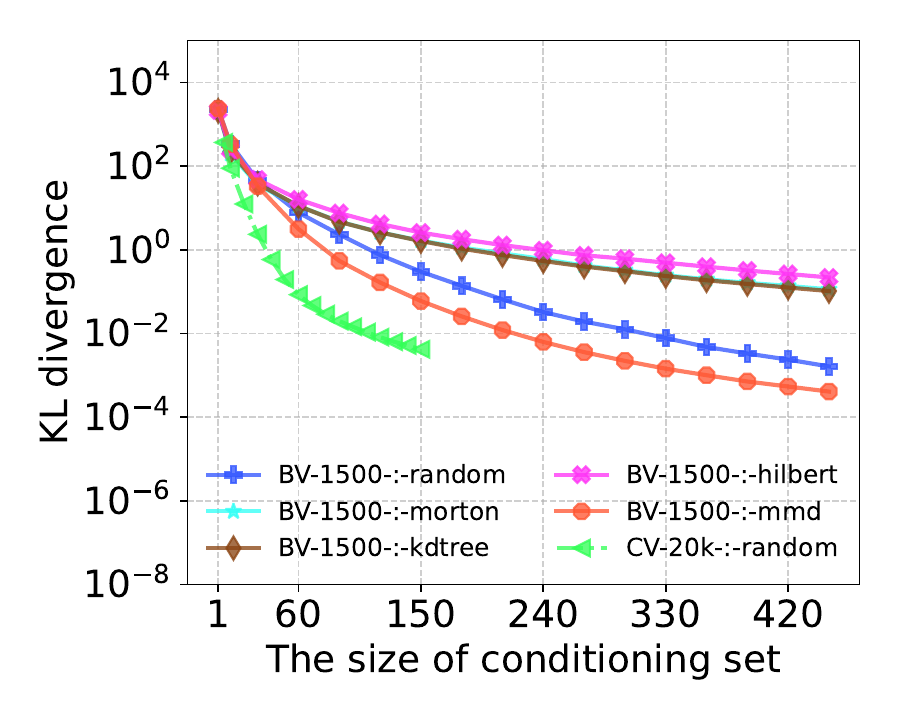}}
    \subfloat[$\beta=0.210158, \nu = 0.5$]{\includegraphics[width=0.33\textwidth]{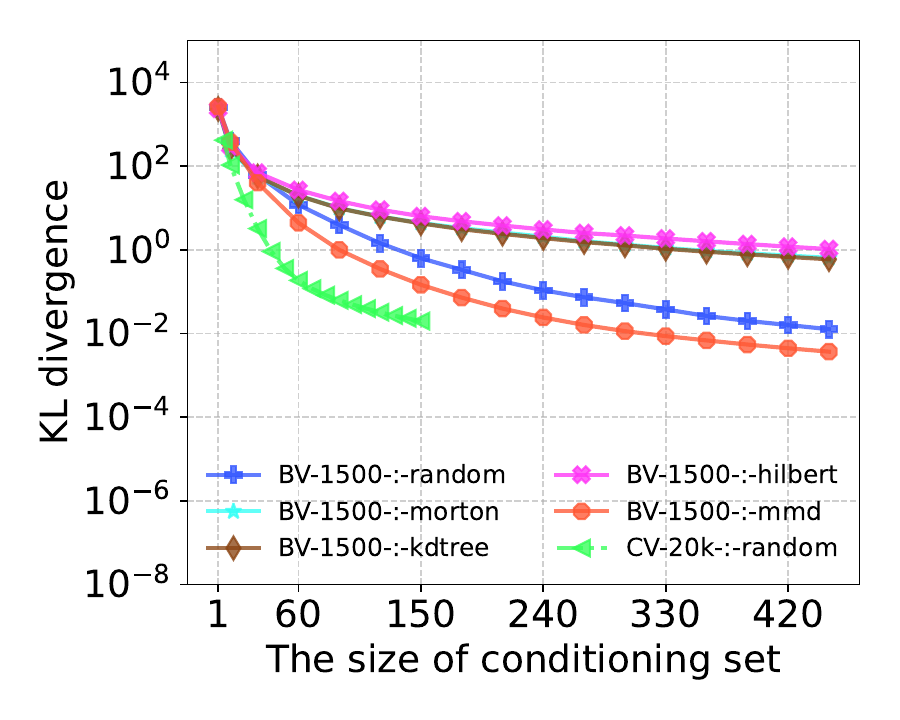}}
    \\
    \subfloat[$\beta=0.017512, \nu = 1.5$]{\includegraphics[width=0.33\textwidth]{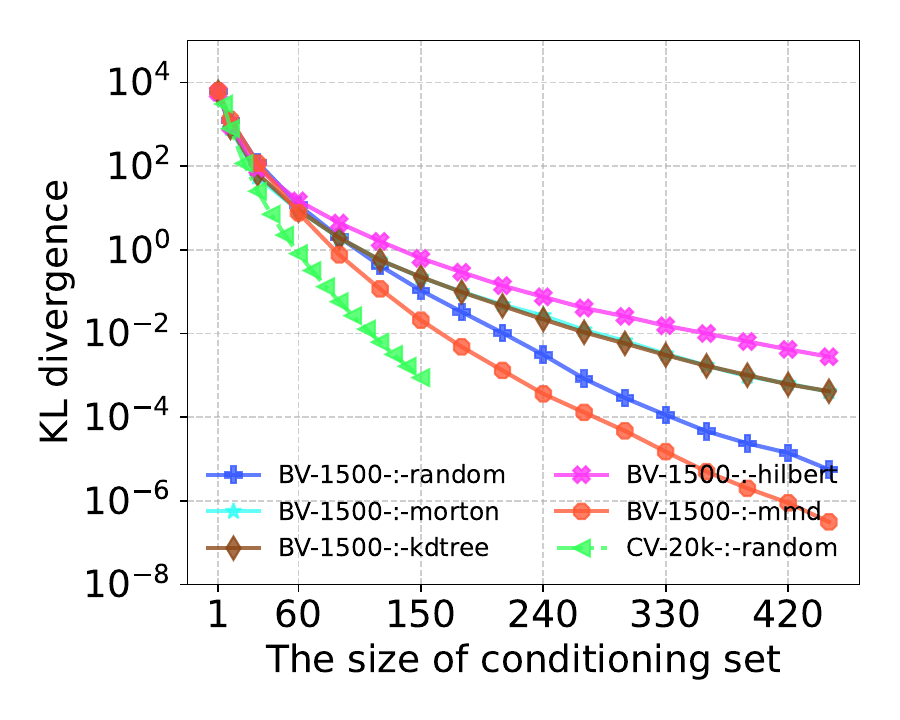}}
    \subfloat[$\beta=0.052537, \nu = 1.5$]{\includegraphics[width=0.33\textwidth]{fig/20k-kl-1500/0.052537_1.500000.pdf}}
    \subfloat[$\beta=0.140098, \nu = 1.5$]{\includegraphics[width=0.33\textwidth]{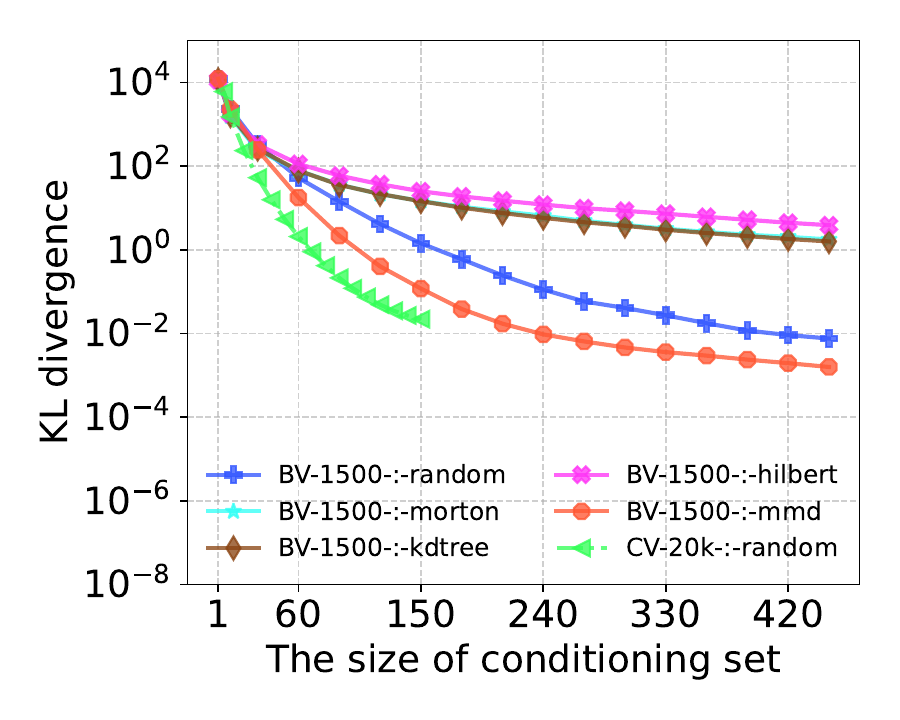}}
    \\
    \subfloat[$\beta=0.014290, \nu = 2.5$]{\includegraphics[width=0.33\textwidth]{fig/20k-kl-1500/0.014290_2.500000.pdf}}
    \subfloat[$\beta=0.042869, \nu = 2.5$]{\includegraphics[width=0.33\textwidth]{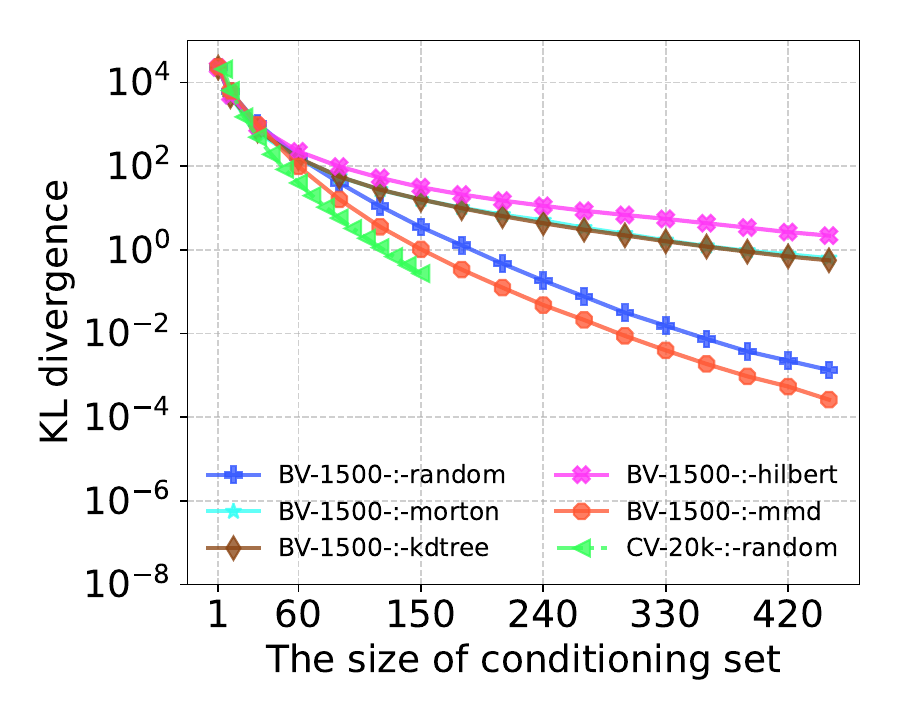}}
    \subfloat[$\beta=0.114318, \nu = 2.5$]{\includegraphics[width=0.33\textwidth]{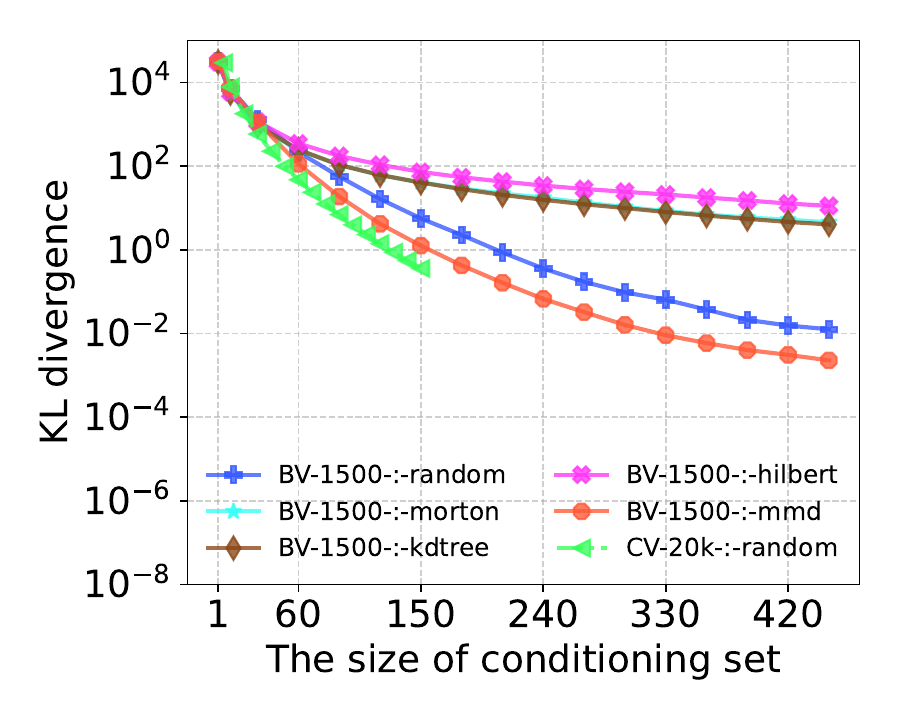}}
    \caption{KL divergence and conditioning size under 20K locations with log10 scale under $bc = 1500$. 
    }
    \label{fig:20k-kl-1500-appendix}
\end{figure}

\textcolor{black}{We investigate the numerical accuracy of the block Vecchia algorithm using Mat\'ern kernel. Figure \ref{fig:20k-kl-1500-appendix} shows that the block Vecchia method improves KL divergence as smoothness increases, narrowing the gap between the block Vecchia and classic Vecchia and achieving higher overall accuracy with larger conditioning set.}

\newpage
\subsection{Additional Results on Block Vecchia Accuracy and Time}
\label{spp:faster}

\begin{figure}[htbp]
    \centering
    \subfloat[$\beta=0.026270, \nu = 0.5$]{\includegraphics[width=0.33\textwidth]{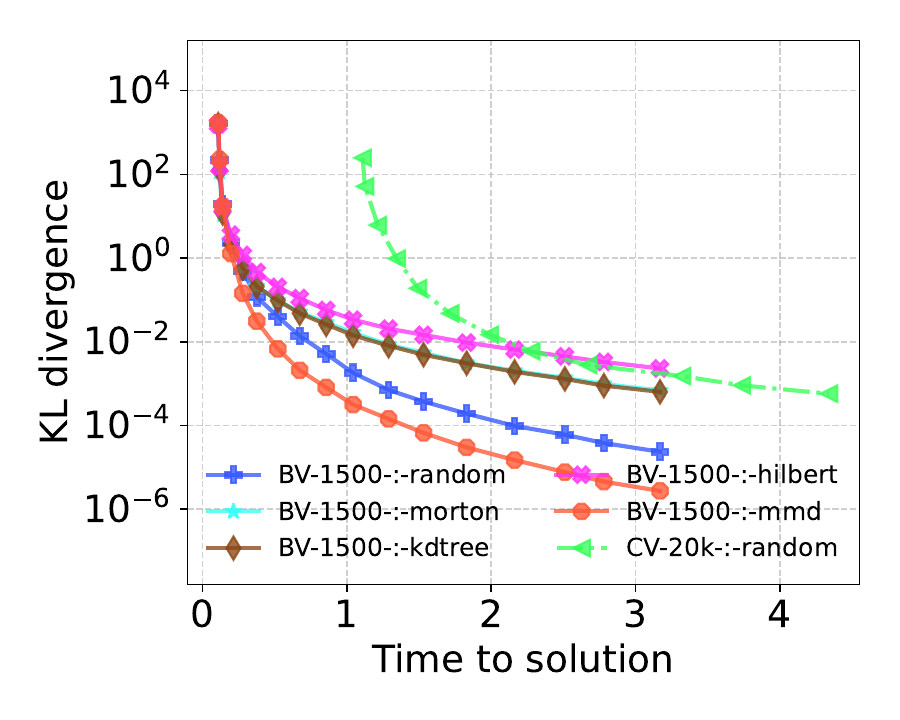}}
    \subfloat[$\beta=0.078809, \nu = 0.5$]{\includegraphics[width=0.33\textwidth]{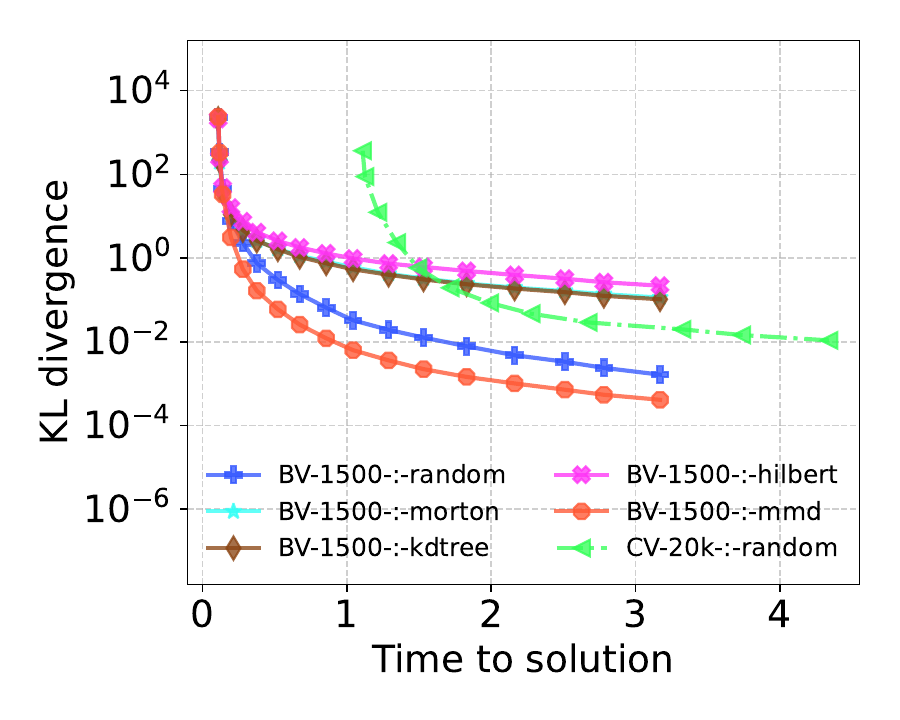}}
    \subfloat[$\beta=0.210158, \nu = 0.5$]{\includegraphics[width=0.33\textwidth]{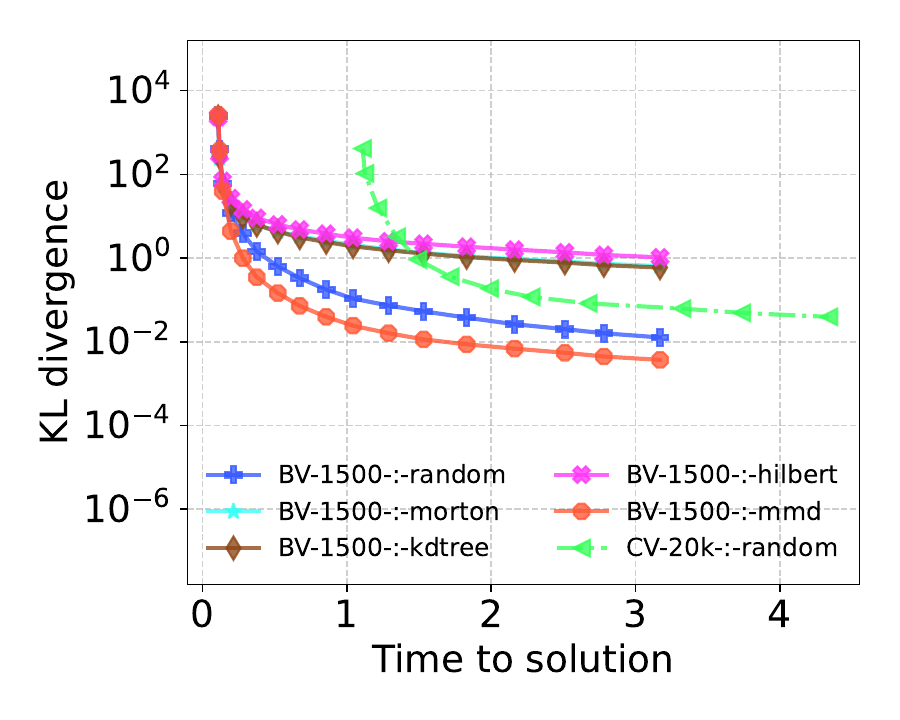}}
    \\
    \subfloat[$\beta=0.017512, \nu = 1.5$]{\includegraphics[width=0.33\textwidth]{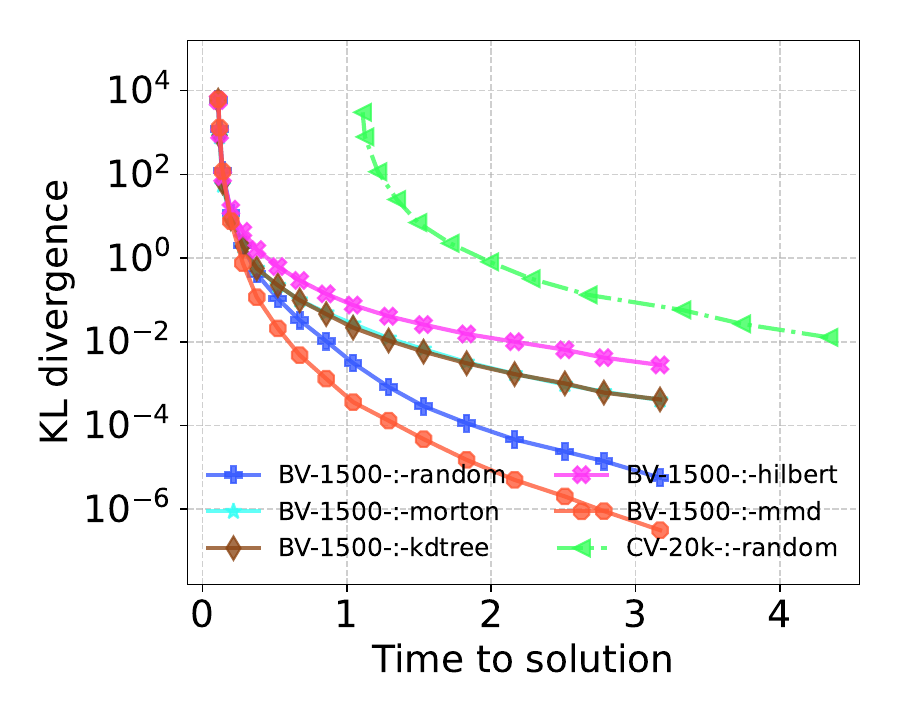}}
    \subfloat[$\beta=0.052537, \nu = 1.5$]{\includegraphics[width=0.33\textwidth]{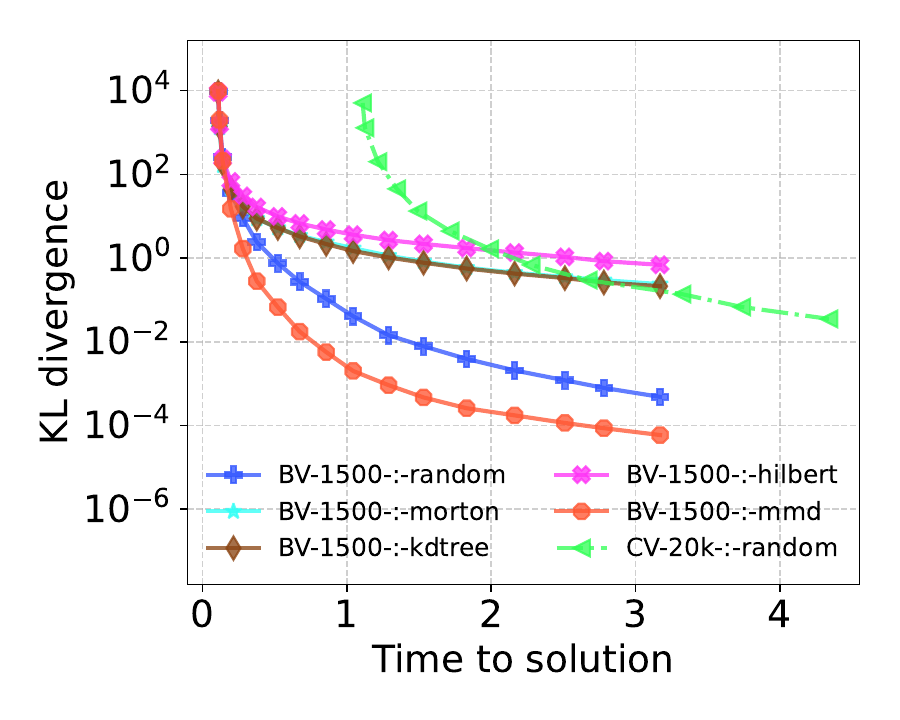}}
    \subfloat[$\beta=0.140098, \nu = 1.5$]{\includegraphics[width=0.33\textwidth]{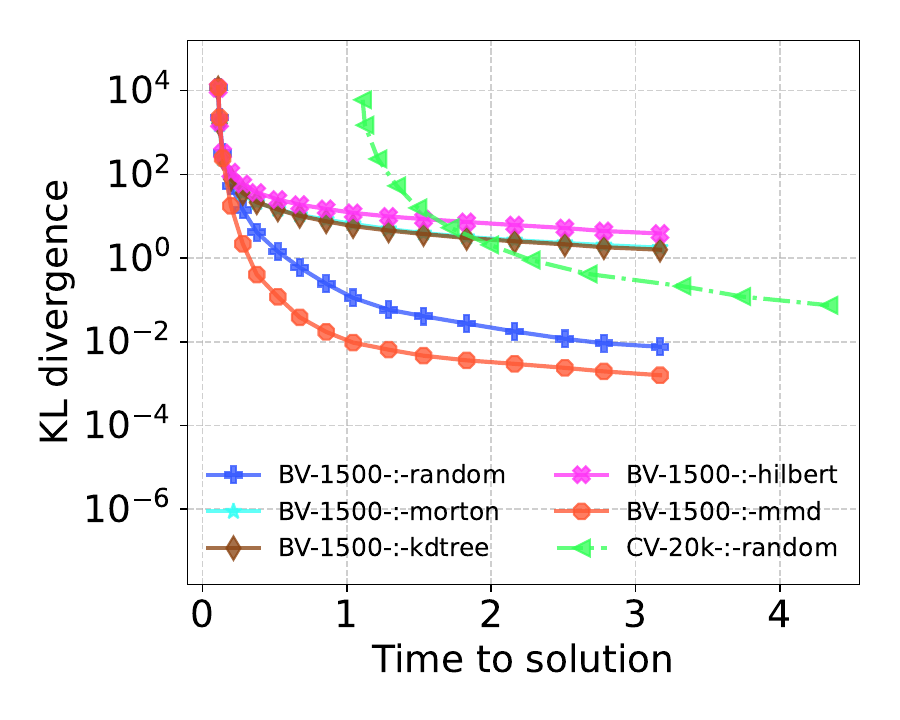}}
    \\
    \subfloat[$\beta=0.014290, \nu = 2.5$]{\includegraphics[width=0.33\textwidth]{fig/20k-kl-1500-time/0.014290_2.500000.pdf}}
    \subfloat[$\beta=0.042869, \nu = 2.5$]{\includegraphics[width=0.33\textwidth]{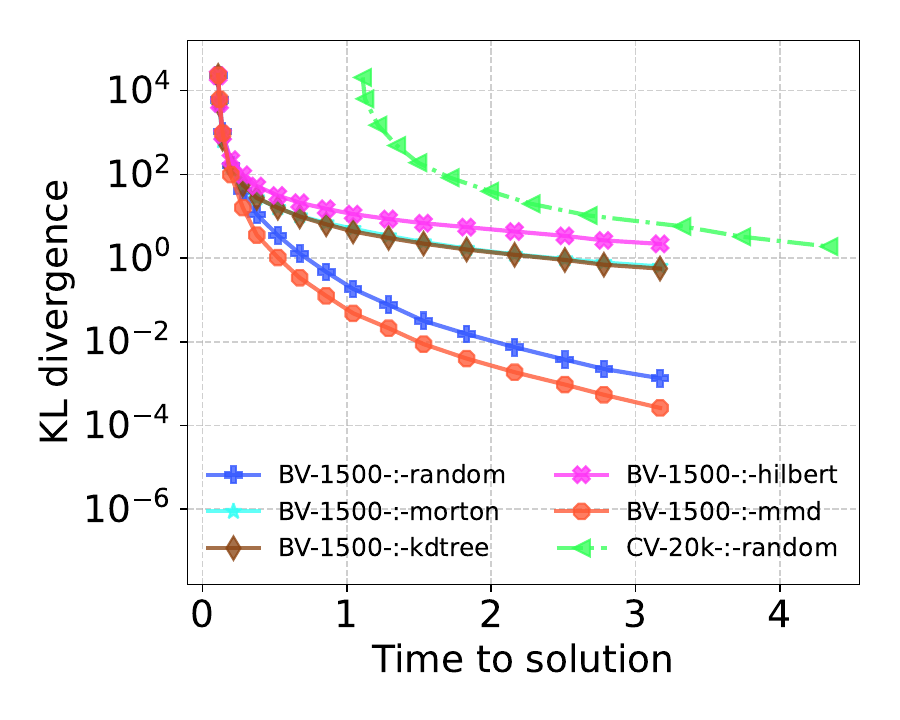}}
    \subfloat[$\beta=0.114318, \nu = 2.5$]{\includegraphics[width=0.33\textwidth]{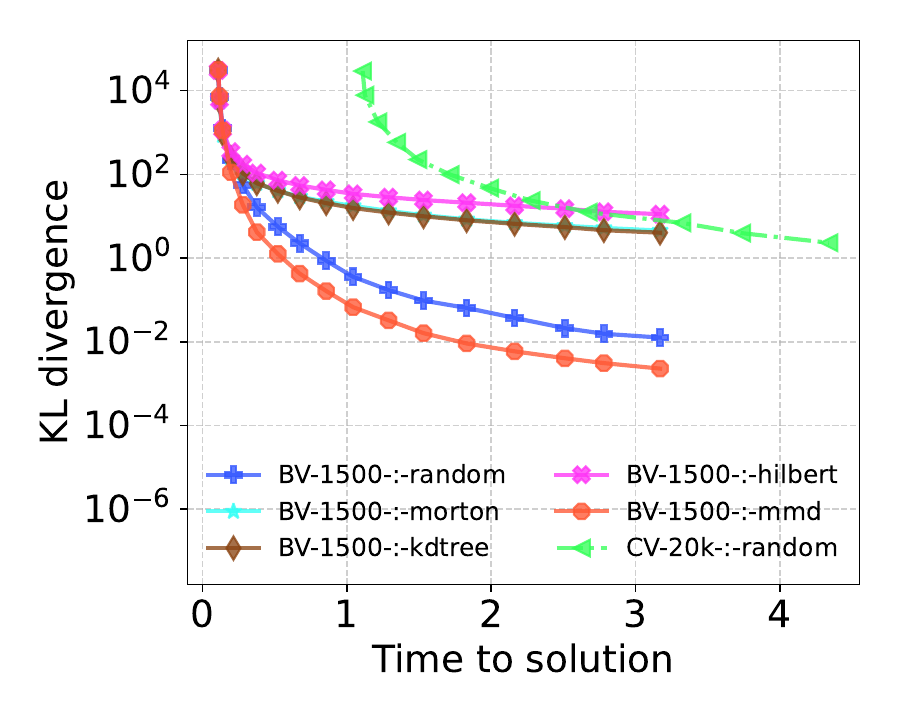}}
    \caption{KL divergence and time-to-solution under 20K locations with log10 scale under $\nu=2.5$ (y-axis is KL divergence).
    }
    \label{fig:20k-kl-1500-time-appendix}
\end{figure}

\textcolor{black}{We analyze the time efficiency and accuracy of the block Vecchia algorithm in relation to the parameters of the Matérn kernel. Figure \ref{fig:20k-kl-1500-time-appendix} demonstrates that the block Vecchia method achieves a more accurate KL divergence in a shorter time.}

\newpage
\subsection{Impact of Increasing the Number of Locations}
\label{spp:increasingnumber}

\textcolor{black}{We investigate the behavior of the block Vecchia as the number of locations increases. For this analysis and simplification, we chose a middle smoothness and range context 
$\beta = 0.052537, \nu = 1.5)$ and fixed the number of clusters at 2500. We focused only on max-min and random orderings, excluding other orderings due to their negligible impact on accuracy. In the following figures, the abbreviation 
\textit{BV-random-20k} represents the block Vecchia method with random ordering applied to 20,000 locations.}

\textcolor{black}{
Our findings indicate that approximation accuracy declines as the number of locations increases while keeping the number of clusters fixed. Specifically, a larger number of clusters and a higher count of nearest neighbors necessarily improve accuracy, which satisfies our intuition. Additionally, more time is required to compute the log-likelihood as the number of locations increases.}

\begin{figure}[htbp]
    \centering
    \subfloat[Accuracy with increasing conditioning size]{\includegraphics[width=0.45\textwidth]{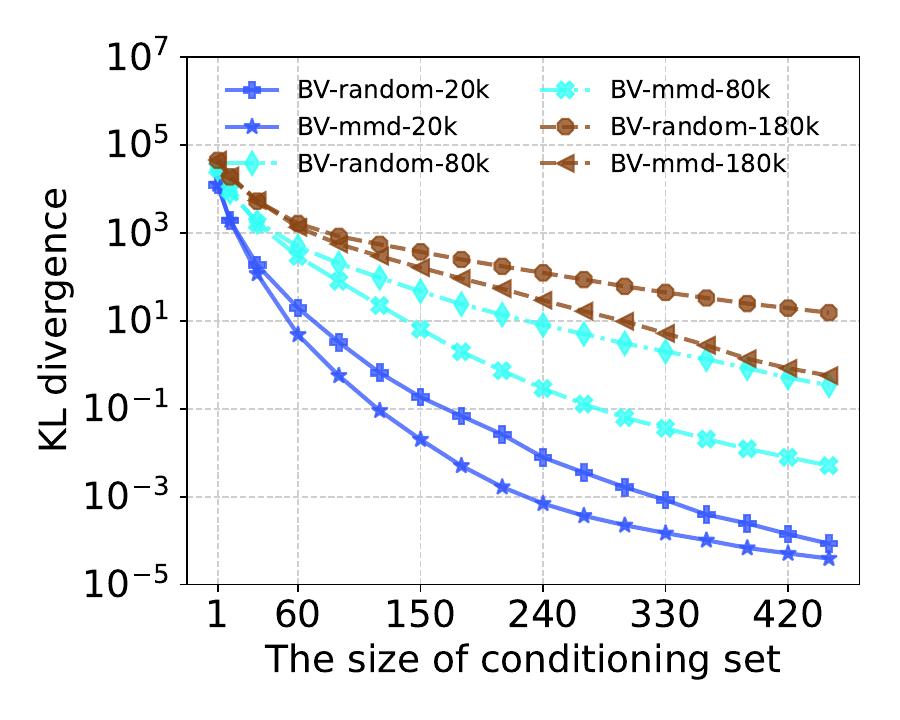}}
    \subfloat[Time (seconds)]{\includegraphics[width=0.45\textwidth]{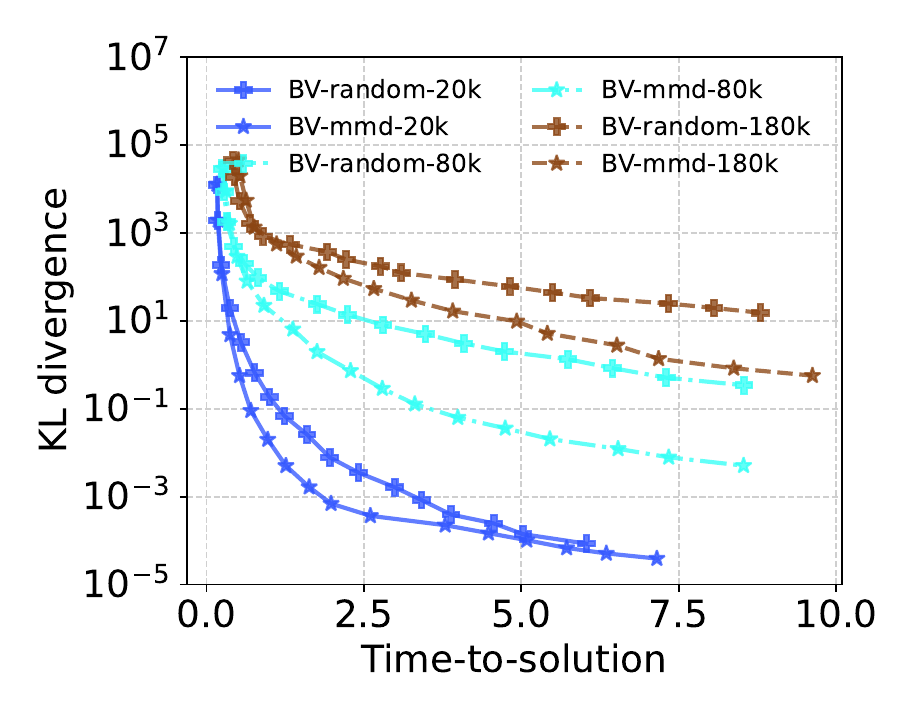}}
    \caption{Accuracy and time with a growing number of locations. (the number of clusters is fixed at 2500) }
    \label{fig:append-increa-locations}
\end{figure}

\newpage
\subsection{Simulation Study}
\label{spp:simu}

\begin{figure}[htbp]
    \centering
    \subfloat[$\hat \sigma^2$]{\includegraphics[width=0.33\textwidth]{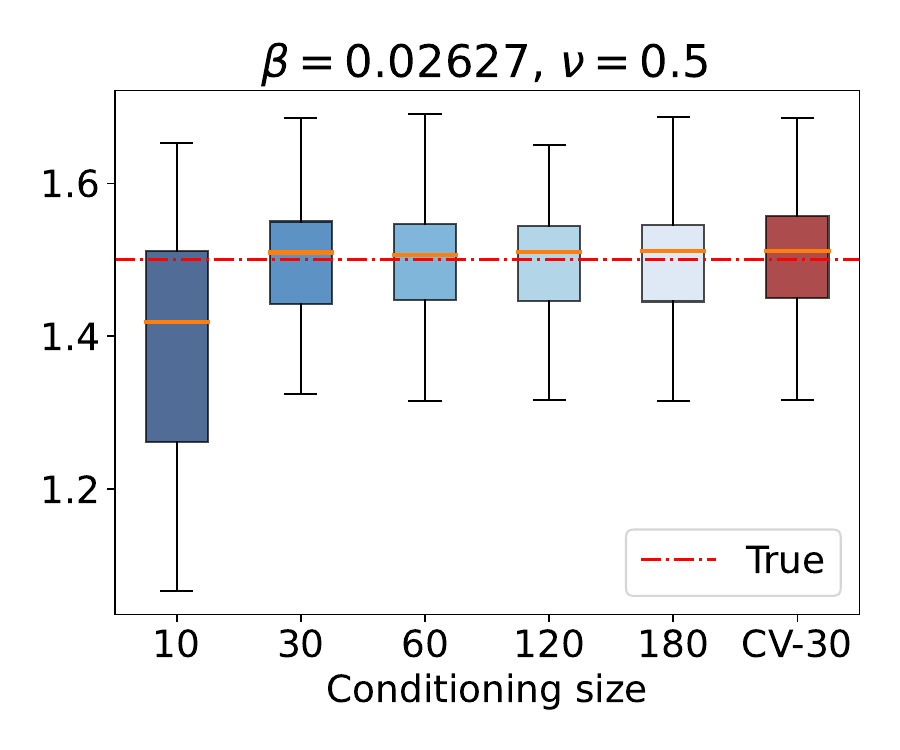}} 
    \subfloat[$\hat \sigma^2$]{\includegraphics[width=0.33\textwidth]{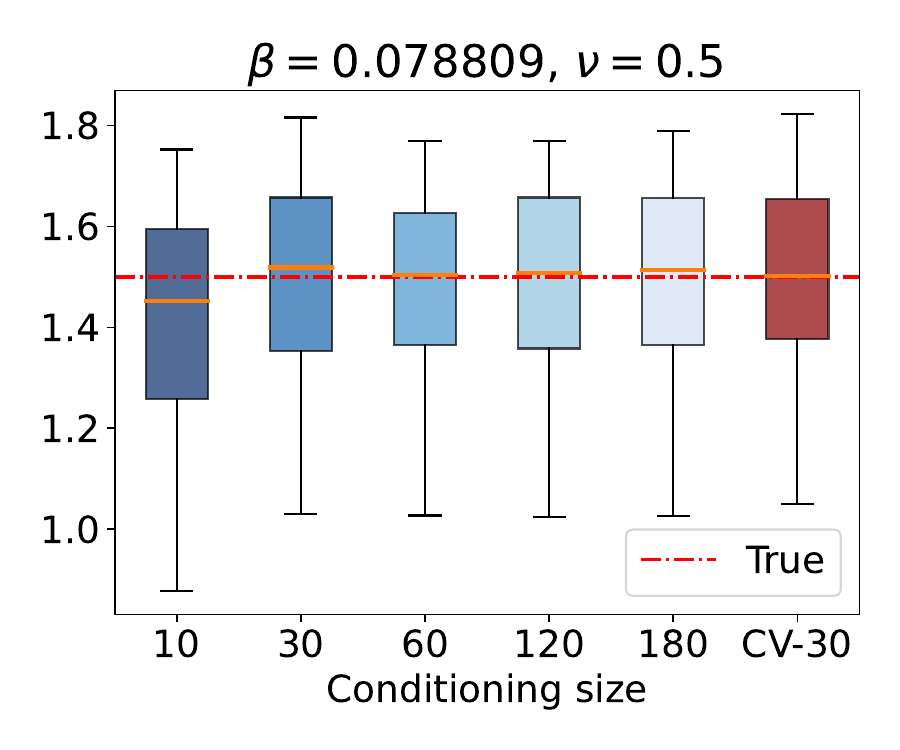}} 
    \subfloat[$\hat \sigma^2$]{\includegraphics[width=0.33\textwidth]{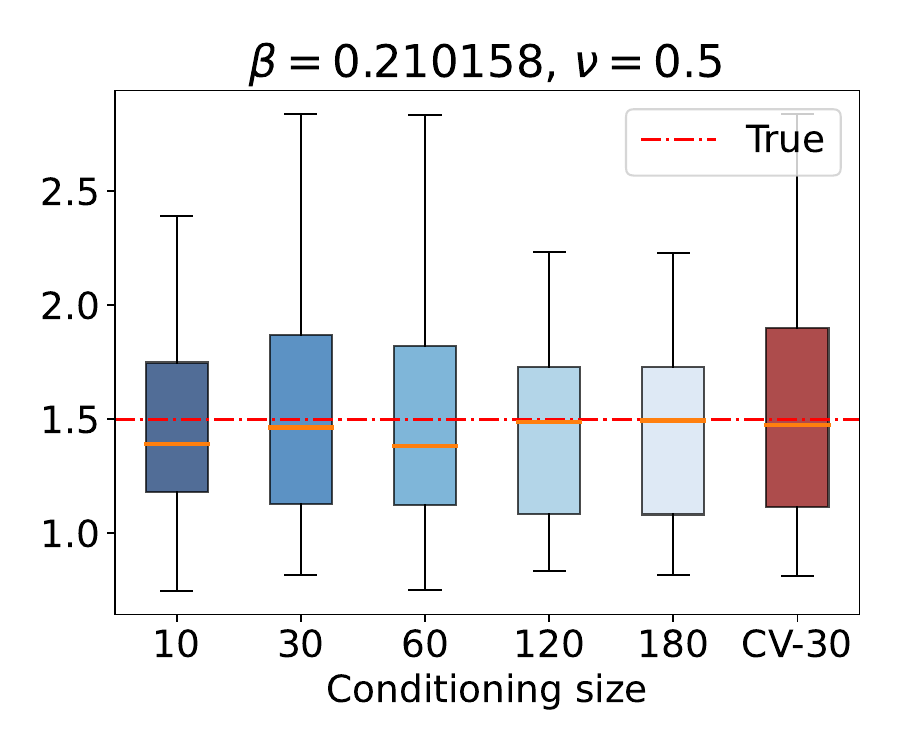}} \\
    \subfloat[$\hat \sigma^2$]{\includegraphics[width=0.33\textwidth]{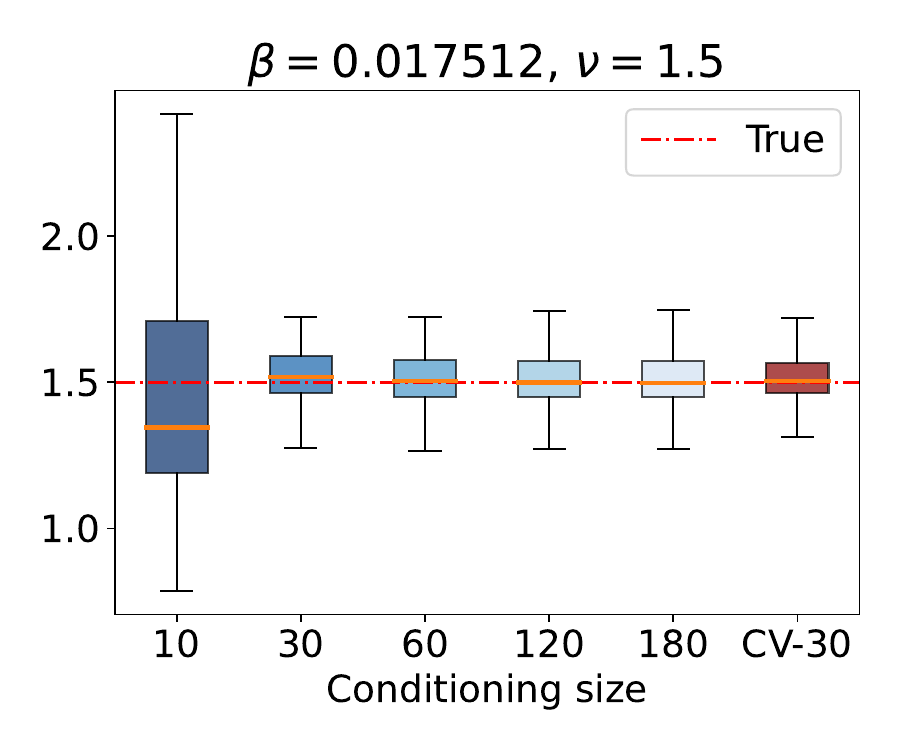}} 
    \subfloat[$\hat \sigma^2$]{\includegraphics[width=0.33\textwidth]{fig/simu-20k/0.052537_1.500000_0.pdf}} 
    \subfloat[$\hat \sigma^2$]{\includegraphics[width=0.33\textwidth]{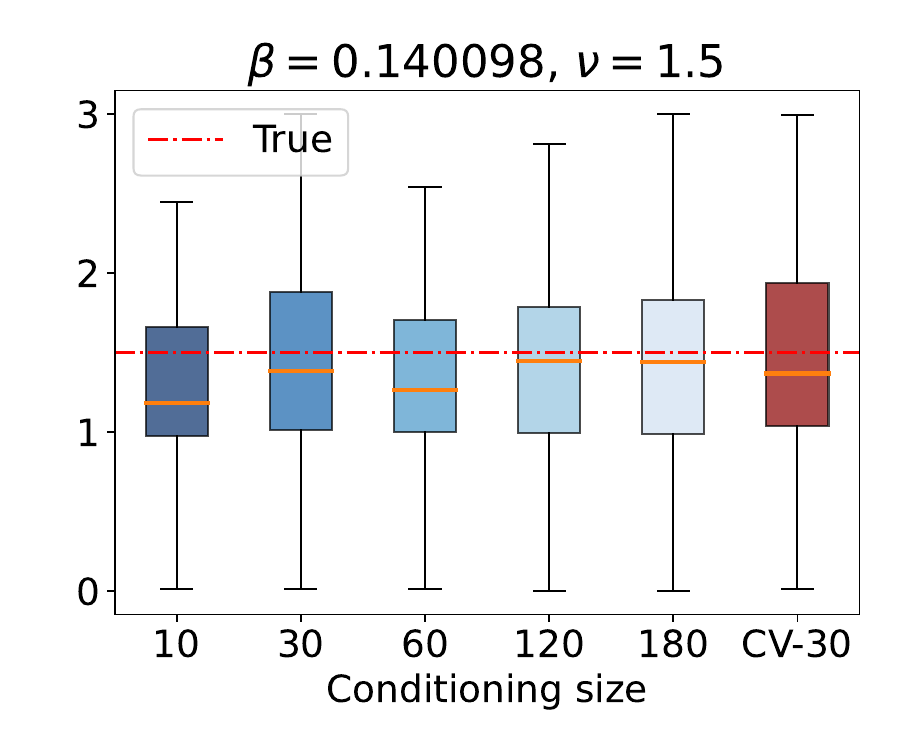}} \\
    \subfloat[$\hat \sigma^2$]{\includegraphics[width=0.33\textwidth]{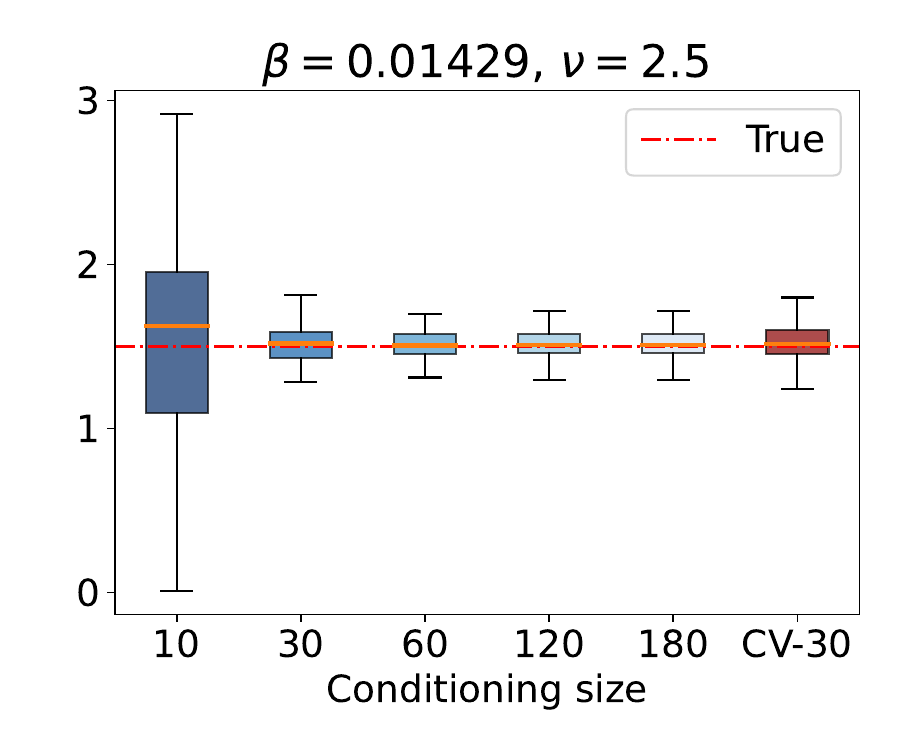}} 
    \subfloat[$\hat \sigma^2$]{\includegraphics[width=0.33\textwidth]{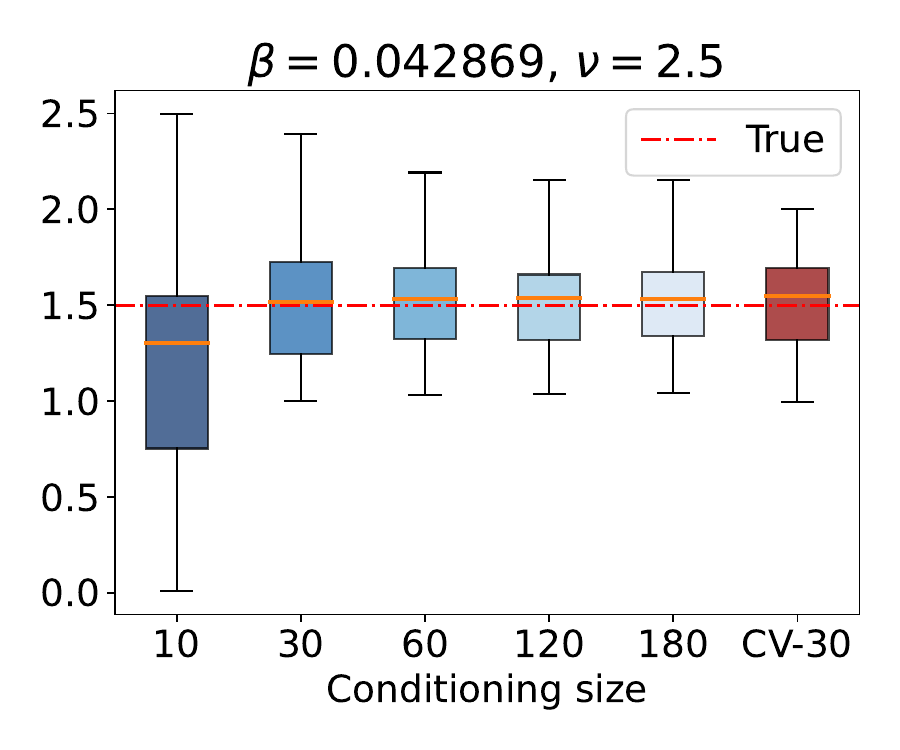}} 
    \subfloat[$\hat \sigma^2$]{\includegraphics[width=0.33\textwidth]{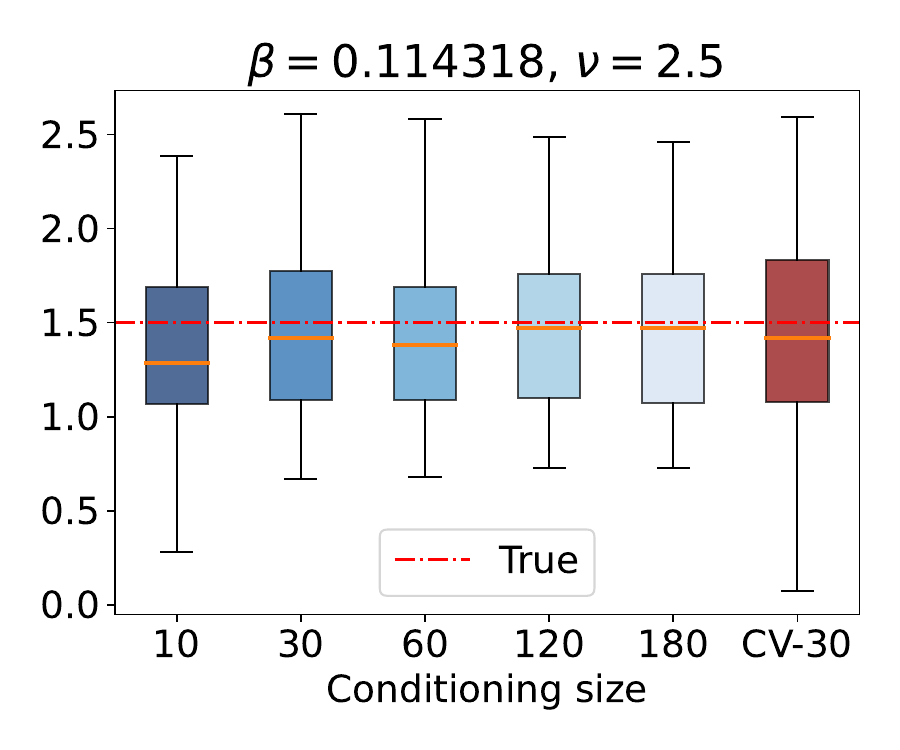}} 
    \caption{Simulations on parameter estimation $\sigma^2$.}
    \label{fig:simu-20k-sigma-appendix}
\end{figure}

\begin{figure}[htbp]
    \centering
    \subfloat[$\hat \beta$]{\includegraphics[width=0.33\textwidth]{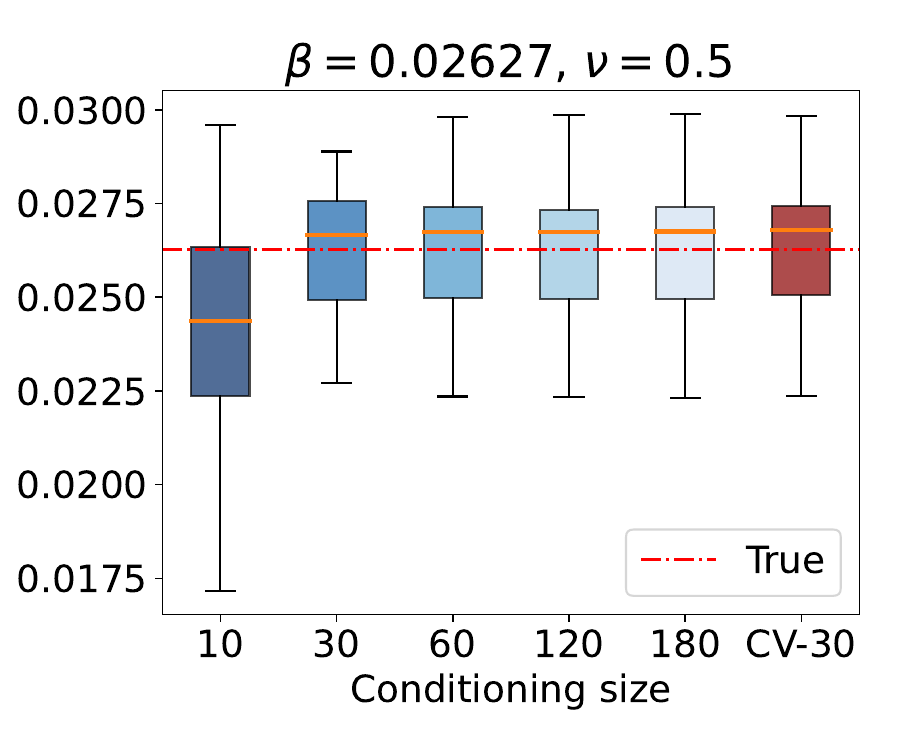}} 
    \subfloat[$\hat \beta$]{\includegraphics[width=0.33\textwidth]{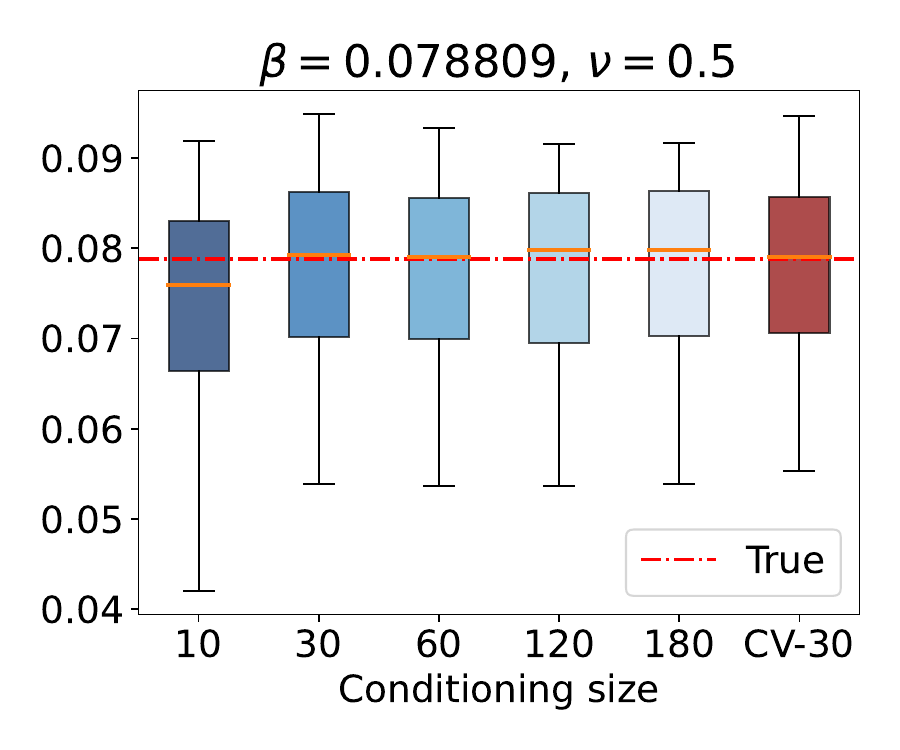}} 
    \subfloat[$\hat \beta$]{\includegraphics[width=0.33\textwidth]{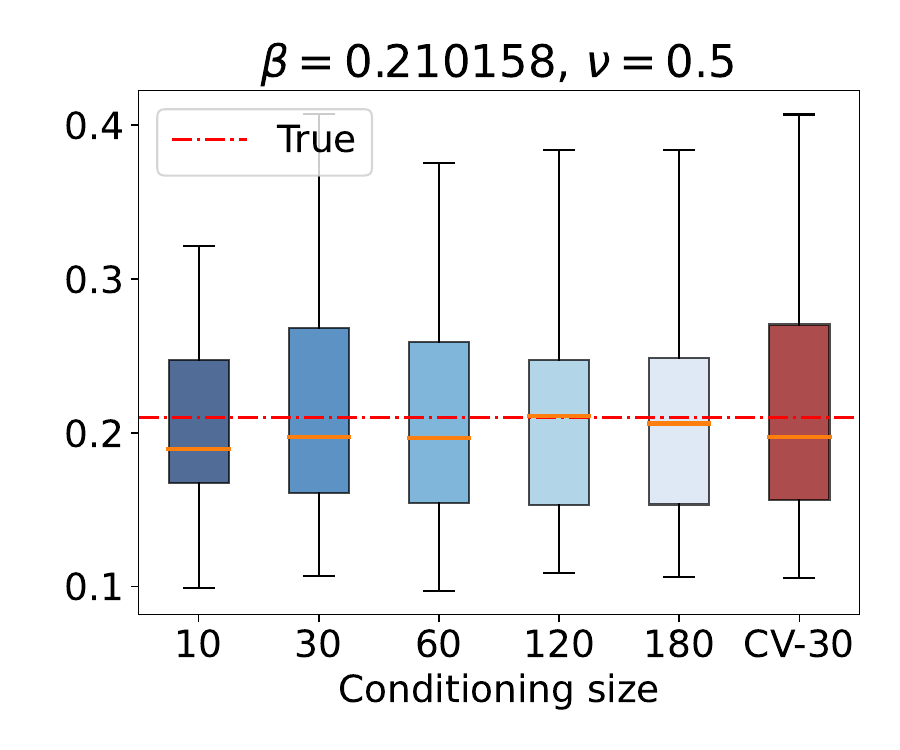}} \\
    \subfloat[$\hat \beta$]{\includegraphics[width=0.33\textwidth]{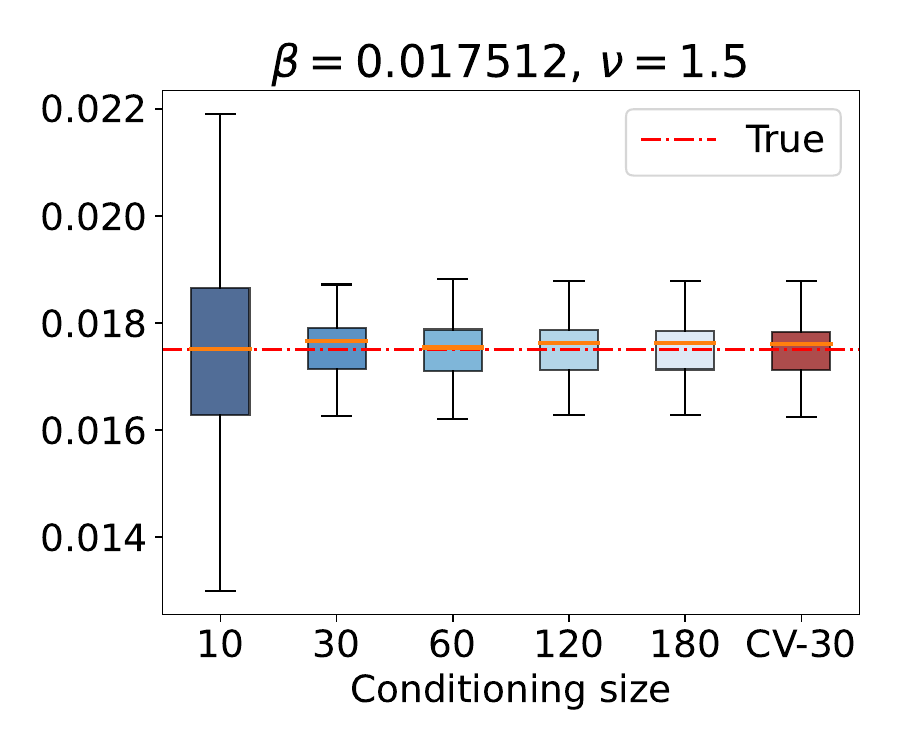}} 
    \subfloat[$\hat \beta$]{\includegraphics[width=0.33\textwidth]{fig/simu-20k/0.052537_1.500000_1.pdf}} 
    \subfloat[$\hat \beta$]{\includegraphics[width=0.33\textwidth]{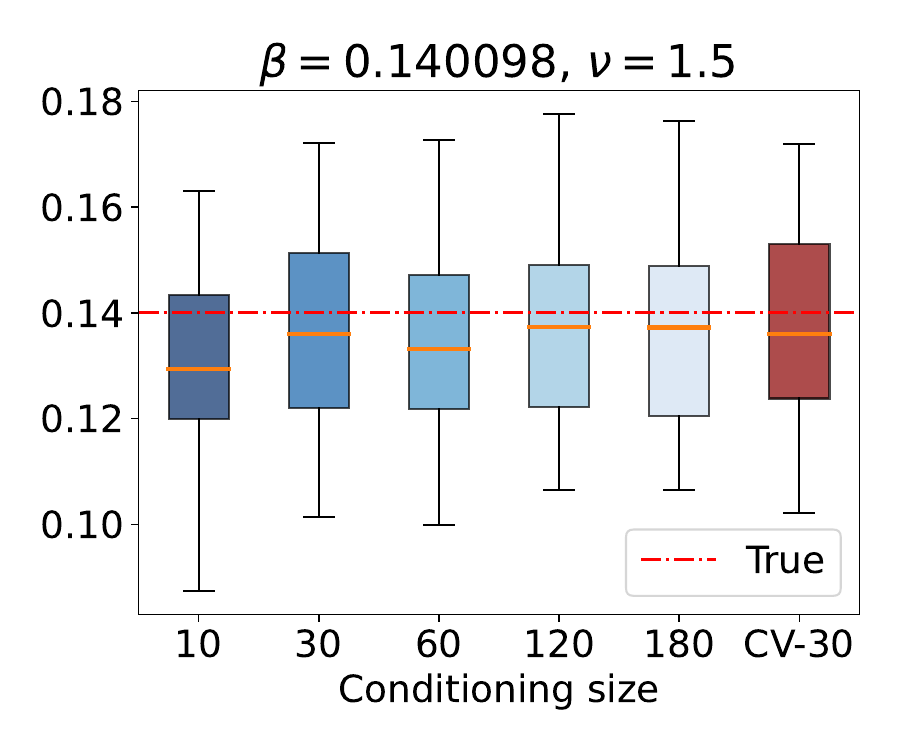}} \\
    \subfloat[$\hat \beta$]{\includegraphics[width=0.33\textwidth]{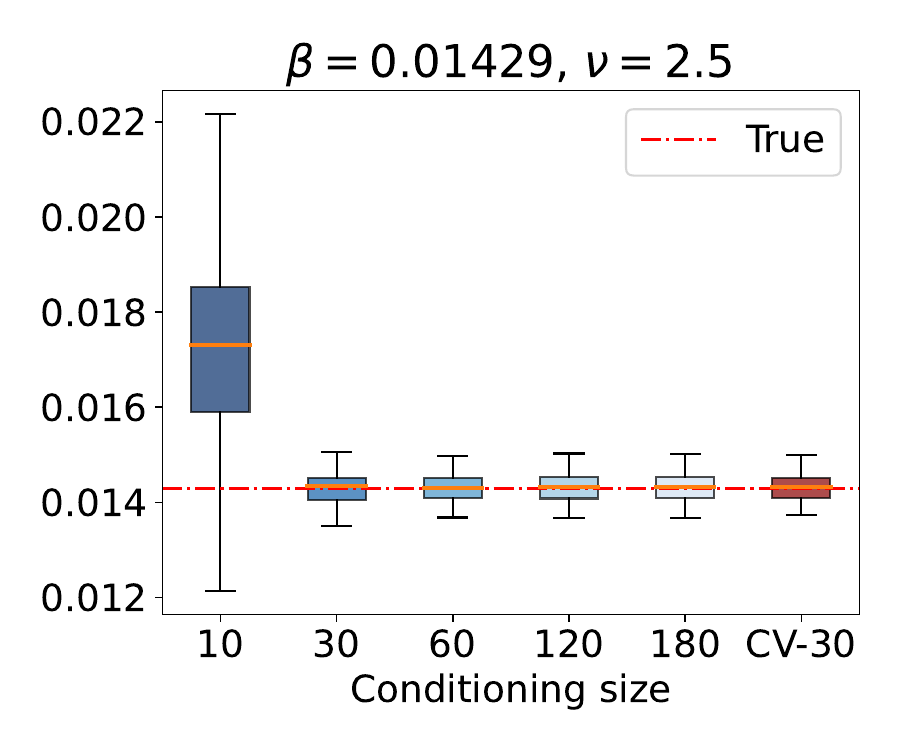}} 
    \subfloat[$\hat \beta$]{\includegraphics[width=0.33\textwidth]{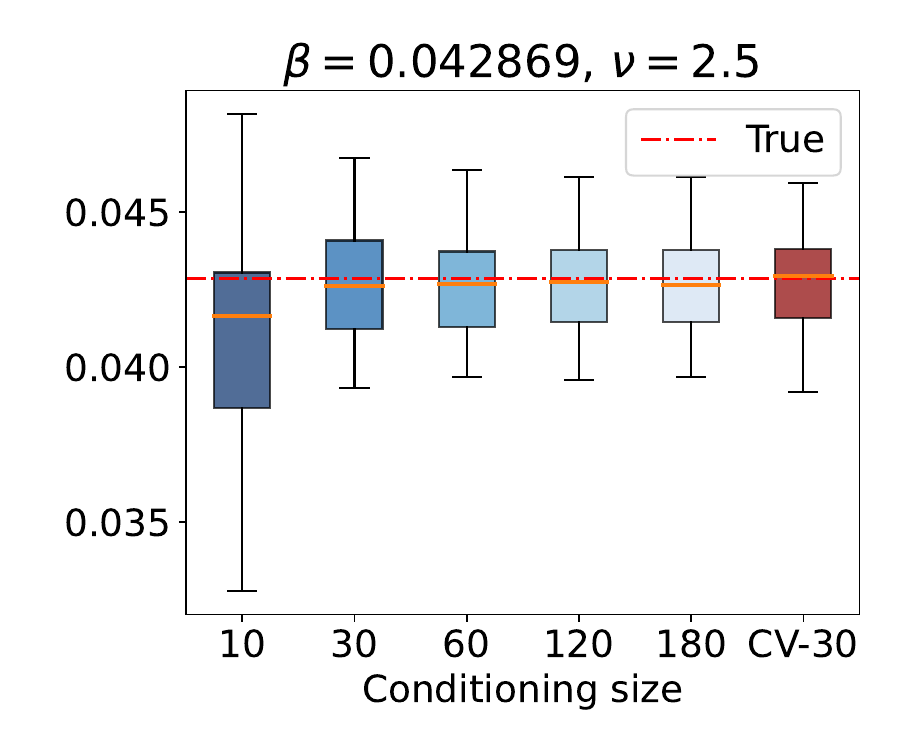}} 
    \subfloat[$\hat \beta$]{\includegraphics[width=0.33\textwidth]{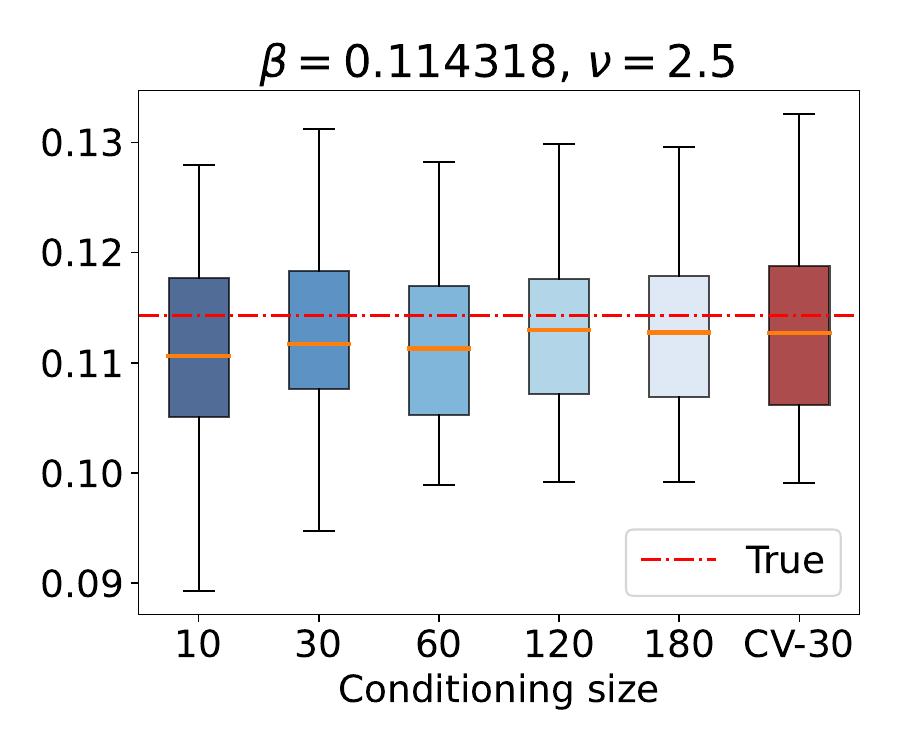}} 
    \caption{Simulations on parameter estimation $\beta$.}
    \label{fig:simu-20k-beta-appendix}
\end{figure}

\begin{figure}[htbp]
    \centering
    \subfloat[$\hat \nu$]{\includegraphics[width=0.33\textwidth]{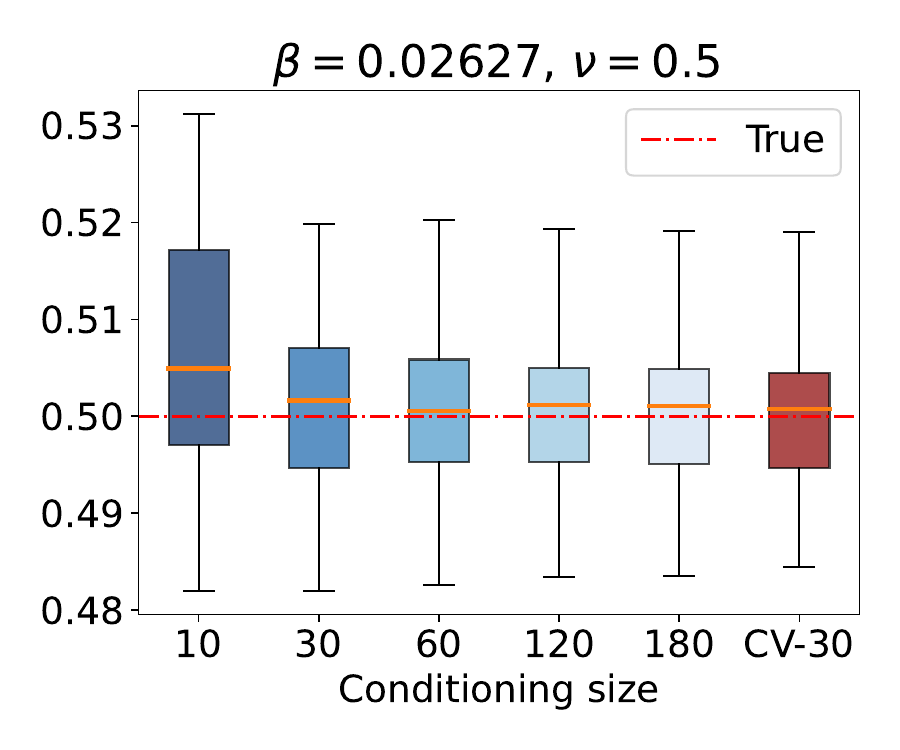}} 
    \subfloat[$\hat \nu$]{\includegraphics[width=0.33\textwidth]{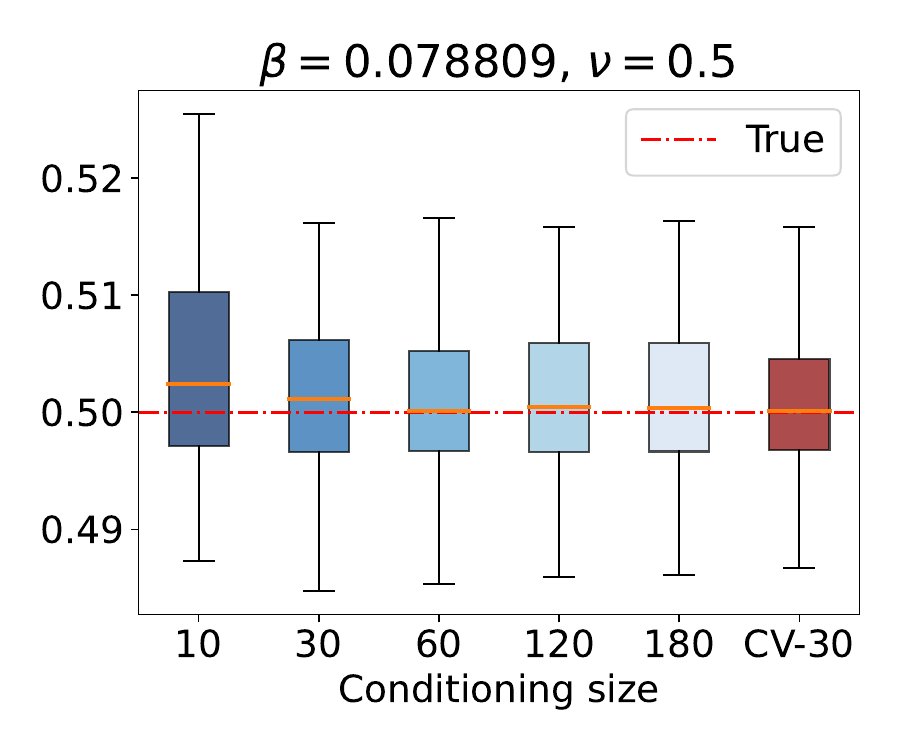}} 
    \subfloat[$\hat \nu$]{\includegraphics[width=0.33\textwidth]{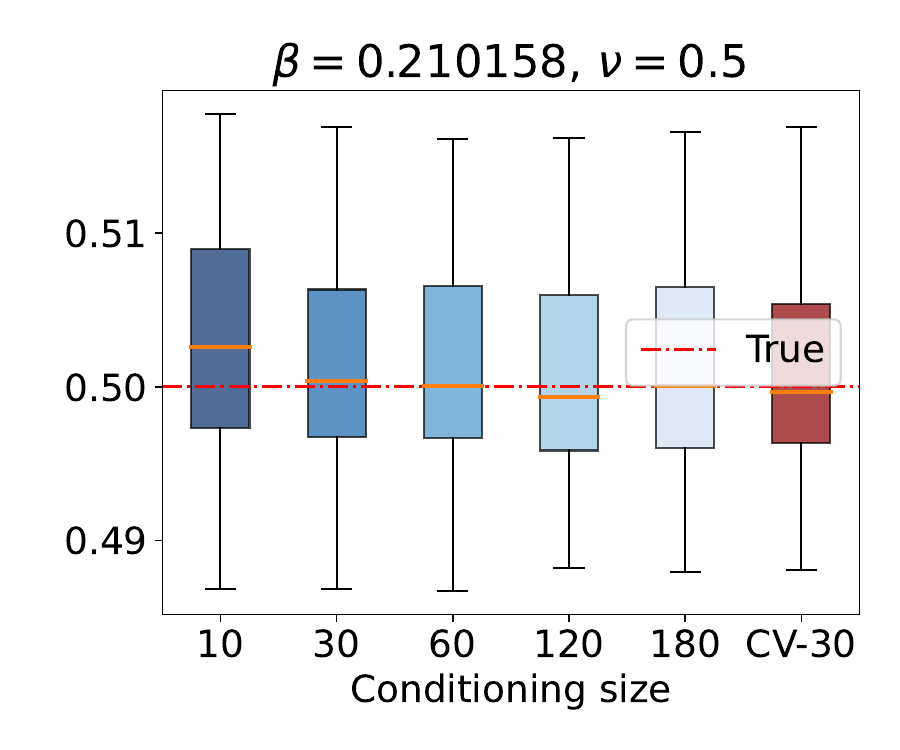}} \\
    \subfloat[$\hat \nu$]{\includegraphics[width=0.33\textwidth]{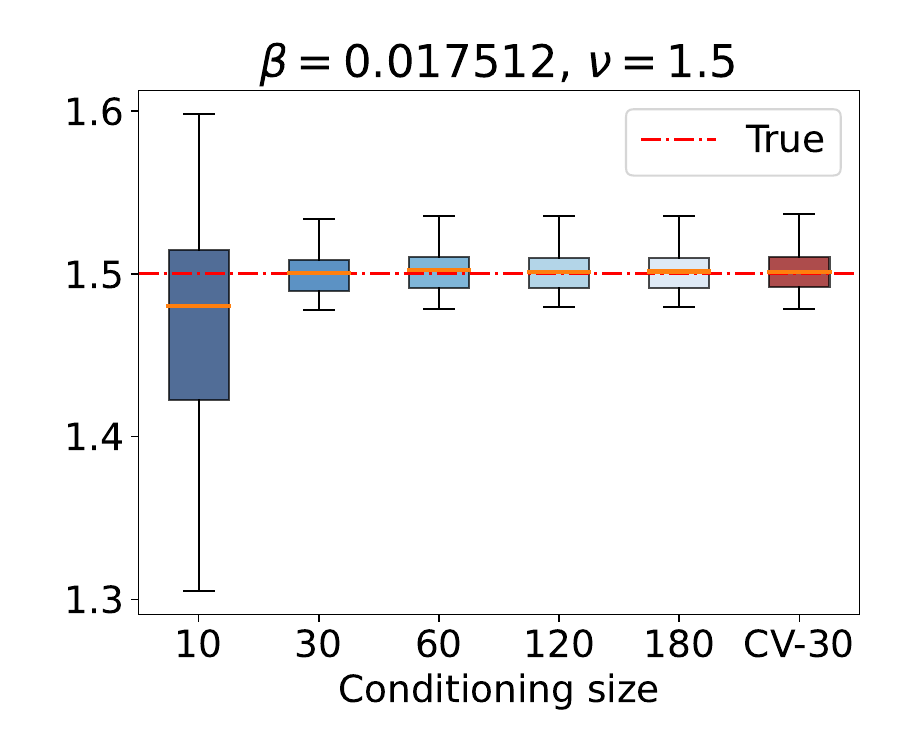}} 
    \subfloat[$\hat \nu$]{\includegraphics[width=0.33\textwidth]{fig/simu-20k/0.052537_1.500000_2.pdf}} 
    \subfloat[$\hat \nu$]{\includegraphics[width=0.33\textwidth]{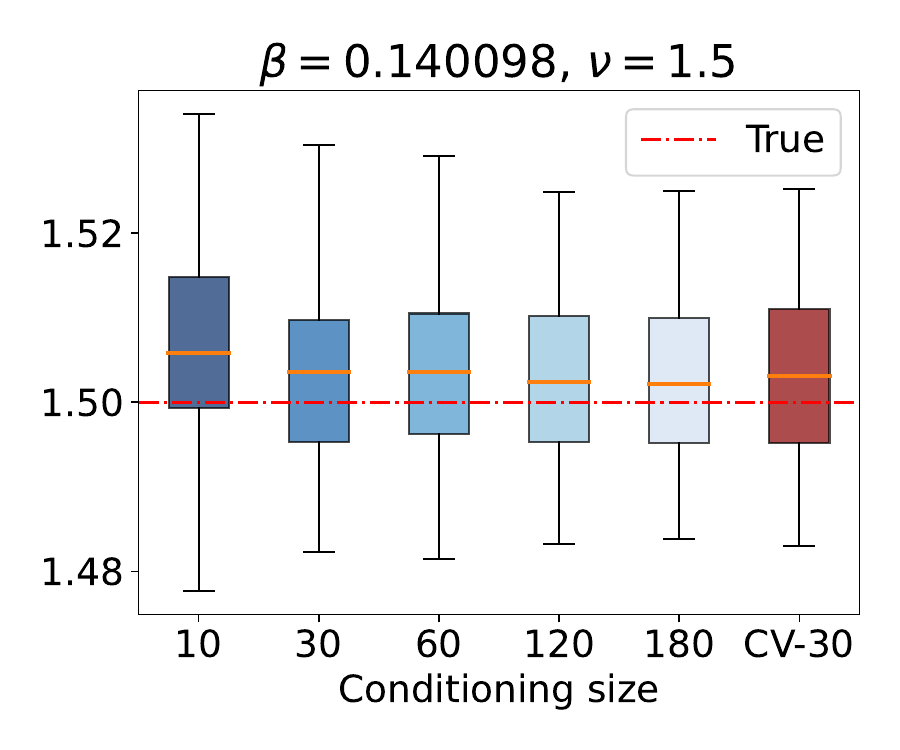}} \\
    \subfloat[$\hat \nu$]{\includegraphics[width=0.33\textwidth]{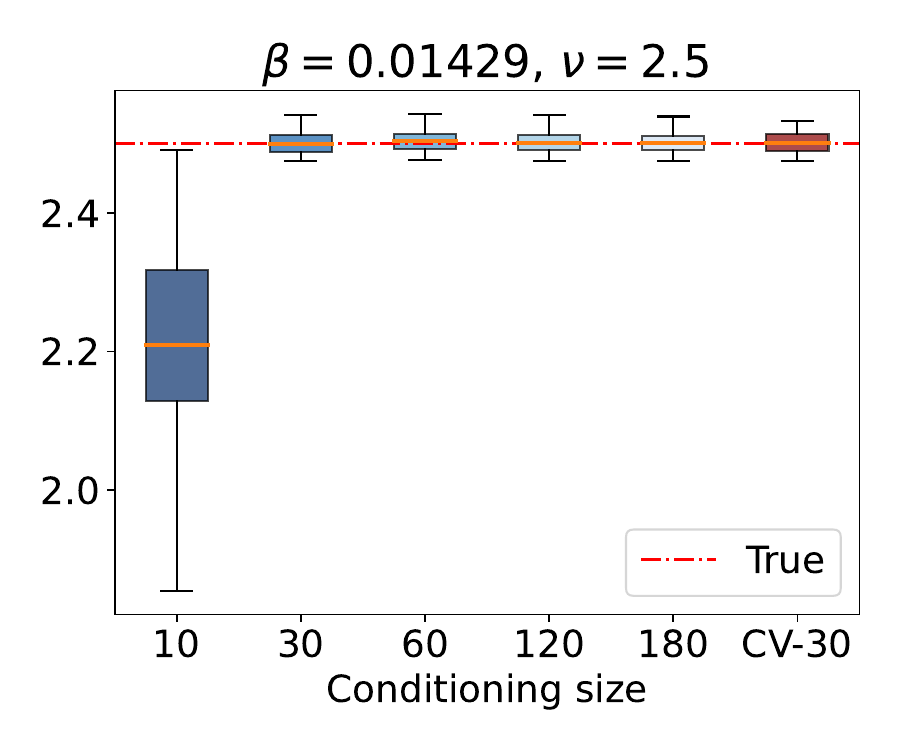}} 
    \subfloat[$\hat \nu$]{\includegraphics[width=0.33\textwidth]{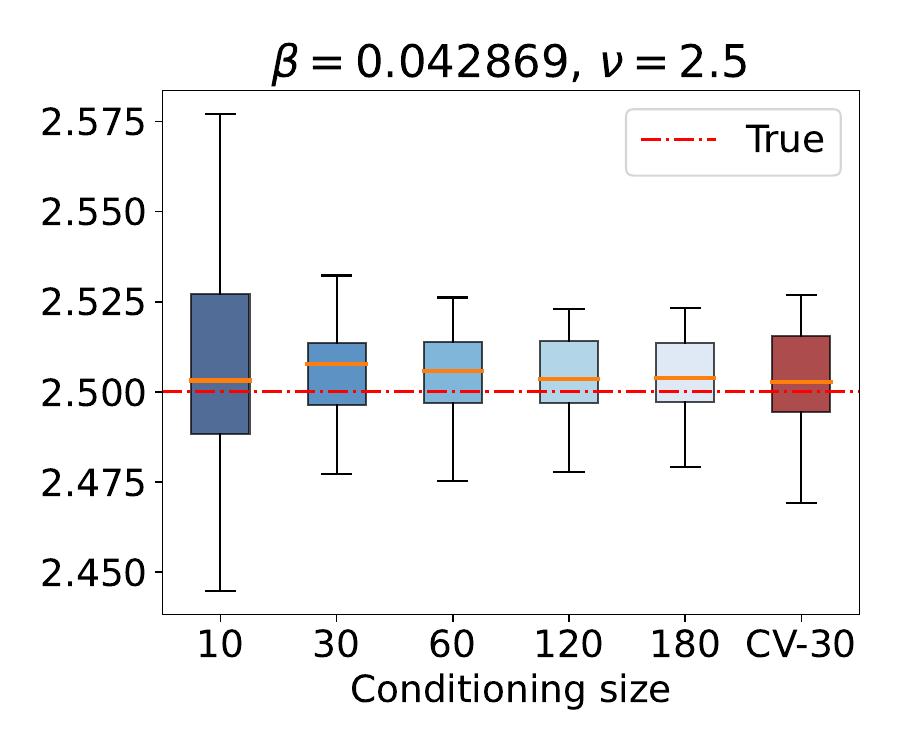}} 
    \subfloat[$\hat \nu$]{\includegraphics[width=0.33\textwidth]{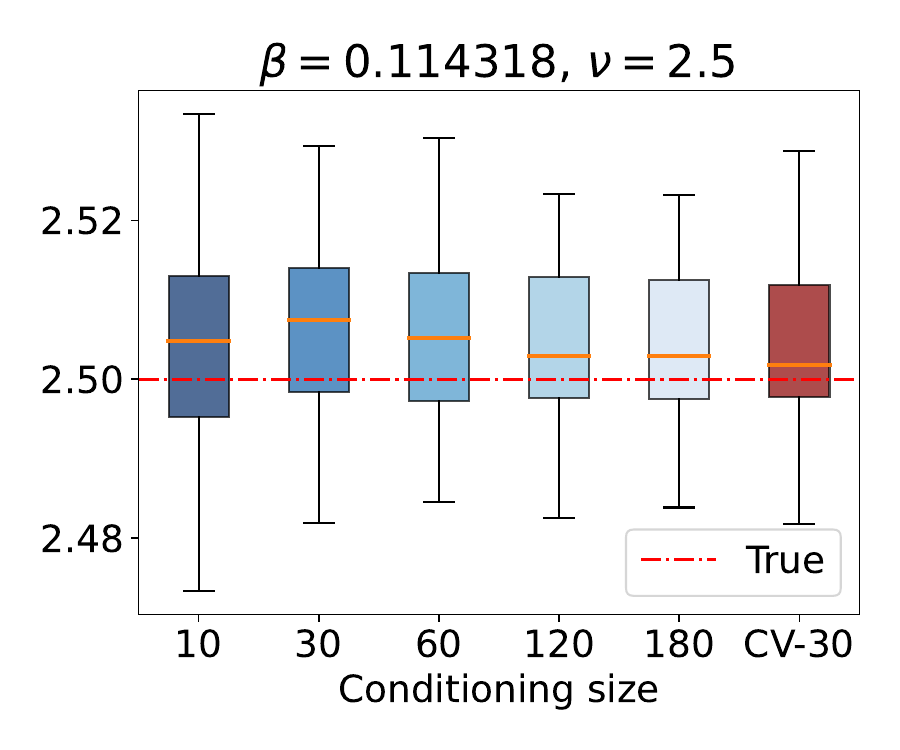}} 
    \caption{Simulations on parameter estimation $\nu$.}
    \label{fig:simu-20k-nu-appendix}
\end{figure}

\begin{figure}
    \centering
    \subfloat[$\nu=0.5$]{\includegraphics[width=0.33\linewidth]{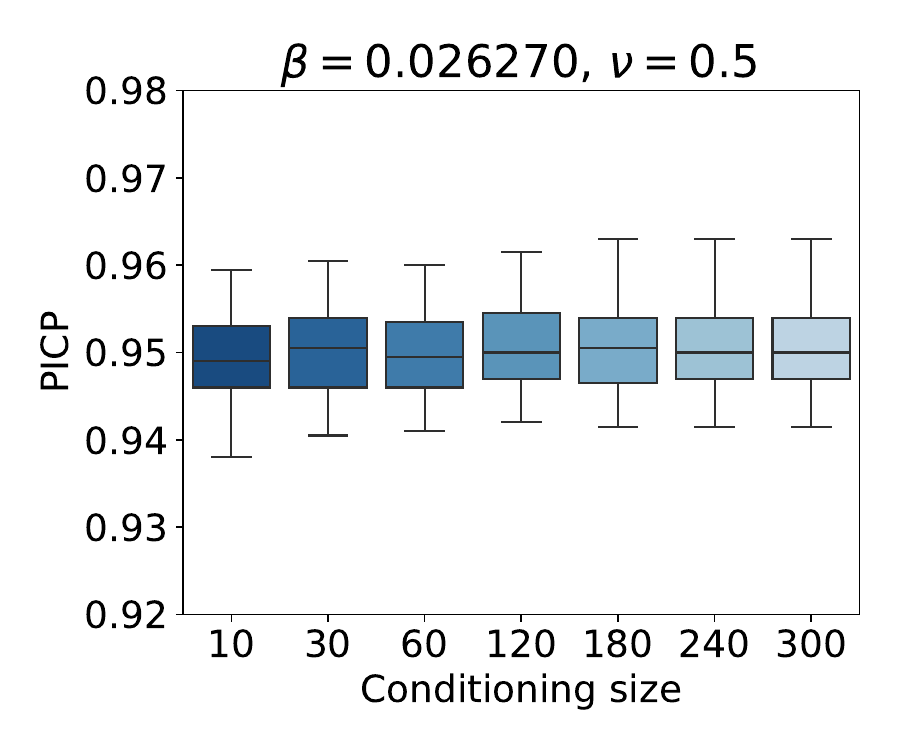}}
    \subfloat[$\nu=1.5$]{\includegraphics[width=0.33\linewidth]{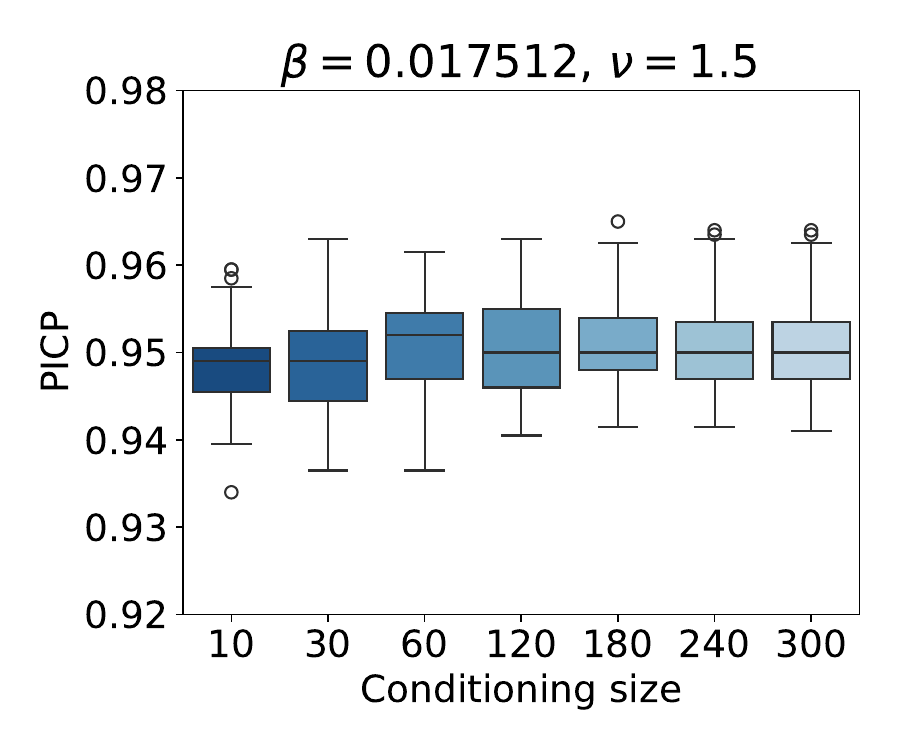}}
    \subfloat[$\nu=2.5$]{\includegraphics[width=0.33\linewidth]{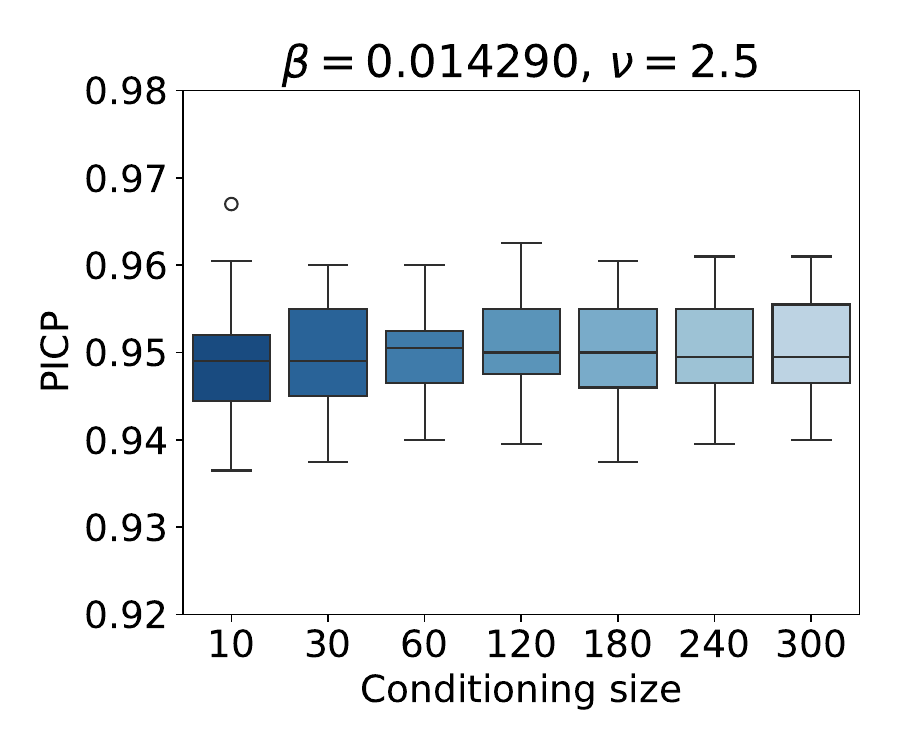}}
    \\
    \subfloat[$\nu=0.5$]{\includegraphics[width=0.33\linewidth]{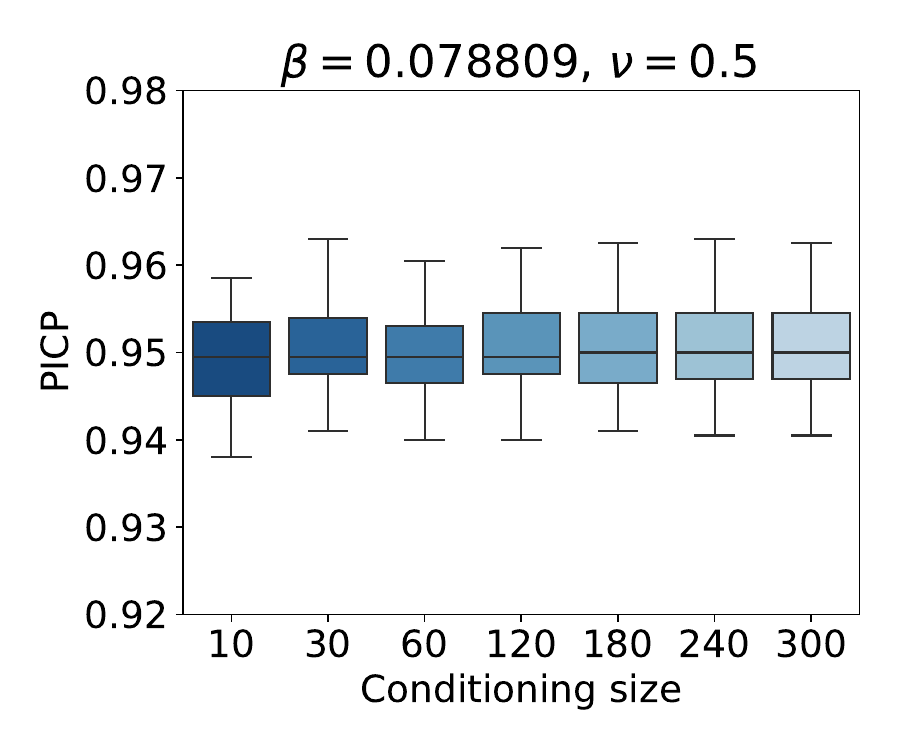}}
    \subfloat[$\nu=1.5$]{\includegraphics[width=0.33\linewidth]{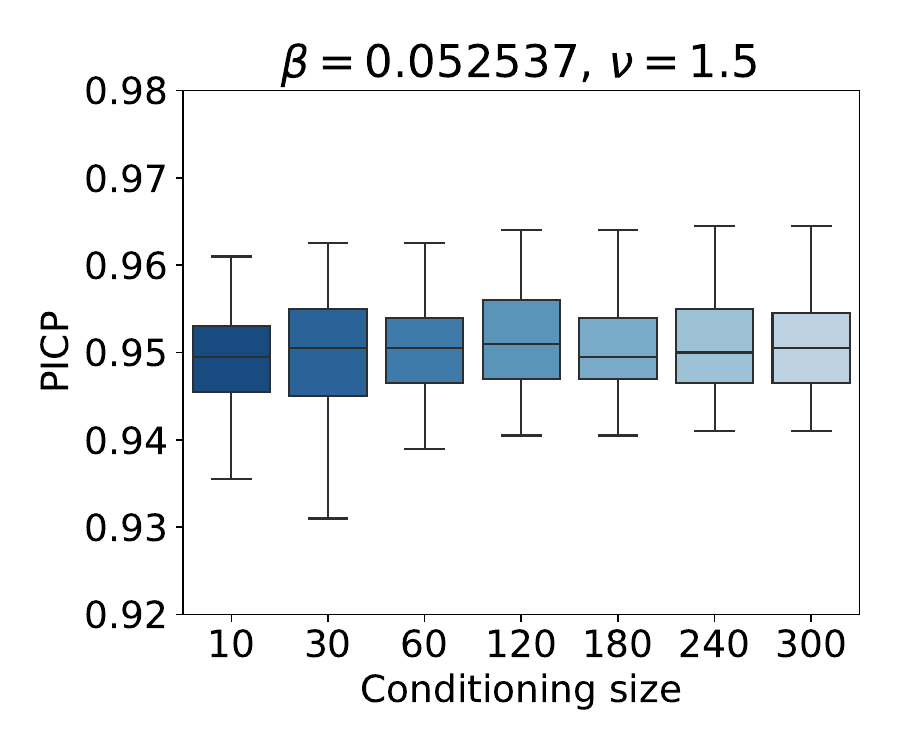}}
    \subfloat[$\nu=2.5$]{\includegraphics[width=0.33\linewidth]{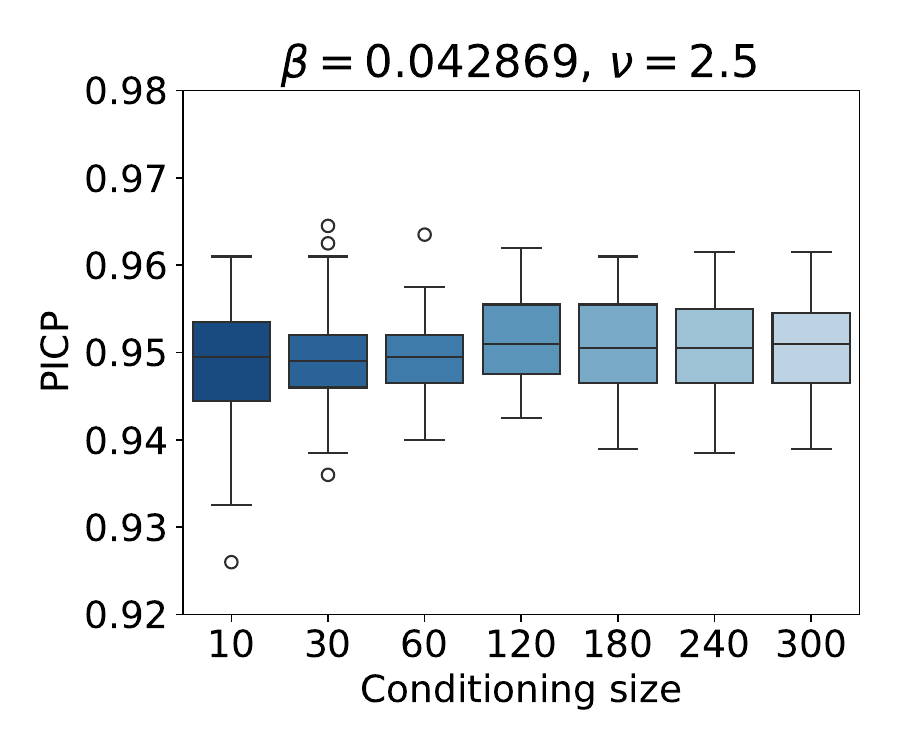}}
    \\
    \subfloat[$\nu=0.5$]{\includegraphics[width=0.33\linewidth]{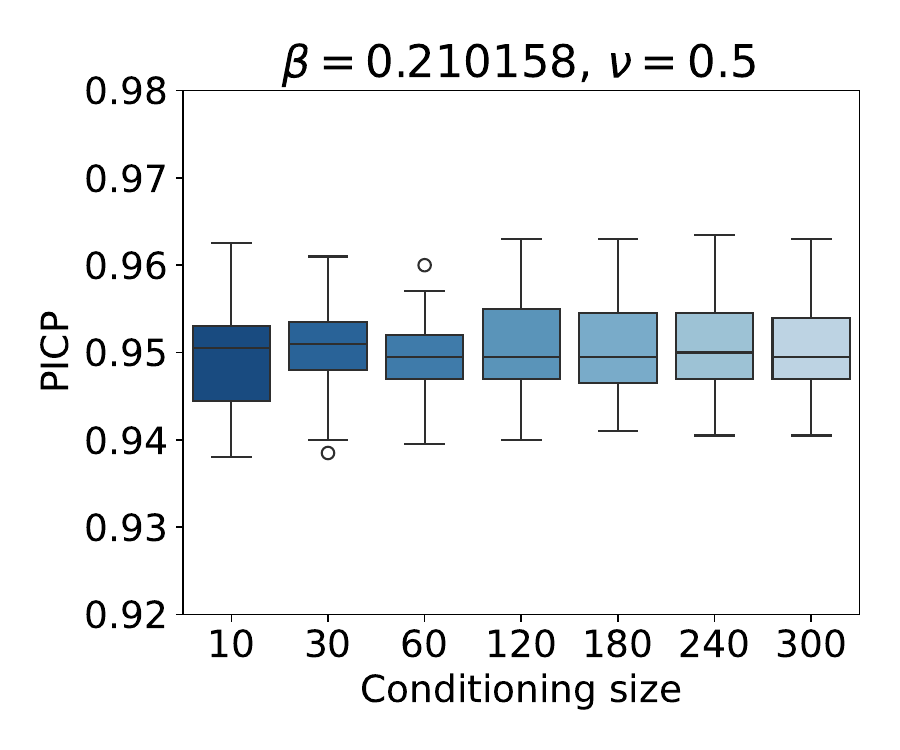}}
    \subfloat[$\nu=1.5$]{\includegraphics[width=0.33\linewidth]{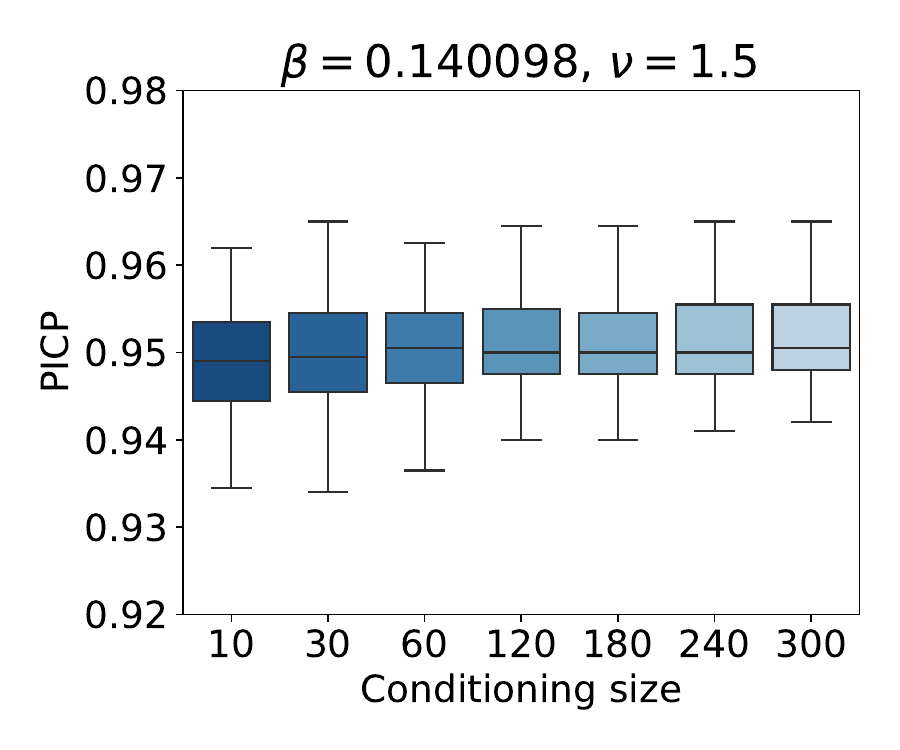}}
    \subfloat[$\nu=2.5$]{\includegraphics[width=0.33\linewidth]{fig/predict/coverage_rate_boxplot_0.114318_2.500000.pdf}}
    \caption{PICP for simulated datasets.}
    \label{fig:spp:predrate}
\end{figure}

\newpage
\subsection{Analysis of Real Applications}
\label{spp:realdataset}

This section provides the visualization of residuals of 3D windspeed and focuses on modeling high-resolution 2D soil moisture. The scalability of the block Vecchia method is evaluated on a single GPU (NVIDIA V100 with 32 GB memory) with problem sizes at the million level. The entire soil moisture dataset is used here, i.e., 1.8 million data locations in a 2D profile. The rest of the experimental configuration remains consistent with Section \ref{sec:realdataset}, and we then apply the block Vecchia approximation to estimate the parameters.

\begin{figure}[htbp]
    \centering
    \includegraphics[width=0.6\linewidth]{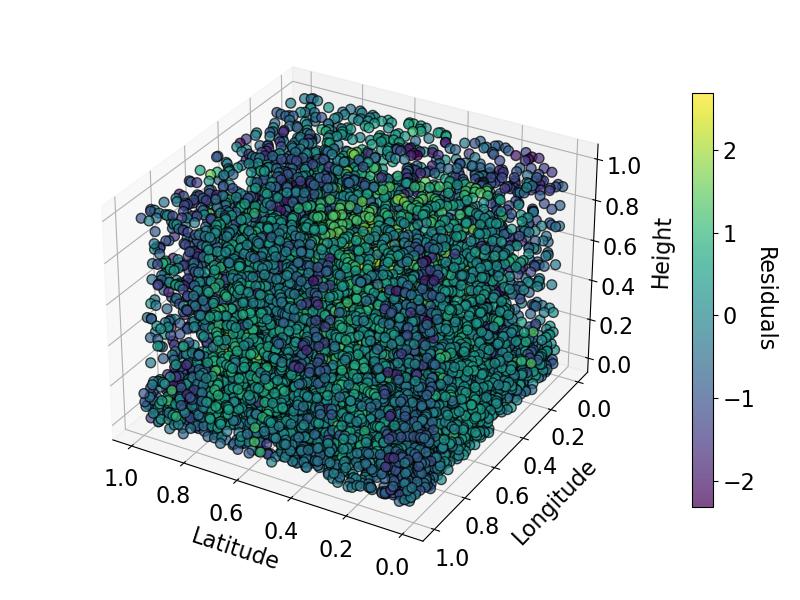}
    \caption{Residual of wind speed in 3D profile.}
    \label{sppfig:3dprofile}
\end{figure}

\begin{figure}[htbp]
    \centering
    \subfloat[Soil $\hat\sigma^2$]{\includegraphics[width=0.33\textwidth]{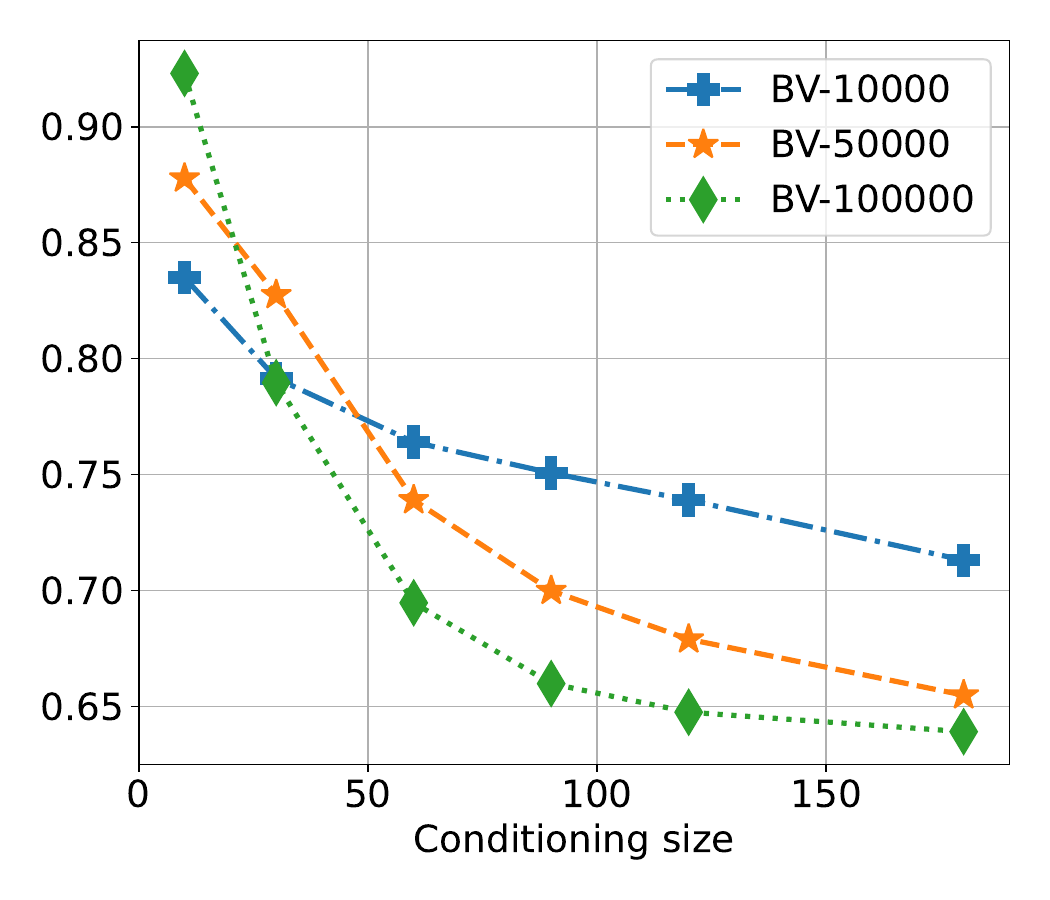}}
    \subfloat[Soil $\hat\beta$]{\includegraphics[width=0.33\textwidth]{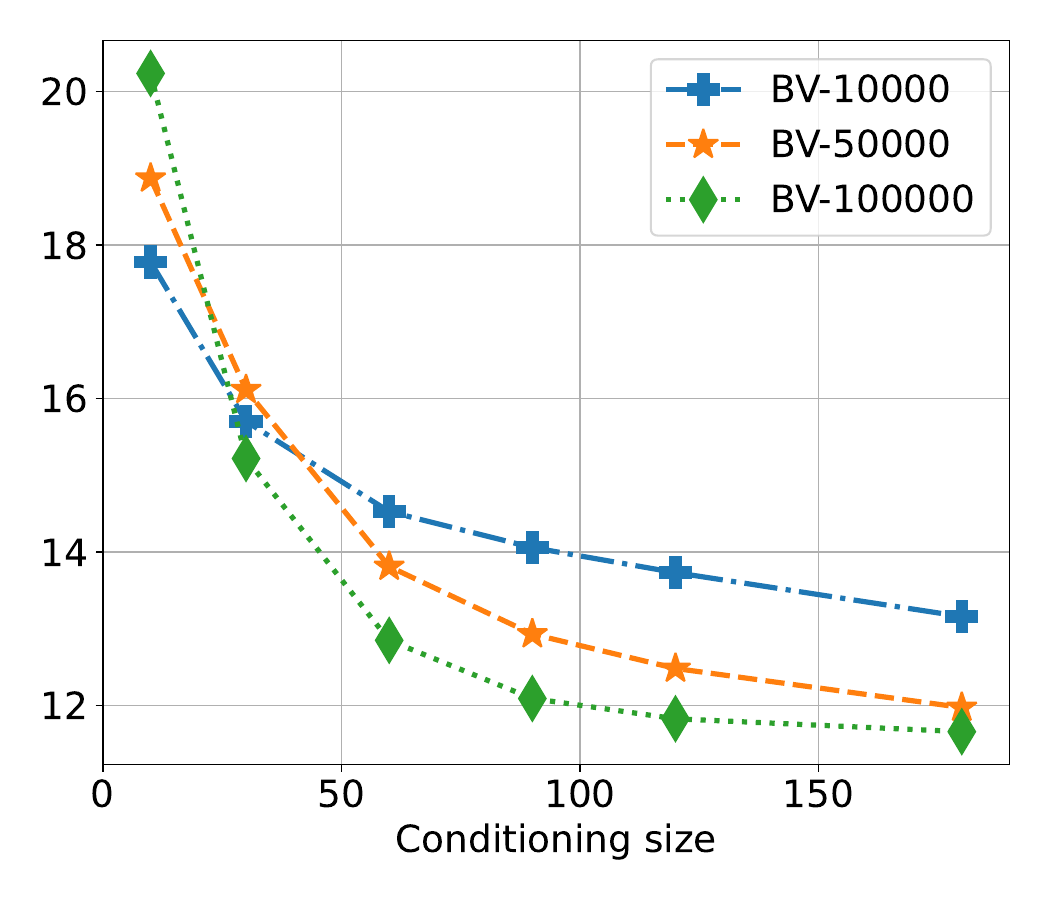}}
    \subfloat[Soil $\hat\nu$]{\includegraphics[width=0.33\textwidth]{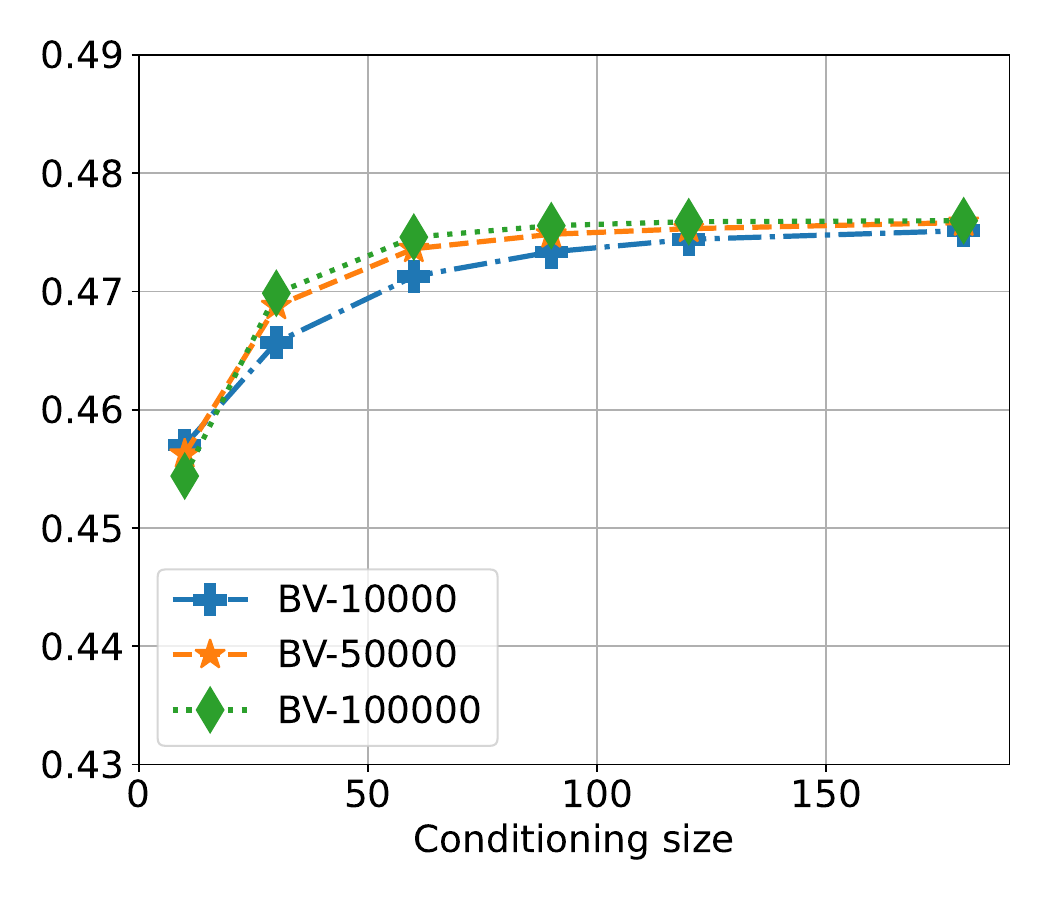}}
    \caption{The estimated parameters using block Vecchia with different block counts. (The range parameters are scaled.)}
    \label{fig:spp:real-million-soil}
\end{figure}

\textcolor{black}{
Figure \ref{fig:spp:real-million-soil} presents the results of parameter estimation for modeling the residuals of 2D soil moisture. The number of blocks increases (e.g., 10K, 50K, and 100K), reflecting our interest in improving accuracy in the approximation of the block Vecchia method. The findings indicate that: 1) parameter estimation gradually converges to a specific value as the conditioning size increases; 2) the convergence rate improves with a larger block count; and 3) the estimation for 2D data is more accurate than for 3D data, i.e., the range parameter has a slower convergence rate in 3D wind speed, suggesting that more data may be required for parameter estimation in the 3D context, which encourages extending the Vecchia approximation to larger-scale problems. These results demonstrate that larger conditioning sizes and block counts enhance the accuracy of parameter estimation. More importantly, the block Vecchia method can handle much larger problem sizes than the classic Vecchia, facilitating large-scale modeling.
}

\begin{figure}
    \centering
    \includegraphics[width=0.4\linewidth]{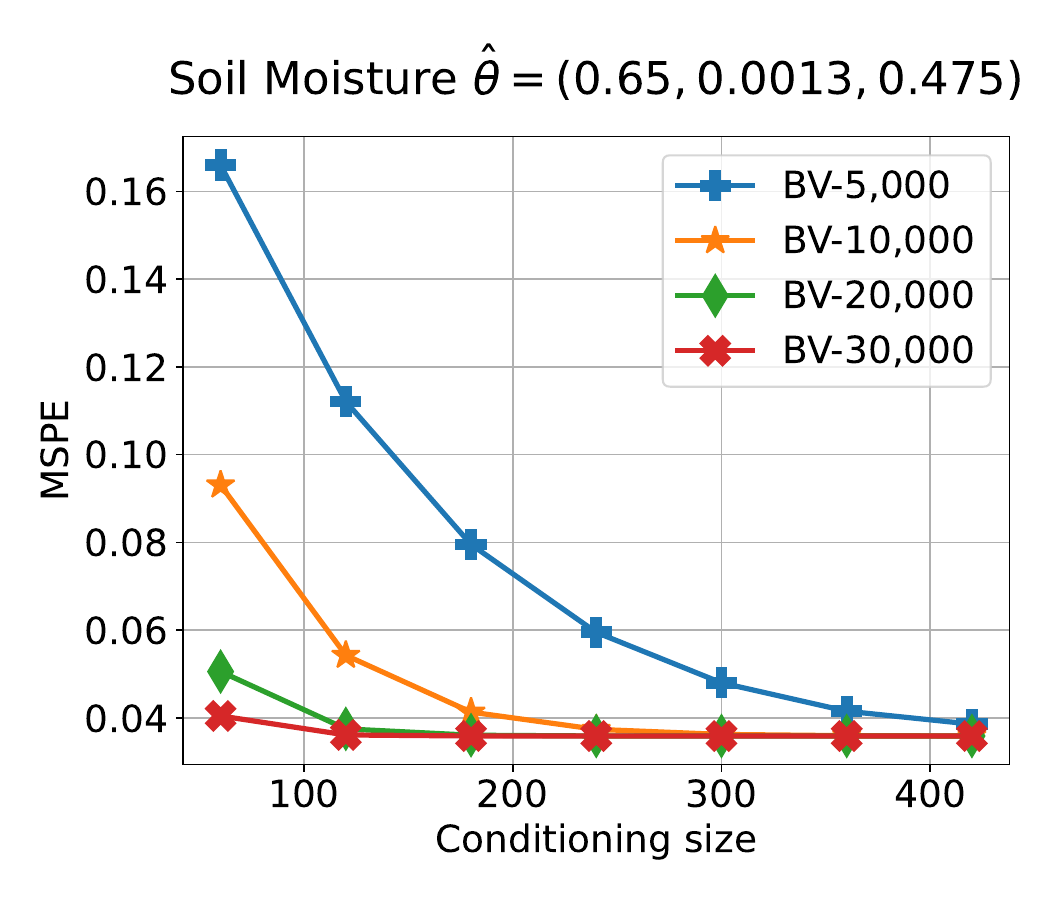}
    \caption{The MSPE of block Vecchia with different block counts.}
    \label{fig:spp:sd-predicted}
\end{figure}

\textcolor{black}{
We also assessed the predictive ability of the block Vecchia method. We plug in the converged estimated parameters $\hat \theta$ into the block Vecchia approximation, then conduct 1000 rounds of the conditional simulations, and finally report the MSPE and the standard deviation. We could derive the same conclusions in Section \ref{sec:realdataset}.
}

\begin{figure}[htbp]
    \centering
    \subfloat[BV-5000 with conditioning size 60]{\includegraphics[width=0.5\linewidth]{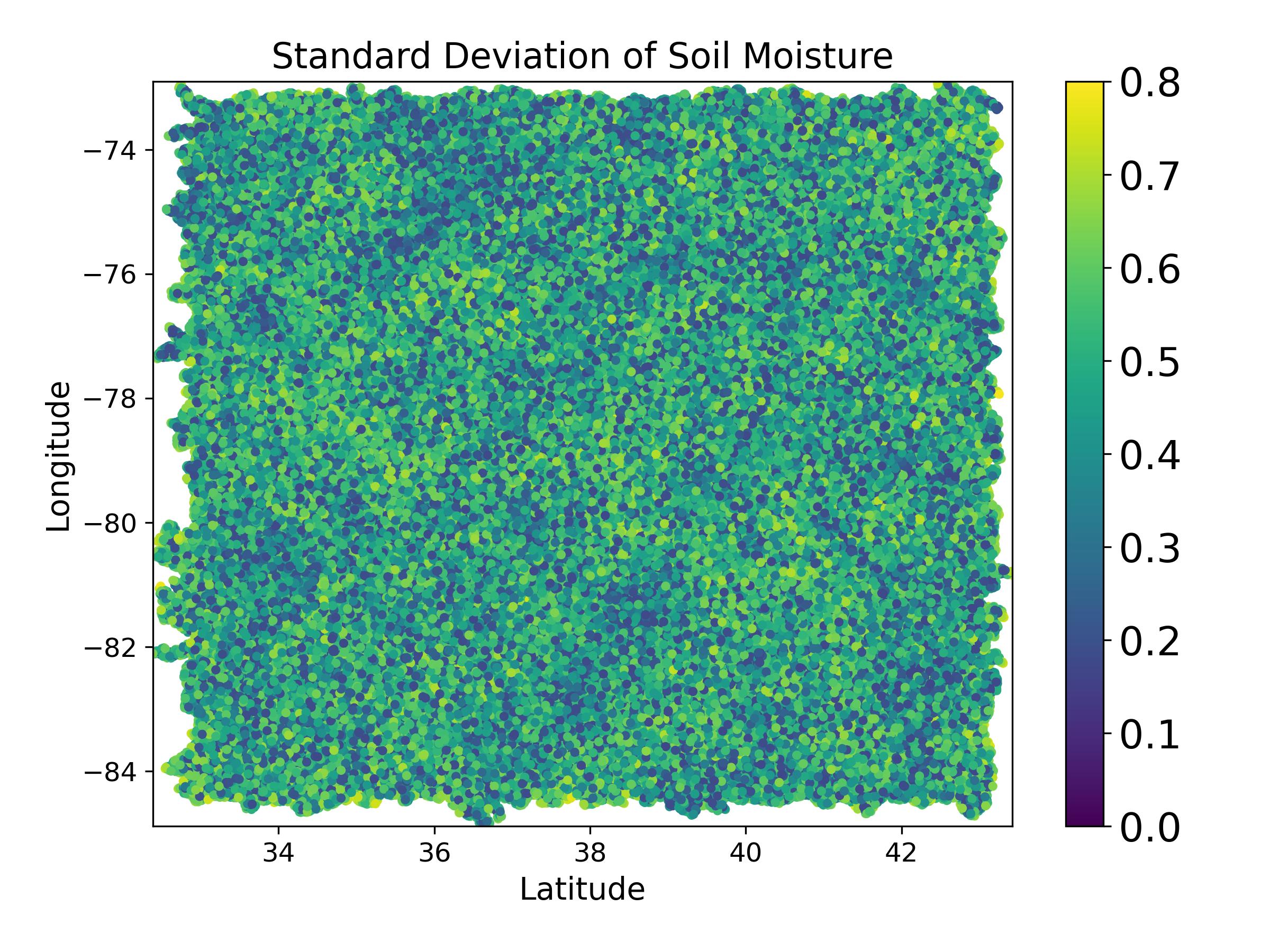}}
    \subfloat[BV-30000 with conditioning size 420]{\includegraphics[width=0.5\linewidth]{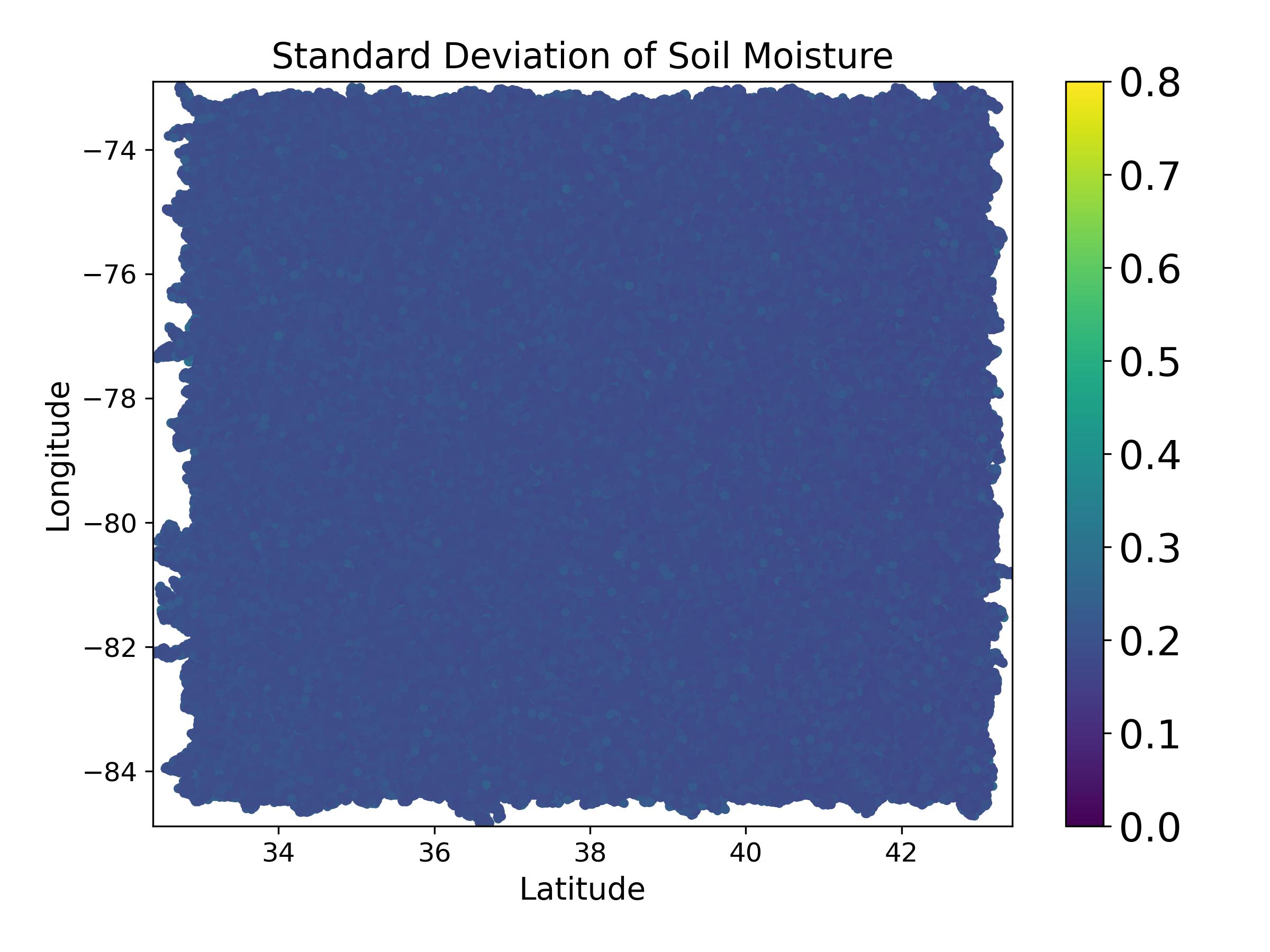}}
    \caption{Prediction standard deviations of conditional simulations for soil moisture.}
    \label{fig:spp:conditional-simu-sd}
\end{figure}

\textcolor{black}{Moreover, Figure \ref{fig:spp:conditional-simu-sd} shows the conditional simulated prediction standard deviations for soil moisture using two block Vecchia setups, with the left plot representing BV-5000 with a conditioning size of 60 and the right plot representing BV-30000 with a conditioning size of 420. It is observed that BV-5000 (conditioning size 60) has higher prediction uncertainty, with standard deviations up to 0.8, and more spatial variability in the error across the domain.
BV-30000 (conditioning size 420) has a lower prediction uncertainty, with standard deviations under 0.1, resulting in more consistent and reliable predictions across the spatial region. This demonstrates the benefit of increasing the block count and the conditioning size in the block Vecchia approximations for soil moisture predictions. This is essential for applications requiring reliable and accurate predictions over large spatial domains.}

\newpage
\subsection{Numerical Study with Increased Block Count: A Small-Scale Study}
\label{spp:numerical-smaller}

\begin{figure}[htbp]
    \centering
    \subfloat[$\beta=0.052537,bc=200$]{\includegraphics[width=0.33\textwidth]{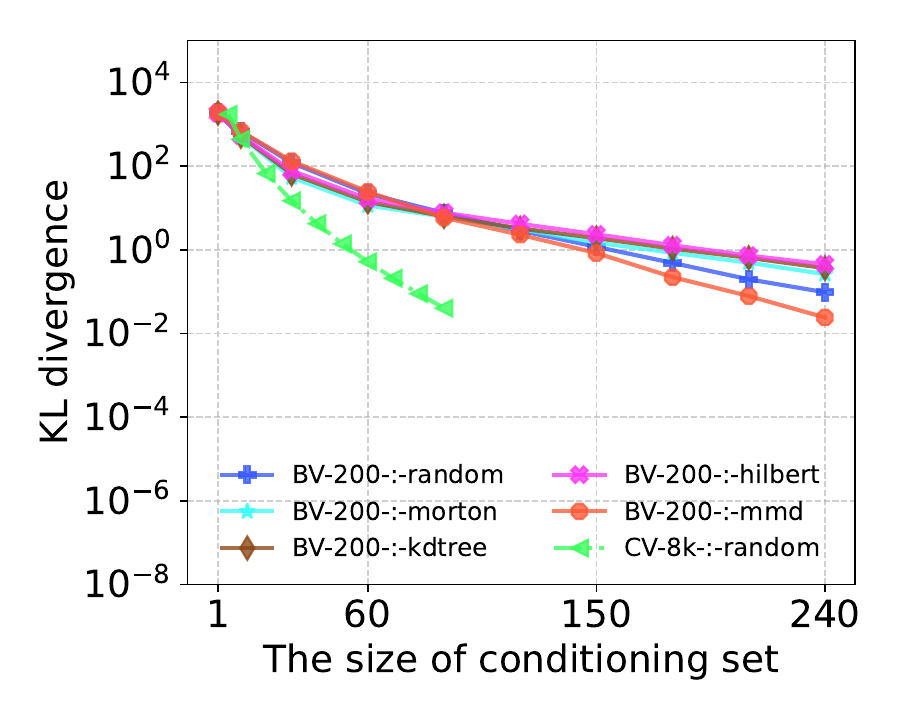}}
    \subfloat[$\beta=0.052537,bc=1500$]{\includegraphics[width=0.33\textwidth]{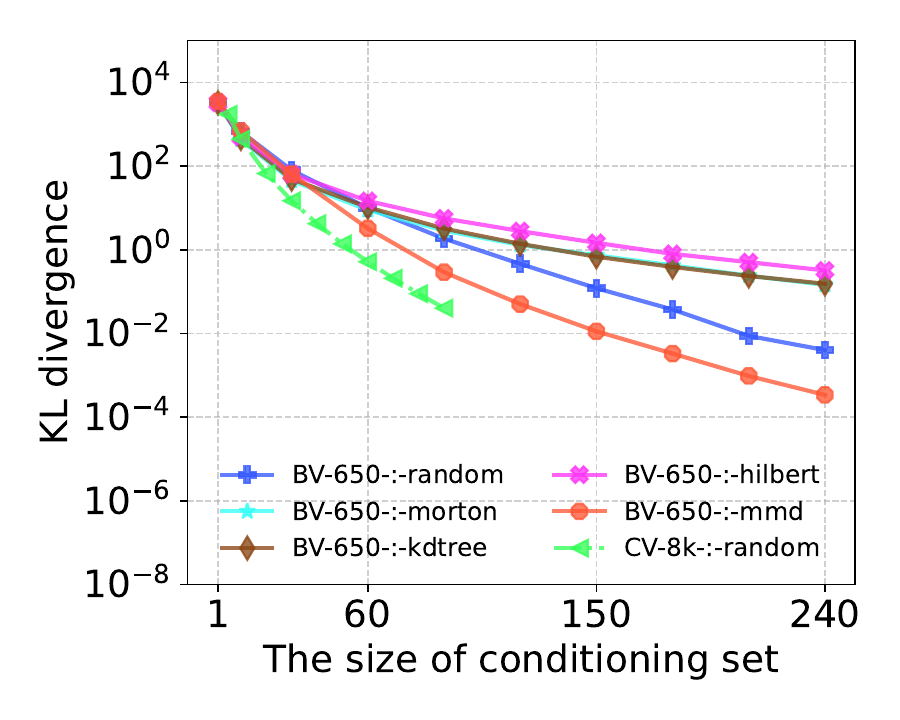}}
    \subfloat[$\beta=0.052537,bc=2500$]{\includegraphics[width=0.33\textwidth]{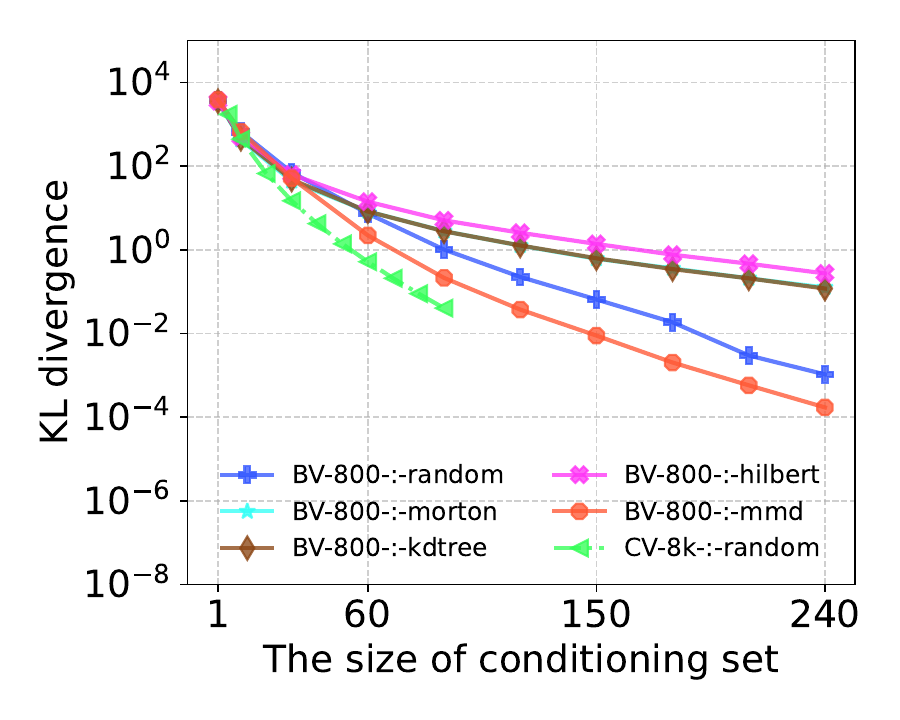}}
    \caption{KL divergence and conditioning size along with different reorderings under medium range/smoothness $\beta=0.052537/\nu=1.5$ and log10 scale.
    }
    \label{fig:20-kl-bc-15-appendix-small}
\end{figure}

\begin{figure}[htbp]
    \centering
    \subfloat[$\beta=0.026270, \nu = 0.5$]{\includegraphics[width=0.33\textwidth]{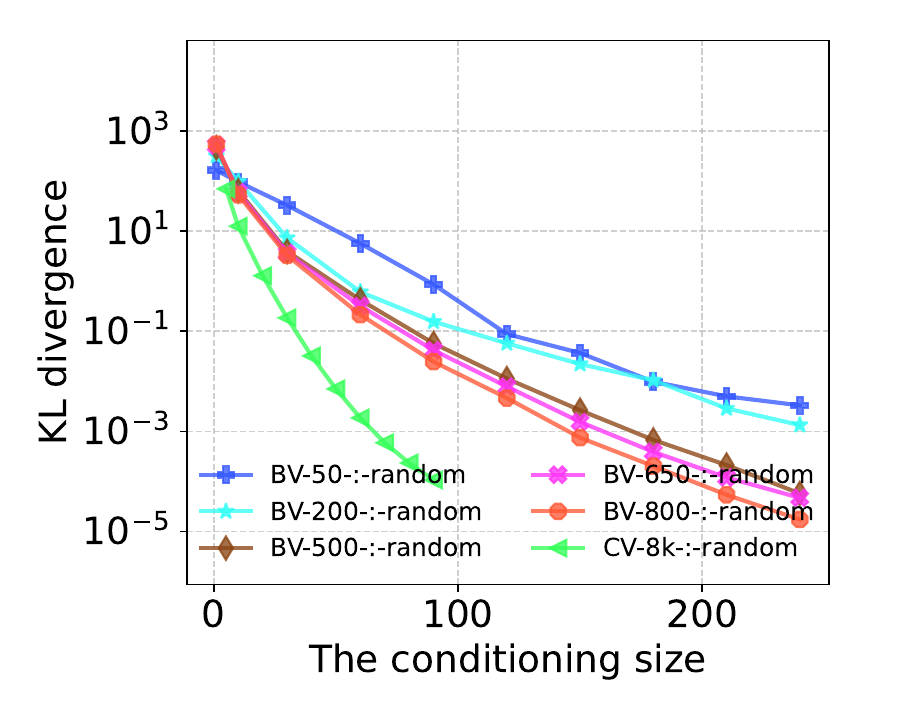}}
    \subfloat[$\beta=0.078809, \nu = 0.5$]{\includegraphics[width=0.33\textwidth]{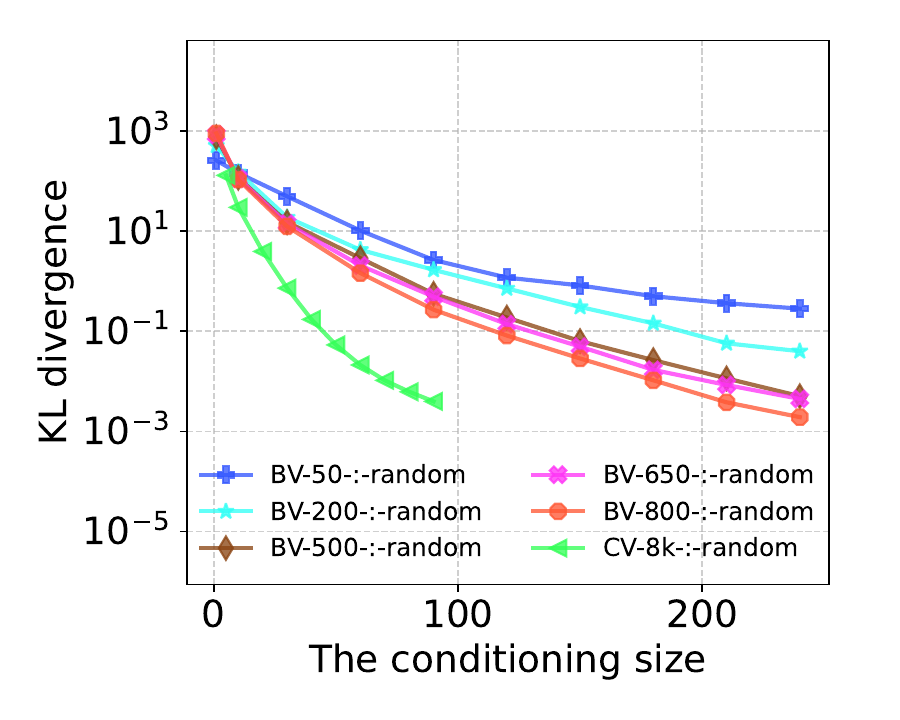}}
    \subfloat[$\beta=0.210158, \nu = 0.5$]{\includegraphics[width=0.33\textwidth]{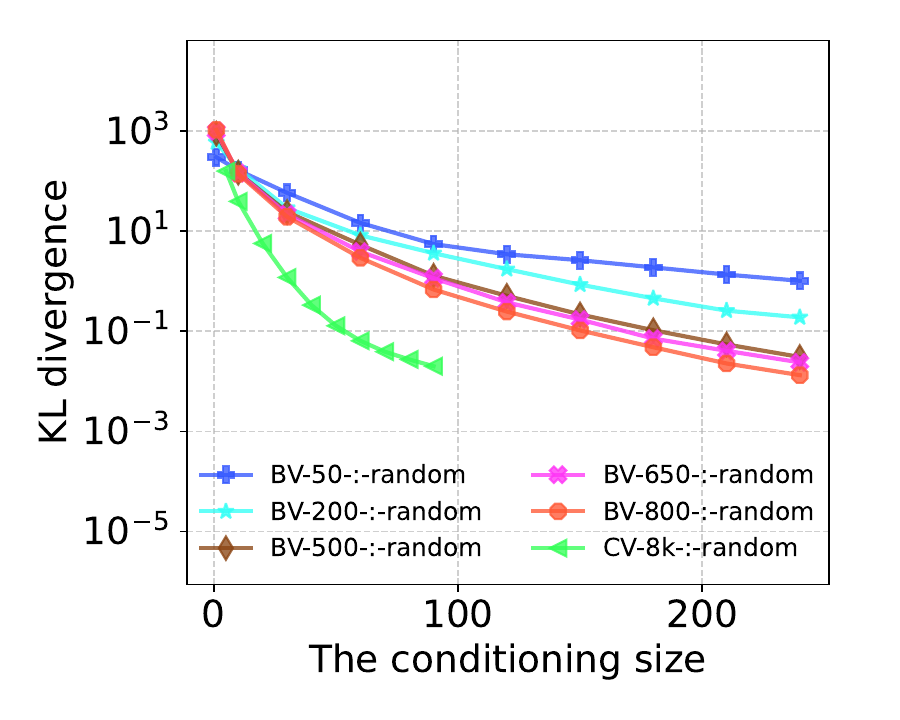}}
    \\
    \subfloat[$\beta=0.017512, \nu = 1.5$]{\includegraphics[width=0.33\textwidth]{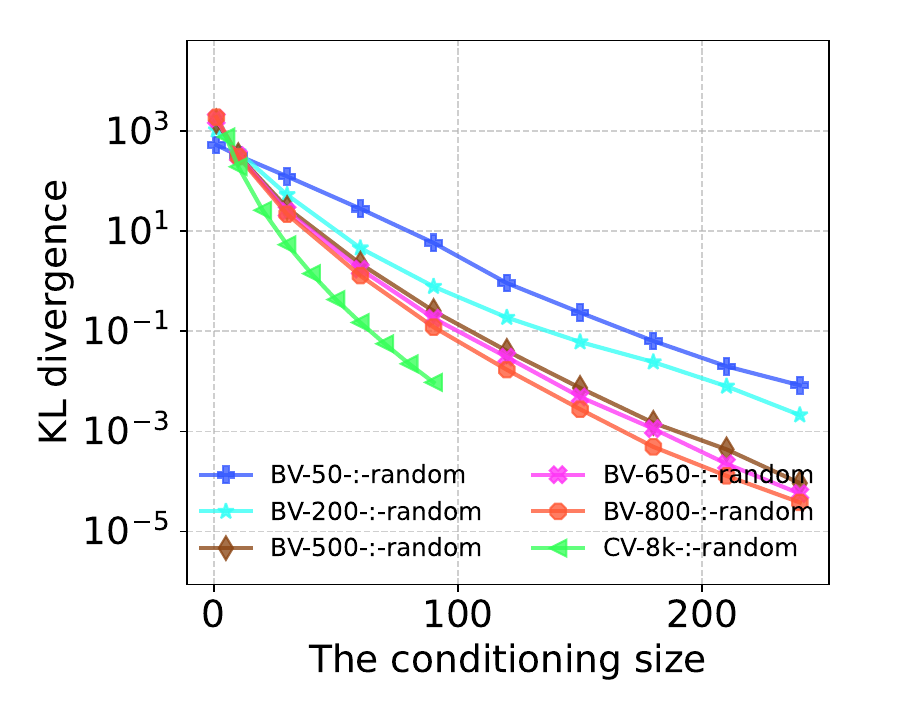}}
    \subfloat[$\beta=0.052537, \nu = 1.5$]{\includegraphics[width=0.33\textwidth]{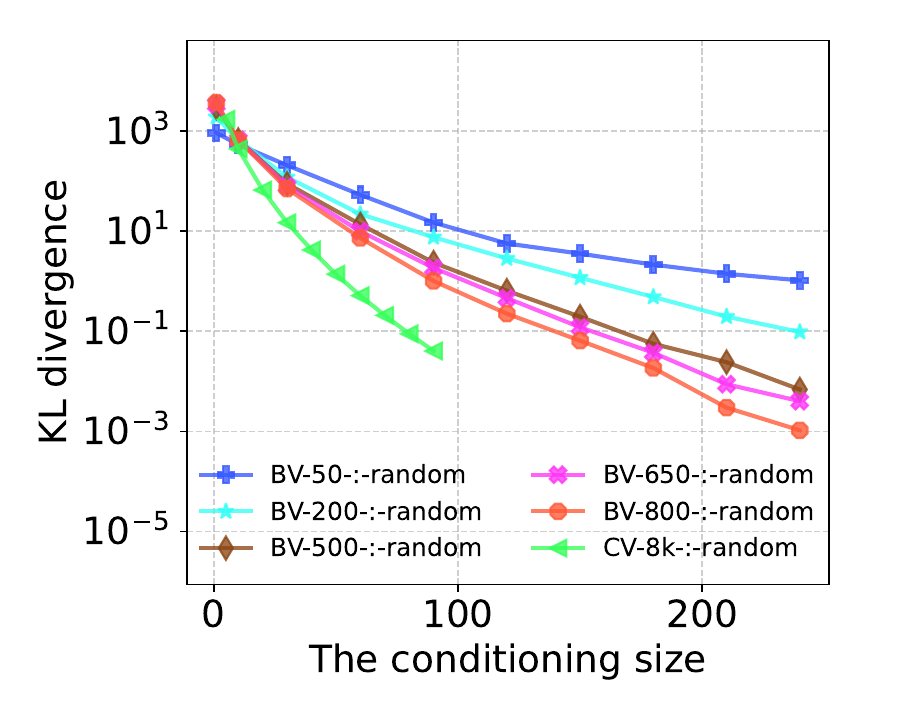}}
    \subfloat[$\beta=0.140098, \nu = 1.5$]{\includegraphics[width=0.33\textwidth]{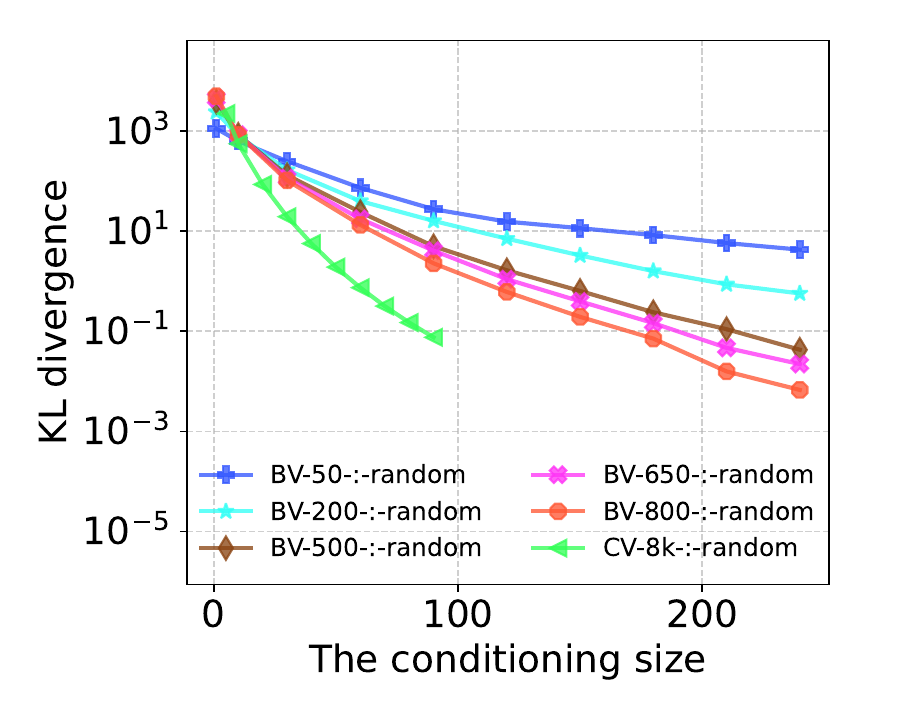}}
    \\
    \subfloat[$\beta=0.014290, \nu = 2.5$]{\includegraphics[width=0.33\textwidth]{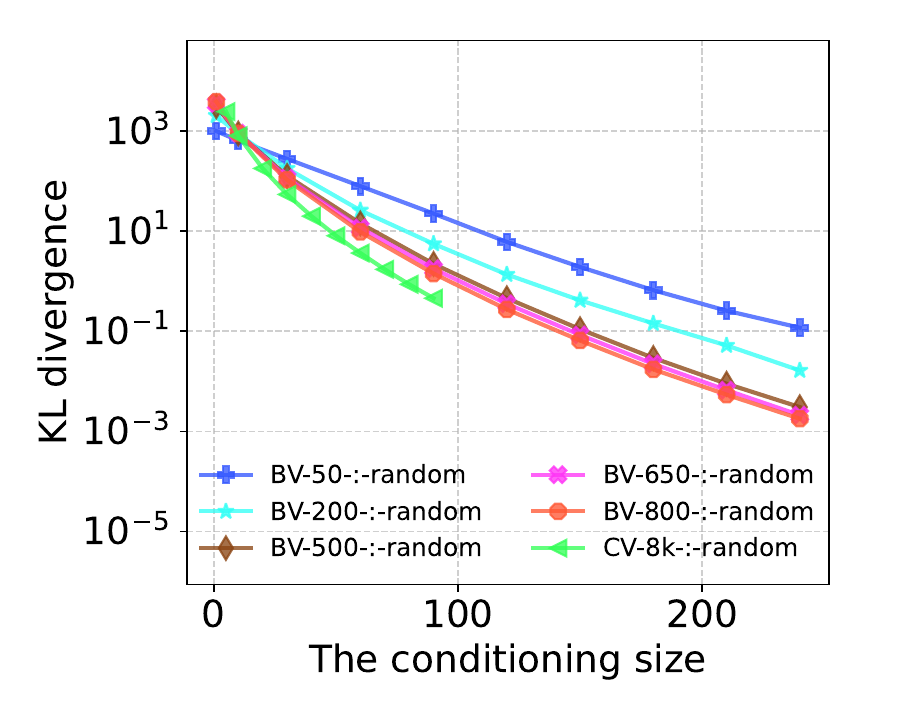}}
    \subfloat[$\beta=0.042869, \nu = 2.5$]{\includegraphics[width=0.33\textwidth]{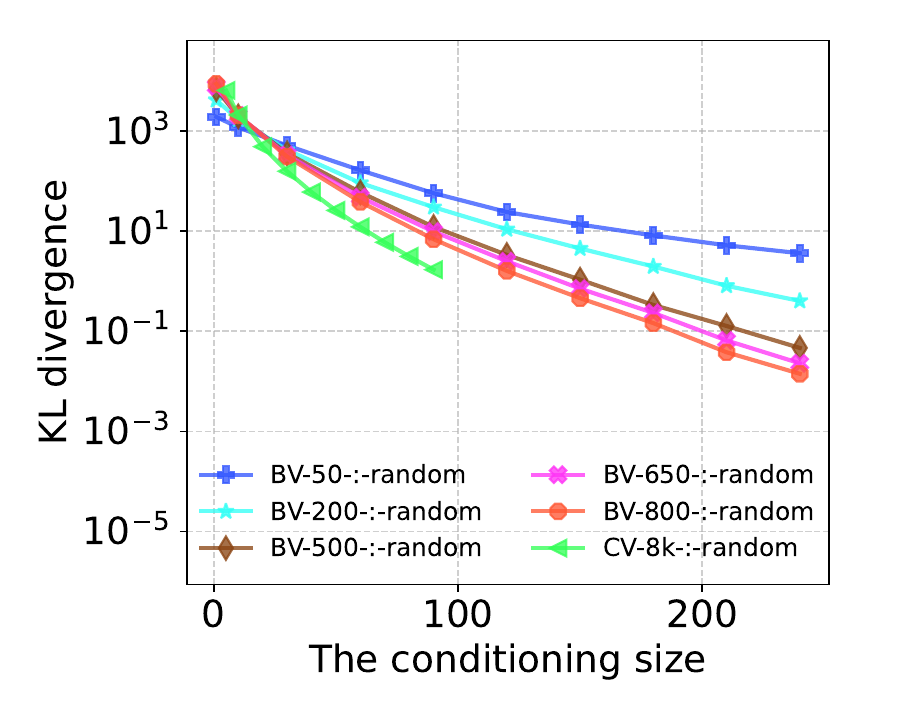}}
    \subfloat[$\beta=0.114318, \nu = 2.5$]{\includegraphics[width=0.33\textwidth]{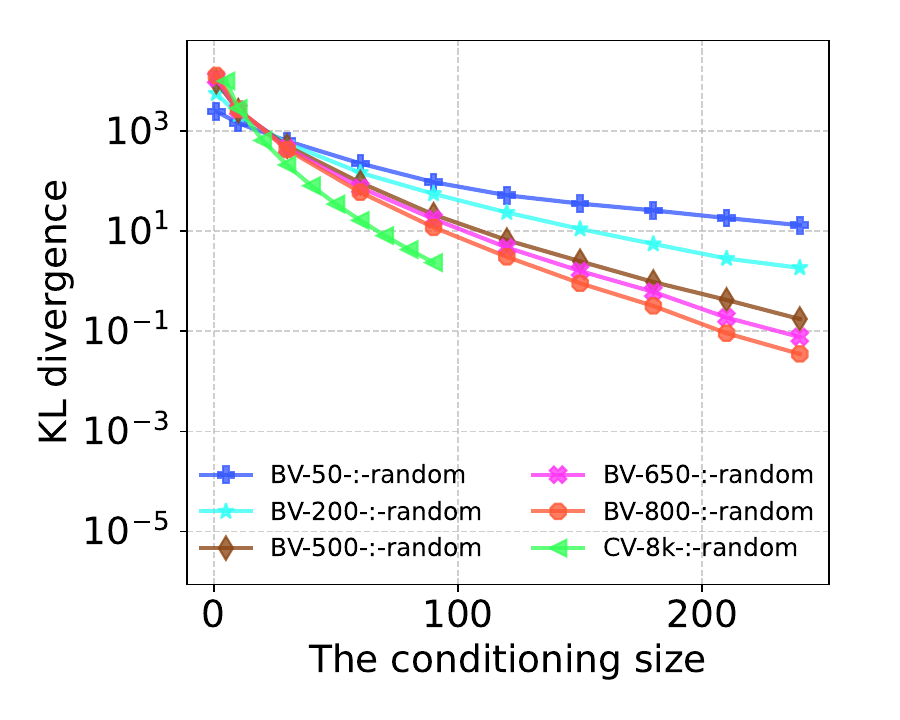}}
    \caption{KL divergence and conditioning size along with increasing block count under log10 scale and random reordering. 
    }
    \label{fig:20-kl-random-appendix-small}
\end{figure}

\textcolor{black}{
We investigate the numerical accuracy of the block Vecchia algorithm, focusing on the impact of block count and different reorderings in a smaller problem, $n=8000$. Figure \ref{fig:20-kl-bc-15-appendix-small} illustrates the KL divergence as the block count increases across three different configurations, with fixed parameters $\beta=0.052537$ and $\nu=1.5$. Our findings indicate that the maxmin reordering (mmd in figures) achieves the highest accuracy, random reordering yields near-optimal accuracy, while other reorderings fail to produce promising results. Additionally, a higher block count improves the accuracy of the block Vecchia method for both random and maxmin reorderings. In Figure \ref{fig:20-kl-random-appendix-small}, we use random reordering as the default, recognizing that it achieves near-optimal accuracy while demanding fewer computational resources than the maxmin reordering. Across various parameter settings, the KL divergence is plotted against different block counts, consistently confirming that a larger block count leads to more accurate results.
}

\newpage
\subsection{Accuracy of Block Vecchia at Varying Smoothness Levels: A Small-Scale Study}
\label{spp:soomthness-smaller}

\begin{figure}[htbp]
    \centering
    \subfloat[$\beta=0.026270, \nu = 0.5$]{\includegraphics[width=0.33\textwidth]{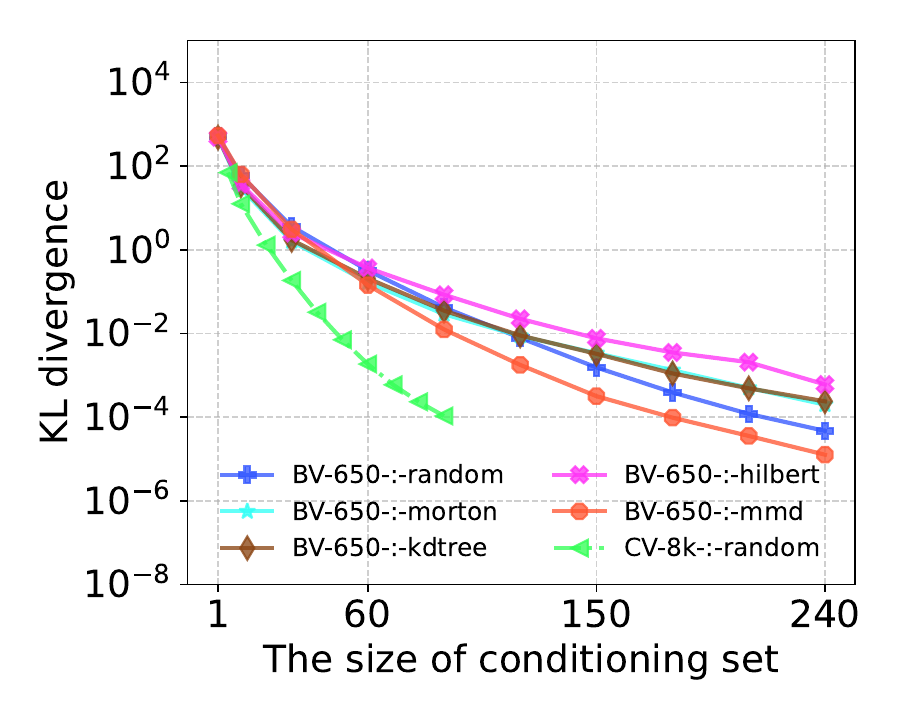}}
    \subfloat[$\beta=0.078809, \nu = 0.5$]{\includegraphics[width=0.33\textwidth]{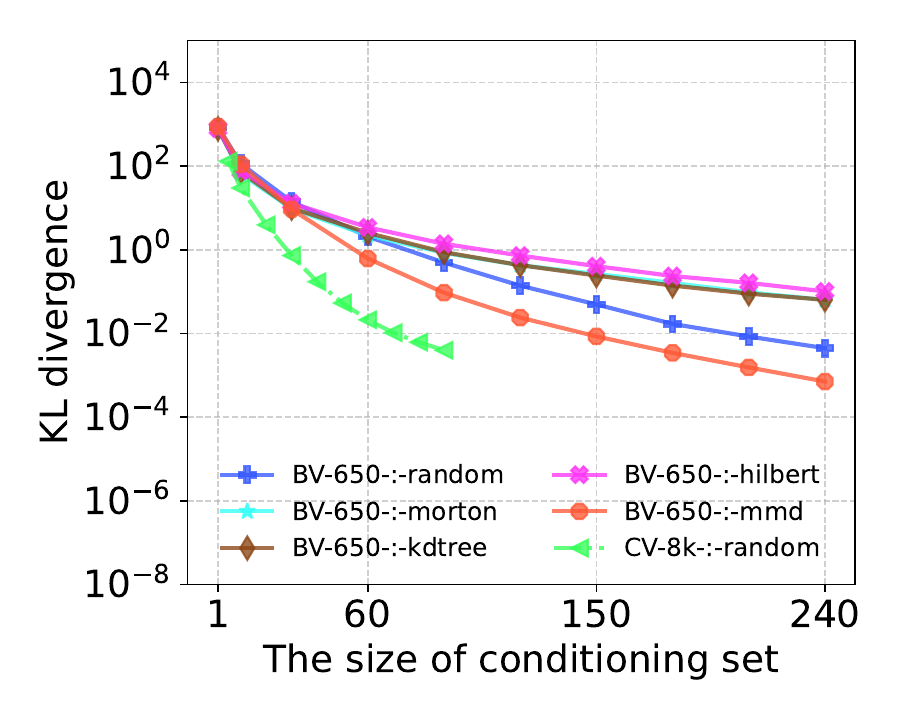}}
    \subfloat[$\beta=0.210158, \nu = 0.5$]{\includegraphics[width=0.33\textwidth]{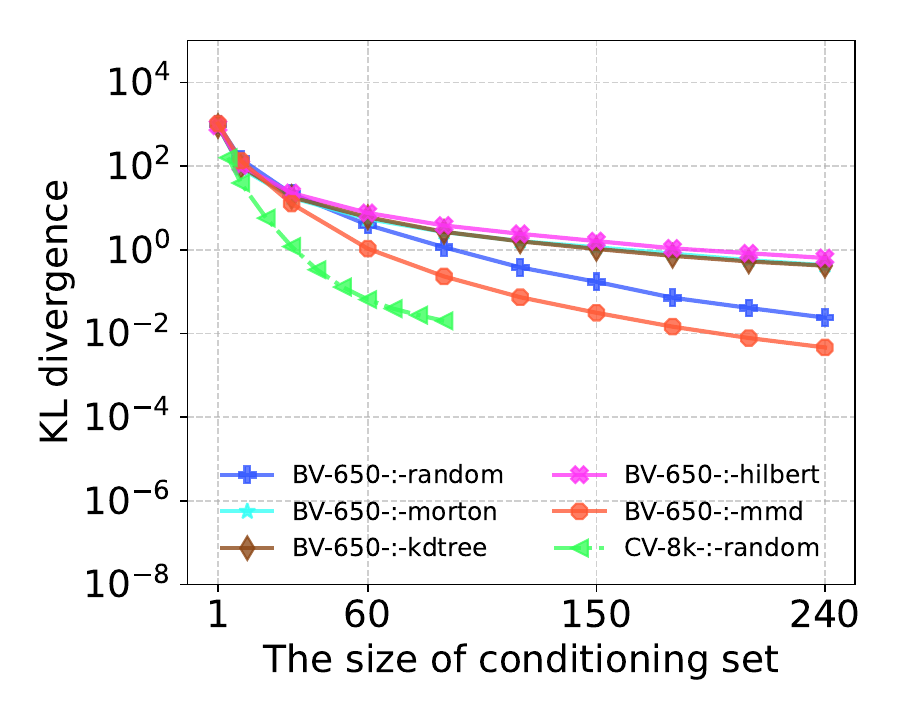}}
    \\
    \subfloat[$\beta=0.017512, \nu = 1.5$]{\includegraphics[width=0.33\textwidth]{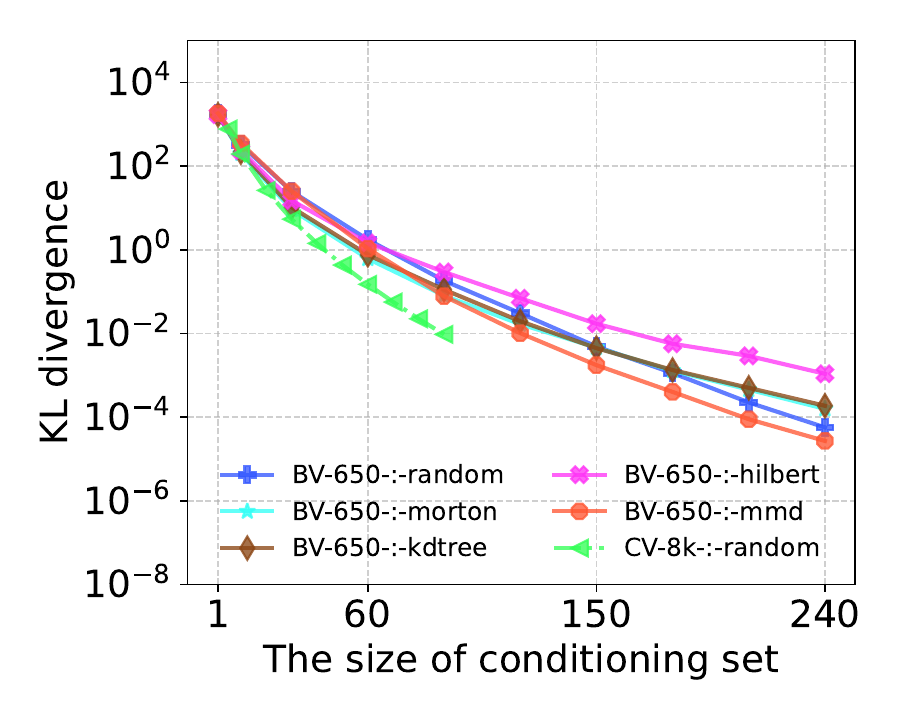}}
    \subfloat[$\beta=0.052537, \nu = 1.5$]{\includegraphics[width=0.33\textwidth]{fig/8k-kl-bc650/0.052537_1.500000.pdf}}
    \subfloat[$\beta=0.140098, \nu = 1.5$]{\includegraphics[width=0.33\textwidth]{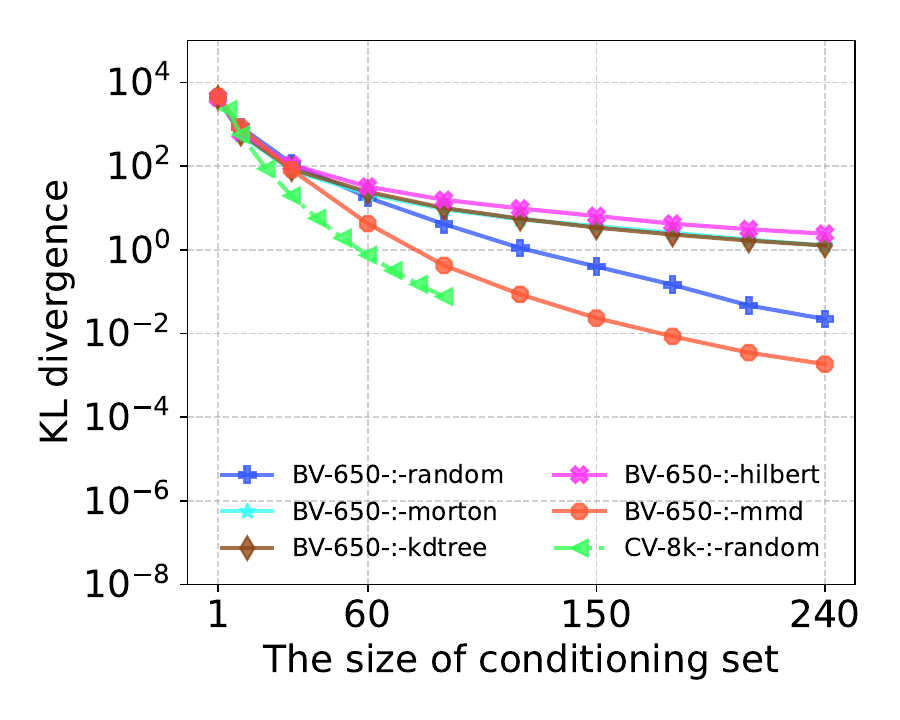}}
    \\
    \subfloat[$\beta=0.014290, \nu = 2.5$]{\includegraphics[width=0.33\textwidth]{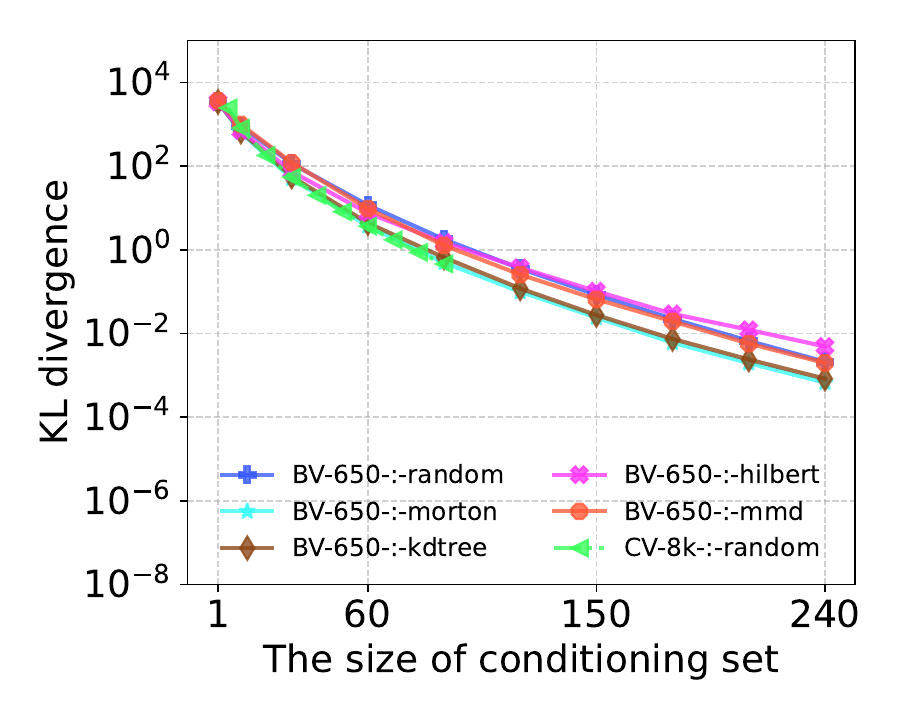}}
    \subfloat[$\beta=0.042869, \nu = 2.5$]{\includegraphics[width=0.33\textwidth]{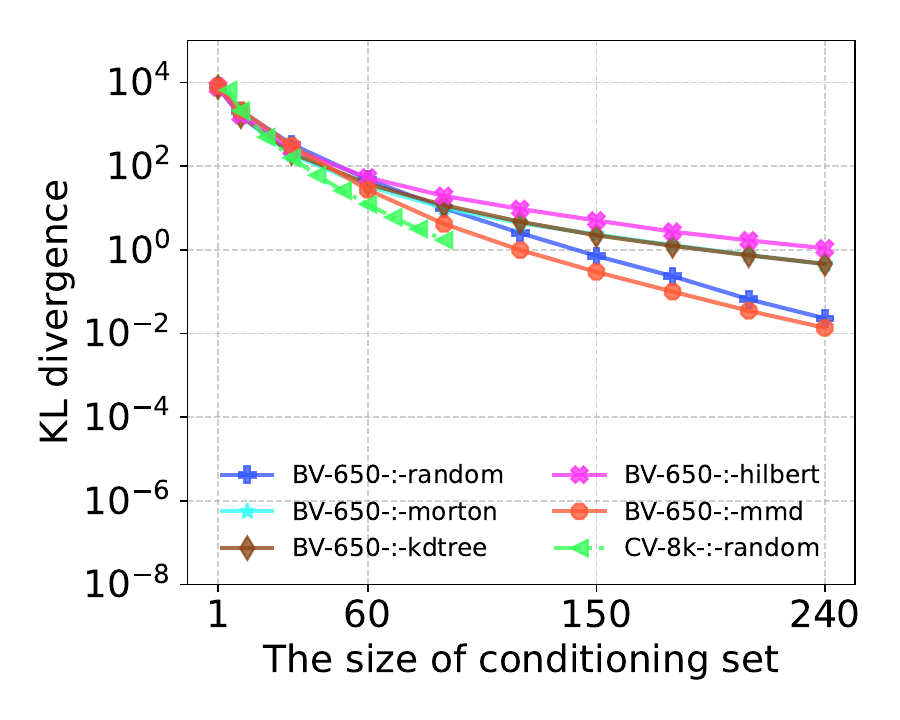}}
    \subfloat[$\beta=0.114318, \nu = 2.5$]{\includegraphics[width=0.33\textwidth]{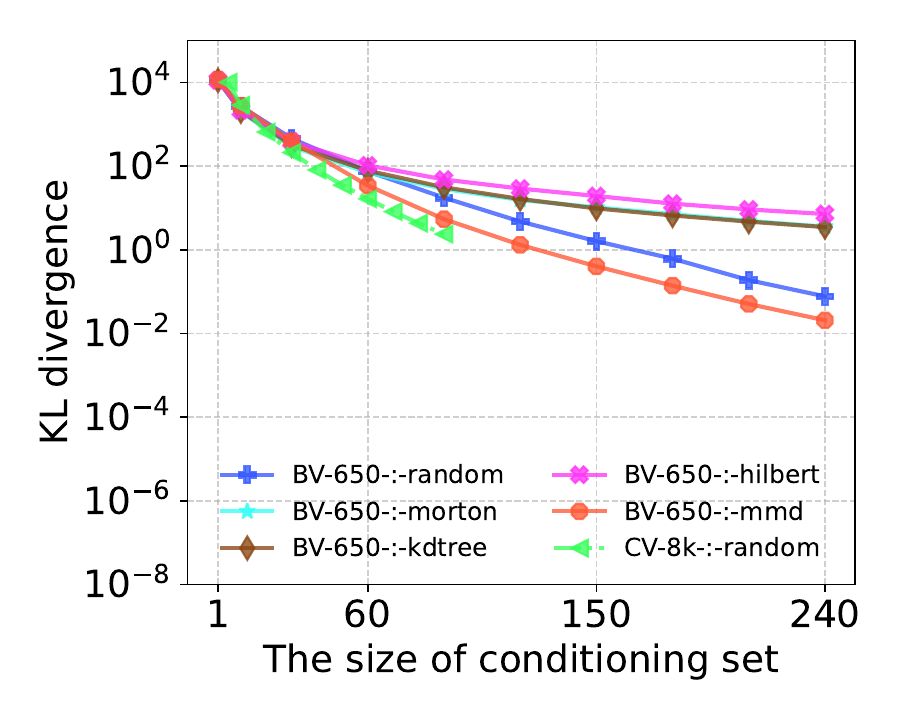}}
    \caption{KL divergence and conditioning size under 8K locations with log10 scale under $bc = 650$. 
    }
    \label{fig:8k-kl-bc650-appendix-small}
\end{figure}

\textcolor{black}{We investigate the numerical accuracy of the block Vecchia algorithm in terms of parameters in the Mat\'ern kernel at $n=8000$. Figure \ref{fig:8k-kl-bc650-appendix-small} indicates that the block Vecchia improves KL divergence as smoothness increases, narrowing the gap between the block Vecchia and classic Vecchia and achieving higher overall accuracy with larger conditioning set.}

\newpage
\subsection{Experiment on Competition Spatial Dataset}
\label{spp:experiments}

\textcolor{black}{\citet{hazra2024exploring} explored various statistical and deep learning methods to address the challenges of analyzing large spatial datasets. They compared the following popular methods: Vecchia approximation, Block Composite Likelihood (BCL), fixed rank kriging (FRK), Gapfill, local approximate Gaussian Process (laGP), Stochastic Partial Differential Equation (SPDE), Nearest Neighbor Gaussian Process (NNGP), and deep learning approaches, e.g., DeepGP and DeepKriging. Their results demonstrated that statistical methods outperformed other approaches across various datasets, particularly BCL, the Vecchia approximation (GpGp), and its zero-mean variant (GpGp0). These methods achieved lower prediction errors and narrower uncertainty intervals. 
We added the proposed BV approximation algorithm to the comparison using only one dataset from this study (dataset 3). We have compared the BV and laGP to the existing methods in \cite{hazra2024exploring}. Specifically, we set the block count as $3000$, the conditioning size as $270$ for the block Vecchia, and the total size of the local designs in the laGP as $60$, the same as the conditioning size in GpGP0. The results are summarized in the last two columns of Table \ref{tab:comparison}, with the results of the first nine models copied from \cite{hazra2024exploring}. The block Vecchia approximation is observed to be faster and achieve accuracy comparable to GpGp0, GpGp, and BCL, with these four methods yielding the best accuracy.  
}


\begin{table}[ht]
\centering
\caption{Model comparison \citep{hazra2024exploring} for the five datasets of Sub-competition 2a \citep{hong2023third} on different metrics.}
\resizebox{\textwidth}{!}{%
\begin{tabular}{lccccccccccc}
\toprule
Models & BCL & FRK & Gapfill & GpGp & GpGp0 & SPDE & NNGP & DeepKriging & DeepGP & laGP & BV\\ \midrule
\textbf{Dataset 1} \\
MSPE & \textbf{0.274} & \textbf{0.274} & 0.282 & \textbf{0.274} & \textbf{0.274} & \textbf{0.274} & 0.278 & 0.318 & 0.317 & 0.277 & \textbf{0.274} \\
MAPE & \textbf{0.417} & \textbf{0.417} & 0.424 & \textbf{0.417} & \textbf{0.417} & \textbf{0.417} & 0.420 & 0.449 & 0.448 & 0.420 & \textbf{0.417}\\
PICP & 0.951 & 0.951 & 0.964 & 0.953 & 0.953 & 0.951 & 0.950 & 0.939 & 0.961 & 0.942 & 0.951\\
MPIW & {2.049} & 2.050 & 2.258 & 2.075 & 2.074 & 2.050 & 2.058 & 2.144 & 2.310 & 2.023 & 2.056 \\
Time (s) & 372 & 209 & 8 & 547 & 595 & 10 & 618 & 392 & 2197 & 158 & 462 \\ \hline
\textbf{Dataset 2} \\
MSPE & \textbf{0.274} & 0.276 & 0.347 & 0.275 & 0.275 & 0.276 & 0.279 & 0.313 & 0.297 & 0.278 & \textbf{0.274}\\
MAPE & \textbf{0.417} & 0.419 & 0.470 & 0.418 & 0.418 & 0.419 & 0.421 & 0.441 & 0.421 & 0.420 & \textbf{0.417} \\
PICP & 0.951 & 0.950 & 0.983 & 0.953 & 0.952 & 0.951 & 0.952 & 0.942 & 0.961 & 0.944 & 0.951\\
MPIW & {2.053} & {2.058} & 3.246 & {2.076} & {2.076} & {2.058} & {2.076} & 2.159 & 2.330 & 2.040 & 2.059 \\
Time (s) & 403 & 156 & 8 & 510 & 478 & 8 & 568 & 393 & 1951 & 159 & 377 \\ \hline
\textbf{Dataset 3} \\
MSPE & \textbf{0.297} & 0.489 & 0.983 & \textbf{0.297} & \textbf{0.297} & 0.419 & 0.298 & 0.380 & 0.829 & 0.337 &\textbf{0.297} \\
MAPE & \textbf{0.434} & 0.557 & 0.788 & \textbf{0.434} & \textbf{0.434} & 0.515 & 0.436 & 0.490 & 0.721 & 0.462 & \textbf{0.434} \\
PICP & 0.951 & 0.951 & 0.974 & 0.952 & 0.952 & 0.949 & 0.954 & 0.951 & 0.963 & 0.946 & 0.951\\
MPIW & {2.137} & 2.744 & 4.391 & {2.147} & {2.147} & 2.543 & {2.177} & 2.418 & 3.830 & 2.265 & {2.143}\\
Time (s) & 555 & 129 & 8 & 514 & 521 & 9 & 626 & 381 & 2233 & 163 & 380 \\ \hline
\textbf{Dataset 4} \\
MSPE & \textbf{0.275} & 0.277 & 0.346 & \textbf{0.275} & \textbf{0.275} & 0.277 & 0.279 & 0.319 & 0.296 & 0.278 & \textbf{0.275} \\
MAPE & \textbf{0.418} & 0.420 & 0.468 & \textbf{0.418} & \textbf{0.418} & 0.419 & {0.421} & 0.447 & 0.433 & 0.420 & \textbf{0.418}\\
PICP & 0.951 & 0.951 & 0.967 & 0.953 & 0.953 & 0.951 & 0.951 & 0.951 & 0.955 & 0.944 & 0.951 \\
MPIW & {2.054} & {2.062} & 2.556 & {2.077} & {2.077} & 2.060 & 2.071 & 2.141 & 2.200 & 2.040 & {2.062} \\
Time (s) & 475 & 145 & 8 & 528 & 494 & 8 & 630 & 379 & 1759 & 173 & 273\\ \hline
\textbf{Dataset 5} \\
MSPE & \textbf{0.302} & 0.540 & 1.004 & \textbf{0.302} & \textbf{0.302} & 0.455 & {0.304} & 0.409 & 0.956 & 0.358 & \textbf{0.302}\\
MAPE & \textbf{0.438} & 0.586 & 0.797 & \textbf{0.438} & \textbf{0.438} & 0.537 & 0.440 & 0.509 & 0.779  & 0.477 & \textbf{0.438}\\
PICP & 0.951 & 0.950 & 0.976 & 0.951 & 0.952 & 0.951 & 0.955 & 0.962 & 0.964 & 0.943 & 0.951 \\
MPIW & {2.154} & 2.885 & 4.576 & {2.162} & {2.162} & 2.648 & {2.206} & {2.493} & 4.088 & 2.316 & {2.160}\\
Time (s) & 340 & 116 & 8 & 565 & 585 & 7 & 648 & 380 & 1630 & 169 & 323 \\ 
\bottomrule
\end{tabular}%
}
\label{tab:comparison}
\end{table}

\newpage
\subsection*{Code Archive}

The Block Vecchia algorithm provides a Github repository, https://github.com/paper-code1/BV-Gaussian. You can follow the README.md to install the package and reproduce the results in the paper. 

\end{document}